\documentclass[10pt, letterpaper]{article}
\usepackage[protrusion=true,expansion=true]{microtype}
\usepackage[margin=0.9in]{geometry}
 \pdfoutput=1

\usepackage{amsmath,amsthm} 
\usepackage{authblk}

\usepackage{color}
\usepackage{epsfig}
\usepackage{algorithm}
\usepackage[noend]{algorithmic}
\usepackage{multirow}
\usepackage{xspace}
\usepackage{graphicx,subfigure}
\usepackage{wrapfig}
\usepackage{color}
\usepackage[normalem]{ulem}

\newcommand{\spara}[1]{\smallskip\noindent{\bf #1.}}
\newcommand{\para}[1]{\smallskip\noindent{\bf #1.}}

\newcounter{packednmbr}

\newenvironment{packeditemize}{\begin{list}{$\bullet$}{\setlength{\itemsep}{0pt}\addtolength{\labelwidth}{0pt}\setlength{\leftmargin}{\labelwidth}\setlength{\listparindent}{\parindent}\setlength{\parsep}{3pt}\setlength{\topsep}{0pt}}}{\end{list}}

\newtheorem{property}{Property}
\newtheorem{theorem}{Theorem}

\newtheorem{example}{Example}
\newtheorem{lemma}{Lemma}

\newcommand{\BigO}[1]{\ensuremath{\operatorname{O}(#1)}}
\newcommand{\E}{\ensuremath{\operatorname{E}}}

\newcommand{\eat}[1]{}
\renewcommand{\sout}[1]{}

\newcommand{\IC}{IC model\xspace}
\newcommand{\nips}{ConTinEst\xspace}
\newcommand{\dtime}{DNP\xspace}
\newcommand{\model}{CNP\xspace}
\newcommand{\algo}{ShardSampling\xspace}

\newlength{\figwidth} 
\newlength{\figfour} 
\newcommand{\techrep}[2]{#1}

\begin{document}

\title{Modeling Non-Progressive Phenomena for Influence Propagation}

 \eat{
\numberofauthors{3} 
\author{
\alignauthor
Vincent Yun Lou\\
       \affaddr{Stanford University}\\
       \email{yunlou@stanford.edu}
\alignauthor
Smriti Bhagat\\
       \affaddr{Technicolor, Palo Alto}\\
       \email{smriti.bhagat@technicolor.com}
\alignauthor
Laks V. S. Lakshmanan Sharan Vaswani\\
       \affaddr{University of British Columbia}\\
              \email{laks@cs.ubc.ca} 
}

\numberofauthors{1}
\author{
\begin{tabular}{cclr}
Vincent Yun Lou & Smriti Bhagat & Laks V.S. Lakshmanan & Sharan Vaswani \\
\affaddr{Stanford University} & \affaddr{Technicolor} & \multicolumn{2}{c}{\affaddr{University of British Columbia} } \\
\affaddr{yunlou@stanford.edu} & \affaddr{smriti.bhagat@technicolor.com} & \multicolumn{2}{c}{\affaddr \{laks,sharanv\}@cs.ubc.ca}
\end{tabular}
}}
\author{
Vincent Yun Lou$^\dagger$, Smriti Bhagat$^\S$, Laks V.S. Lakshmanan$^\ddagger$, Sharan Vaswani$^\ddagger$ \\
$^\dagger$ Stanford University, $^\S$ Technicolor , $^\ddagger$University of British Columbia \\
$^\dagger$ yunlou@stanford.edu, $^\S${smriti.bhagat@technicolor.com} , $^\ddagger$\{laks,sharanv\}@cs.ubc.ca
}

\date{}

\maketitle
\begin{abstract}
Recent work on modeling influence propagation focus on progressive models, i.e., 
once a node is influenced (active) the node stays in that state and cannot become
 inactive. However, this assumption is unrealistic in many settings where nodes 
 can transition between active and inactive states. For instance, a user of a social 
 network may stop using an app and become inactive, but again activate when 
 instigated by a friend, or when the app adds a new feature or releases a new version. 
In this work, we study such non-progressive phenomena and propose an efficient 
model of influence propagation. Specifically, we model influence propagation 
as a continuous-time Markov process 
with 2 states: active and inactive. Such a model is both highly scalable (we evaluated on 
graphs with over 2 million nodes), 17-20 times faster, and more accurate for estimating the spread
 of influence, as compared with state-of-the-art progressive 
 models for several applications where nodes may switch states.
\end{abstract}

\eat{
\category{H.2.8}{Database Applications}[data mining] 
\category{J.4}{Social and Behavioral Sciences}[Sociology] 

\terms{Algorithms, Performance, Experimentation}
}

\makeatletter{}
\section{Introduction}
\label{sec:intro}

Study of information and influence propagation over social networks has 
attracted significant research interest over the past decade, driven by applications such as viral
marketing \cite{kempe03,Domingos}, social feed ranking \cite{Schenkel}, 
contamination detection~\cite{Leskovec07,Ostfeld04,Ostfeld06},
and spread of innovation \cite{Valente05} to name a few. A prototypical problem that has received wide
attention is \emph{influence maximization}: given a social network along with pairwise influence
probabilities between peers, and a number $k$, find $k$ seed nodes such that
activating them at start will eventually lead to the largest number of activated nodes in
the network in the expected sense. Following the early work of Domingos and Richardson
\cite{Domingos} and Kempe et al. \cite{kempe03}, there has been a burst of activity in this
area (e.g., see \cite{Budak,Chen09,Chen10,Goyal10,Rodriguez,Du13,Goyal11}). 
While the majority of previous studies employ propagation models with discrete time, 
in recent work, continuous time models have been shown to be more accurate 
at modeling influence propagation phenomena~\cite{Du13,Rodriguez,Du13Arxiv}.
We refer the reader to the book \cite{chen-etal-2013} for a comprehensive survey and a detailed 
discussion of recent advances in influence maximization. 

As discussed in
\cite{kempe03}, the propagation models can be classified into \emph{progressive} and
\emph{non-progressive} (NP) models. In progressive models, an inactive node can
become active, but once active, a node cannot become inactive. Non-progressive
models relax this restriction and allow nodes to repeatedly transition between active and
inactive states. 

Indeed, an overwhelming majority of studies of information propagation
have confined themselves to progressive models. 
For applications such as \eat{adopting innovation or }
buying a product, the progressive assumption makes perfect sense: buying a product 
is not easily reversible in many cases. 
On the other hand, there are real applications which are not naturally captured by progressive models. For example, consider a user adopting a mobile app. Over time, its appeal may fade and her usage of the app may decline over time. Her interest in the app may be rejuvenated by a friend telling her about a new cool feature being added to the app at which point, she decides to try the app again and may continue using it once again. Alternatively, whenever a new version of the app is released, the user feels tempted to try it again and may, with some probability, decide to continue using it again. 
As a second example, it is well known that fashion follows cycles. Choices that are in fashion at the moment may fall out of fashion and may again become fashionable in the future, as it has been recognized that social choices follow cyclic trends \cite{sreenivas-boredom-wsdm-2013}.
\eat{However, settings where customers subscribe to one of several
competing service providers, and have the opportunity and incentive to switch between
providers cannot be accurately modeled using progressive models. 
As a concrete example, a customer may switch from a mobile carrier
such as Sprint to AT\&T since she finds the calling plans offered by the latter
more suited to her needs. When she switches subscription from Sprint to another
company, she becomes inactive from Sprint's perspective. If she switches back to
Sprint at a later time, she becomes active again. 
Similar situations abound in the context of other service providers 
such as internet, cable etc. 
It is quite common for people to talk to family and friends and find out about better rates or
plans or options offered by a competing provider compared to their own, based on
which they may decide to switch subscriptions. Modeling
the propagation of influence, or more precisely the spread of adoption or
subscription in such applications calls for non-progressive
models. 
since from the perspective of any one provider, users can start or
terminate subscriptions any time. }As a third example, there are many applications where users may become active and stay in that 
state for a period of time before deactivating, such as, 
adopting a feature on a content sharing site where the feature 
may be the ``like'' or ``favorite'' button for a post, filters (sepia, sketch, outline) for photo
editing, or ``check-in'' to a location or a show. Finally, in epidemiology, it is well known that an infected person may recover from a disease but not necessarily acquire lifelong immunity from the disease, thus being susceptible to the disease. In all the above examples, the phenomena in question are subject to spreading via influence. As we will show with experiments on real datasets in this paper, the use of progressive models for capturing such   phenomena leads to considerable error. There is a clear need for a non-progressive model for studying these phenomena. 

\techrep{We next briefly review related work on non-progressive models. A more detailed comparison 
appears in the next section. }{}In their seminal paper, Kempe et al. \cite{kempe03} propose a non-progressive model 
and show that it can be reduced to a progressive model by replicating each 
node for every timestamp in the time horizon under consideration, 
and connecting each node to its neighbors in the previous timestamp. They show  
that this reduction preserves equivalence, which implies all techniques
developed for progressive models can in principle be applied to non-progressive
models. However, replicating a large network for each timestamp over a large
time  horizon will clearly make this approach impractical for large social
networks containing millions of nodes. Thus, this approach is largely of theoretical interest. 
\techrep{Fazli et al.~\cite{fazli12}  study a simple non-progressive model based
on deterministic linear threshold, where the threshold is  given by strict
majority. Their focus is on finding a minimum perfect target set, i.e., a set of
seeds of the smallest size which leads to the eventual activation of all network
nodes. The goals and approaches are considerably different from those of this paper.
In epidemic modeling, the SIRS (Susceptible-Infected-Recovered-Susceptible)
model \cite{prakash2012threshold, ganesh2005effect} follows the non-progressive paradigm, and explicitly allows for recovered patients to remain susceptible to the disease and become infected again. In the economics literature, there have studies~\cite{Blume,Ellison,Jackson} 
on non-progressive
models. A detailed comparison with these and other related works appears in the next section. }{}

Another related area is \emph{competitive} influence maximization, where competing parties choose seed nodes in order to maximize the adoption of their product or opinion \cite{chen-etal-2013}. Non-progressiveness arises naturally from the perspective of any one party involved in the competition. Our focus in this paper is not competition. As illustrated above, there are several example applications where propagation of information or influence happens in a non-progressive manner and it is our goal to model and study them in this paper.

Influence maximization is known to be a computationally hard problem, even over
the relatively simpler progressive models.
\eat{There are two sources of
complexity: (a)\eat{ for IC and LT models,} computing the expected number of active
nodes, called expected spread in the literature \cite{kempe03}, given a seed set is
\#P-hard, and (b) even assuming the expected spread can be computed in constant
time, the problem of seed selection is NP-hard. By exploiting the fact that the
objective function, the expected spread, is monotone and submodular, Kempe et
al. \cite{kempe03} showed that the simple greedy algorithm of repeatedly picking
the seed that yields the maximum marginal gain guarantees a
$(1-1/e)$-approximation to the optimum. 
They advocated the use of Monte Carlo
simulation for estimating the expected spread, which can be done to an arbitrarily
high degree of precision, but tends to be extremely slow. In subsequent work,
researchers proposed various heuristics and alternative approaches \cite{Chen10,Simpath} and
showed that they scale to very large data sets while providing a quality
comparable to that of the greedy algorithm coupled with Monte Carlo simulation on
many real data sets on which they were tested. }We don't expect influence
maximization to be easier over non-progressive models. 
We face the 
challenge, {\sl whether we can design approximation algorithms for influence
maximization over non-progressive models that scale to large data sets}. 
To this end, we first propose a discrete time non-progressive model called DNP. It will turn out that DNP, while accurate at modeling non-progressive phenomena, does not lead to a scalable solution for estimating influence spread. 
To mitigate this, \techrep{we draw inspiration from the recent observations that continuous time 
models lead to greater accuracy in predicting node 
activations~\cite{Du13,Rodriguez},  and}{} we 
propose a {\em continuous time non-progressive model} (\model), 
which models the underlying influence propagation as a Markov process. 
This model can also capture progressive phenomena by appropriately setting the model parameters. We call this variant CNP-Progressive (CP for short). It is interesting to investigate how CP compares with the state-of-the-art progressive continuous time models such as~\cite{Du13,Rodriguez}. 
\eat{As an alternative, we design a discrete time counterpart, called the DNP 
model, and show that the \model model allows for far more efficient node 
activation prediction that the DNP model. } 

A second challenge centers on the question, what should 
the objective be when
selecting seeds with respect to non-progressive models. As 
opposed to maximizing the {\sl number} of active nodes at some time, 
as done in progressive models, 
we argue that it is more appropriate to maximize the {\sl expected time during which nodes may have
been active}. 
\eat{From a service provider's perspective, it makes good sense to
choose seeds so as to optimize the total expected time during which nodes are
active as this is more directly related to its revenue. } 

\eat{More precisely, should we choose
seeds so that the expected number of nodes that are active at the last timestamp
in the horizon is maximum? Or should it be the expected number of nodes that
are active at any time in the time horizon?  Both of these objective functions
ignore} 

Finally, while example applications demonstrating the value of and need for
non-progressive models exist, to date, no empirical studies have compared non-progressive
models with their progressive counterparts with an aim of calibrating their accuracy for explaining propagation phenomena over real data sets. 
This is partly exacerbated by the fact that real non-progressive data sets are
relatively difficult to obtain. Can we establish the value of non-progressive models using
any publicly available data sets? 

In this paper, we address all the above challenges. Specifically, we make the
following contributions. 

\begin{packeditemize} 
\item We propose a discrete time non-progressive model and implement it without 
graph replication (Section~\ref{sec:discrete}).

\item We propose an efficient continuous time non-progressive   model  (Section~\ref{sec:model}).
 
\item We define the objective
  of influence maximization as choosing seeds so as to maximize the total
  expected activation time of nodes.   
\eat{\item We propose a direct approach for computing the expected total activation
  time of nodes given a seed set. It is efficient and leverages the continuous
  time nature of the model. }
We show that the objective function of total expected activation time is both 
  monotone and submodular. This implies the classic greedy seed selection
  algorithm, combined with our direct approach for computing expected total
  activation time, provides a $(1-1/e)$-approximation to the optimal solution 
  (Section~\ref{sec:InflMax}).

\item Through experiments on synthetic and real datasets, 
   we show that the 
  accuracy of our non-progressive model for estimating expected total activation
  time is much higher than its progressive counterparts, including the 
  recently proposed continuous time model~\cite{Du13}.
  Further, we show that our method is  more than one order
  of magnitude faster than an efficient implementation of the \dtime model, 
  whose accuracy is comparable to that of \model. 
 We also show that on datasets that have no deactivations (i.e., progressive setting), 
 our method using CP is 17-20 times faster than the continuous time progressive model of~\cite{Du13}    (Section~\ref{sec:expn}). 
\eat{
\item By means of extensive experiments, we show that our direct approach to
  influence maximization using our non-progressive model is more than one order
  of magnitude faster than  the approach of \cite{kempe03} of replicating the network over the time horizon, as we demonstrate using the DNP model we propose. }
\end{packeditemize}

We start by presenting related work in Section~\ref{sec:related}, and conclude with 
a summary of the paper and a discussion on future work in Section~\ref{sec:concl}. 
The major bottleneck in scaling influence maximization is in estimating the spread (in our case, expected active time). Our \model model significantly outperforms the competition on this step and it's trivial to see this advantage will carry over to influence maximization.

\makeatletter{}

\section{related work}
\label{sec:related}

Bharathi et al.~\cite{Bharathi} use exponential distribution to model the information propagation 
delay between nodes, and use this to avoid tie-breaking
for simultaneous activation attempts by multiple neighbors. We share with them the use of exponential distribution to model activation delays in our \model model. However, their main goal is designing response strategies to competing cascades rather than
maximizing the spread. 
Considerable work on non-progressive models 
has been done by the economics community \cite{Blume}. But they do not focus on computational issues, especially in relation to influence spread computation and maximization. 

Kempe et al. \cite{kempe03} proposed several propagation models, including non-progressive ones, but all based on discrete time. Indeed, the \dtime model we describe is fashioned after the non-progressive LT model they describe. As we show, our continuous time model \model significantly outperforms \dtime in terms of scalability. 
Our model and contributions are orthogonal to theirs. In particular, our efficient sampling strategy enables a scalable implementation of influence maximization.   
Recently, non-progressive models have received attention from the 
research community~\cite{fazli12,Dar07,Li13,Pathak10}.
As observed in \cite{fazli12}, progressive models are
not  accurate and there is scalability issue with non-progressive models. 
Their model is a simplistic model based on strict majority. 
While theoretically appealing, it's easy to show it's not submodular and no scalable influence 
maximization algorithm is provided. 
Furthermore, they focus on finding a perfect target set, one that ends up activating every 
node, not a realistic goal. 
Maximizing the overall activation times of nodes is more realistic 
goal for a business, which is what we study. 
\techrep{\cite{Dar07} studies a voter model on an unsigned undirected graph and show  that the most effective seeds for maximizing influence over long term are the highest degree nodes. \cite{Li13} study the considerably more complex case of voter model on a signed network with competing opinions. \cite{Pathak10} considers a generalization  of the LT model with $k$ competing cascades and analyzes the steady state distribution of the network using a stochastic graph coloring process. As such the goal and contributions of our paper are orthogonal to all these. }{}\techrep{Prakash et al. \cite{prakash2012threshold} find a condition 
under which an infection will die out in a given network and not cause an epidemic under the SIRS virus propagation model. 
The problem studied is significantly different from influence maximization or spread estimation, the focus of our paper. 
Ayalvadi et al. \cite{ganesh2005effect} examine the topological properties of a network that determine the persistence of epidemics under a continuous time epidemic spread model. They formulate the state transitions (infected/recovered) for nodes
using a Markov process. Again, influence maximization is not their focus.  
Kuhlman et al. \cite{kuhlman2011inhibiting} study a bi-threshold diffusion process, where an inactive node activates if the number of active neighbors exceeds a threshold $\theta_1$  and an active node deactivates if the number of inactive neighbors exceeds $\theta_2$, else the node remains in its previous state. They show the process converges to a steady state and study the problem of finding a ``critical'' set of nodes such that the total cost spent in forcing the these nodes is under a given budget and the number of nodes in the active state at steady state is maximized. While the goal seems similar to influence maximization, the model and objective function are technically very different from ours. Bischi et al. \cite{bischi2010global} study word of mouth rumors. Each individual has an initial state. A subset of individuals meet at each iteration and switches to state (true / false)  according to the majority state in the set. There are multiple such disjoint meetings at each iteration of the diffusion process. They do not consider an explicit network but one can be induced. Their focus is different from influence maximization.}{Other works such as~\cite{prakash2012threshold,ganesh2005effect,kuhlman2011inhibiting,bischi2010global} study related problems where nodes have active and inactive states. However, these are significantly different from 
influence maximization. See~\cite{techreport} for a detailed survey.}

Finally, a continuous-time Markov chain based progressive model was
proposed by Rodriguez et al. \cite{Rodriguez}, and more recently improved upon by Du et al.~\cite{Du13}. 
\techrep{In \cite{Rodriguez}, the authors 
use continuous-time Markov chains to 
analytically compute the spread, i.e., the average total number of nodes reached by a 
diffusion process starting from a set of seed nodes. Their model also uses exponential activation 
time delays on edges and thus the action time of a node is the shortest path distance from any seed node to that node.
However, their methods do not scale well as the time complexity of their 
solution can be exponentially large for ``dense networks'', which the authors define as 
networks with average node degree $>$ 2.5. By that definition, most 
social networks are dense. Although the authors propose speed-ups 
that provide approximate solutions or sparsify the networks, their experiments
are run on small graphs of at most 1000 edges. In comparison, 
we evaluate our model on graphs with nearly 30 million edges. 
Furthermore, it is not easy to directly extend their model to the non-progressive setting. 

The recent paper by Du et al. \cite{Du13} avoids calculating the shortest path distance mentioned above and instead makes use of a randomized  algorithm for estimating the neighborhood size of a single source node  and leverages this for estimating the influence spread within a given time horizon. Another nice feature of this paper is that they don't restrict to exponential distributions for their edge activation delays and allow a broad class of distributions. As such, this approach dominates \cite{Rodriguez}.}{}In our experiments, we compare our \model and its progressive variant CP with the method in \cite{Du13}. On data sets corresponding to progressive phenomena, both \cite{Du13} and CP have a comparable accuracy (which is very high). On data sets corresponding to non-progressive phenomena, both CP and \cite{Du13} suffer from high error rates while \model enjoys a very high level of accuracy. In all experiments, both CP and \model run 17-20 times faster than \cite{Du13}.

\eat{It is interesting to ask just how accurate such progressive models are in dealing with phenomena that are inherently non-progressive. Do they offer a reasonable approximation of the total  activation time? We empirically compare our \model model with this approach in Section~\ref{sec:expn}. 

To evaluate the expected
spread based on the distribution, they get continuous phase-type distributions
for  the nodes and get the . For the influence maximization problem, their model
can select a better seed set than the state of art solution. However, their
model  cannot be extended for non-progressive models easily, because continuous
phase-type distributions will not be valid for non-progressive models. Moreover,
in dense networks (networks where nodes have more edges), the time complexity of
their solution can be exponentially large. They solve this problem by sparsifying
the network. In their largest experiment, they sparsify the graph to only 1000 edges. 
}

\makeatletter{}

\section{Discrete time NP model}
\label{sec:background}
\label{sec:discrete}

There are two popular influence propagation models~\cite{kempe03}: independent cascade (IC)
and linear threshold (LT).
In~\cite{kempe03}, Kempe et al. also described an intuitive non-progressive extension of the discrete time LT model. 
Fundamentally, the models we propose in 
the next sections are close to the IC model.  
To set the proper context, in this section, we describe a discrete time non-progressive model that is
inspired by the framework given in \cite{kempe03}, 
but closer to the framework we will follow for our \model model. 

\techrep{\subsection{\dtime Model}}{}
Let $G = (V,E,P)$ be a weighted, directed graph representing a social network, with 
nodes (users) $V$ and edges (social ties) $E$, with the 
function $P : E \to [0, 1]$ representing the probability of influence along edges: $P(u,v) := P_{u,v}$ on edge $(u,v)\in E$ is the 
probability that node $v$ will be activated at time \begin{math} t+1\end{math}
given that $u$ is active at time \begin{math} t \end{math}. Additionally, the function $q: V \to [0,1]$ associates each node $u\in V$ with a deactivation probability: $q(u) := q_u$ represents the probability that $u$ will deactivate at time $t+1$ given that it's active at $t$. These are the key ingredients of our discrete time non-progressive model. 
Given the social network graph 
and a seed set of nodes $S$ that are active at the start of the propagation process, 
time unfolds in discrete steps. At time $t=0$, nodes in $S$ are active. At any time $t>0$, 
each of the currently active nodes $u$ makes one attempt at activating each of its neighbors $v$
and succeeds with probability $P_{u,v}$. 
At any time, an active node $u$ can deactivate with probability $q_u$. 
We refer to this model as the {\em discrete-time non-progressive} (\dtime) model. 

\techrep{
\para{Progressive implementation using replicated graphs}}{}
In non-progressive models,  nodes can get activated and deactivated infinitely often, so the influence propagation process can continue
indefinitely. Thus, we need to consider a fixed \emph{time horizon} as the time period within 
which we would like to study the propagation process. 
Kempe et al. \cite{kempe03} showed that their non-progressive (LT) model's behavior over a given time horizon 
$T$ can be simulated using a progressive model. 
The key is to replicate the social network graph for each timestamp. 
\techrep{We mimic their steps, and adapt them to the context of our \dtime model 
described above. }{However, a na\"ive implementation 
with replicated graphs is not practical. We describe a space efficient
implementation that avoids graph replication in our tech report~\cite{techreport}. We show that 
the \dtime model still suffers from a serious inefficiency that at each time step, each node needs
to make the decision of whether or not it changes its state. 
Thus, at each time step, $n$ nodes need to sample 
a Bernoulli distribution to determine their state at the next time step. 
Several nodes may stay in their current state for long periods of time. Hence, 
sampling at each time step at each node is extremely inefficient. 
We therefore move to the continuous-time regime for efficiently 
modeling the non-progressive phenomena. }

\techrep{
Essentially, we create a node $u_i^t$ for each node $u_i$ and each timestamp $t$. 
For any edge $(u_i,u_j)\in E$ in the original social network, create an edge $(u_i^{t-1},u_j^t)$, 
the edge weight is same as the probability of edge $(u_i,u_j)$. Also, each node $u_i^t$ is connected to 
$u_i^{t-1}$ (each node is connected to itself at previous timestamp) with edge weight 
$(1- q_{u_i})$. A node $u_i{t+1}$ is active if it has a neighbor which is active at $t$ and the activation attempt succeeds. Note that the deactivation probability $q_u$ has been compiled into the activation probability $(1-q_u)$ from $u^t$ to $u^{t+1}$. Following the footsteps of \cite{kempe03}, the equivalence between the \dtime model and the replicated progressive model  can be shown. 

Clearly, an implementation of the non-progressive model based on the above simulation  
 is not scalable. The graph is replicated for each time stamp, which makes
it memory intensive. 
Consider for example, one of the graphs in our evaluation which has 2.5M nodes and 30M edges, and a time horizon of 365 days. Assuming we store only two integers for the end points of an edge, and one double as the edge weight, the memory required to store just the edge weights would be approximately 160 GB! 
Therefore, a na\"ive implementation with replicated graphs is not practical.

\para{Space efficient implementation}
Here, we propose an efficient implementation of our non-progressive \dtime model above, and use that as our baseline method for comparison purposes. 
We reduce the memory footprint of the implementation described above 
by \emph{avoiding the replication of nodes}. 
Notice that the influence from node $u$ to $v$ is the same for any timestamp, so we only need to 
store the edge weights $P_{u,v}$ once for each edge $(u,v)$ in the original social graph. 
In the simulation, we only need to store the state (active or inactive) of each node at the current and previous 
timestamp. A node $u$ is active at time $t$ if it was active at time $t-1$ and it did not deactivate; or if a 
neighbor $v$ was active at time $t-1$ and it activated node $u$ (with success probability probability $P_{v,u}$).
With appropriate mapping of edge and deactivation probabilities, the activation conditions at each timestamp in \dtime  can be made the same as that in the na\"ive implementation with graph replication. 
As a result, the expected spread will be the same for the two models. 
In the sequel, by \dtime model, we mean this improved implementation.

In spite of the savings achieved by avoiding graph replication, the \dtime model still suffers 
from a serious inefficiency, described next. 
The \dtime model involves each node making a decision at each time step of whether the node 
changes its state. Thus, in an implementation, at each time step, $n$ nodes need to sample 
a uniform distribution to decide their state at the next time step. 
Several nodes may stay in their current state for long periods of time. Hence, 
sampling at each time step at each node is extremely inefficient. 
We therefore move to the continuous-time regime to allow for a much 
more efficient modeling of the non-progressive phenomena. Later, 
in Section~\ref{sec:compare} we show how the discrete-time and 
continuous-time models are related. }{}

\makeatletter{}

\section{Continuous-time NP Model}
\label{sec:model}
\label{sec-modeldef}

\subsection{Model description}
\sloppy
\techrep{In this section, we present a continuous-time 
non-progressive model that permits a far more efficient implementation compared 
to the \dtime model.}{}We model influence propagation as a continuous-time 
Markov process with nodes being in one of two states: \emph{active} and \emph{inactive}. 
As in classical propagation models, in our model, events trigger state 
changes and happen probabilistically. 
We start with a seed set of active nodes. At any time, there are two events 
that may happen at an active node: the node may
activate its neighbor, or may deactivate itself. 
Similarly, for any inactive node, the node may get activated by one of its active neighbors, 
or stay inactive. 
We refer to an event that activates an inactive node as an 
{\em activation event} and one that deactivates an already active node as a 
{\em deactivation event}. It is these deactivation events that allow the model 
to be non-progressive.

More specifically, there are two parameters, one for activation and the other for 
deactivation, both being exponentially distributed random variables. 
\techrep{In Section~\ref{sec:params} we show how these parameters can be learned from data.}{}\eat{Notice that unlike the discrete time counterpart, the model parameters govern the times at which events 
happen as opposed to whether the events will happen. This is a direct consequence of moving to 
continuous time.}Each edge $(u,v)\in E$ has an associated activation rate parameter $\gamma_{+,u,v}$, and 
each node $u$ has a deactivation rate parameter $\gamma_{-, u}$.
We start with a seed set of nodes that are, by definition, active at time $0$. 
For each node $u$ that is activated at time $t$, 
(a) a time $\tau$ sampled according to rate parameter $\gamma_{+,u,v}$ has the semantic 
that $v$ will be activated no later than $t+\tau$, and
(b) a time $\tau^\prime$ sampled according to rate parameter $\gamma_{-, u}$, 
has the semantic that node $u$ will deactivate at time $t+\tau^\prime$.
Notice that another neighbor of $v$ may activate it sooner. In particular, 
an inactive node $v$ that is reachable from one or more active nodes activates 
at a time equal to the shortest path from those active nodes, 
that is shortest in terms of the sum of sampled propagation times of the 
edges forming the path. 
\eat{
We model the events using an exponential distribution: 
at any time, the rate at which an event happens is constant and is
determined by a rate parameter. The time that
the  next event will happen is also exponentially distributed with a rate 
parameter. 
Each node $u$ and each edge $(u,v)\in E$ has an associated rate
parameter \begin{math}
  \gamma_{-, u}  \end{math}  and \begin{math} \gamma_{+,u,v}  \end{math} respectively. For any edge \begin{math} u
  \rightarrow v  \end{math},  when the start node $u$ is active and $v$ is
inactive, $u$ will try to activate $v$. The activation time of $v$ from $u$ 
follows an exponential  distribution with rate parameter  $\gamma_{+, u,v} $.  
When a node is active, it tries to deactivate, and the
deactivation time follows an exponential distribution with rate
parameter  $\gamma_{-, u}$. 
}However, each activation 
or deactivation  with its associated rate parameter is one {\em local} event. 
That is, only the ego-centric network of a node is involved in any event. 
This observation is key to the scalability of our proposed  method. In particular, unlike the recently 
proposed continuous time (but progressive) models \cite{Du13,Rodriguez}, we don't need to compute or even 
estimate the shortest path length directly. 

\subsection{Semantics of the propagation}
\label{sec:semantics}
During an influence propagation cascade, there are multiple activation 
and deactivation events that may happen. In order to model the cascade, 
we need to find the one that happens first and 
update the activation status of the corresponding node. 
For instance, if $u$ is active, it deactivates with some rate parameter, however, it is also trying to 
activate its inactive neighbor $v$ with some rate parameter. 
If $u$ deactivates before activating $v$, then $v$ may not have a chance to 
activate (assuming it has only one neighbor) unless $u$ activates again. 
Further, if there are multiple neighbors trying to activate a node $v$, it will get activated by the
local event that happens first, i.e., by the neighbor that first activates it. Therefore, 
it is important to understand and model the order of events. 
We crucially make use of two key properties of exponential distributions for modeling 
the time and order of events. 
\begin{property}
\label{prop1}
For $n$ different events with rate parameters 
$\gamma_1, \gamma_2\ldots \gamma_n$,
the probability that the \begin{math} i^{th}  \end{math} event will happen first is
$\frac{\gamma_i}{\sum_{i=1} ^{n} \gamma_i}$.
\end{property}

\begin{property}
\label{prop2}
 For different events with rate parameters
\begin{math} \gamma_1, \gamma_2\ldots \gamma_n \end{math}, 
the time of the first event is exponentially distributed with rate parameter:
$ \sum_{i=1} ^{n} \gamma_i$.
\end{property}

We keep track of the current time  $t_{cur}$ during a propagation process.
At each iteration, the categorical distribution 
in Property~\ref{prop1} is sampled to determine the event that happens first (or next). 
Then, the exponential distribution with rate parameter 
$\sum_{i=1} ^{n} \gamma_i$ is sampled (Property~\ref{prop2}) to obtain the time elapsed 
$\tau$ between last event and this event. 
The current time is then updated as $t_{cur} = t_{cur}+ \tau$, and we proceed to the next 
iteration if $t_{cur}<T$, where $T$ is the time horizon, and stop otherwise. In other words, 
even though the model is continuous time, it has a clear  interpretation in terms of discrete 
steps, namely the occurrence of events.

Another way to understand the model semantics is in terms of possible worlds. 
A deterministic possible world for our model can be constructed as follows:
For each edge $(u,v)\in E$ we sample an array of timestamps and
sort it. A timestamp in the array indicates that if $u$ is active at that time, it will
activate node $v$. We call this array the {\em schedule of activations}.
Similarly, for each node $u\in V$, we sample an array of deactivation times. 
If $u$ is active at those timestamps, it will get deactivated. 
We refer to this array as the {\em schedule of deactivations}.
The set of possible worlds for a given instance of our \model model is the set of all 
such edge activation schedules and node deactivation schedules, for every edge and node 
in the given social graph.  
Such a construction of possible world aptly covers all possibilities in our random 
process. 
\techrep{The following example illustrates the notion of schedule of activations. 
The notion of schedule of deactivations works analogously. 
\begin{example} 
For instance, let the time horizon be $T=6$ and the sampled array 
for the edge$(u,v)$ be $[1, 3.2, 5.8]$. Let and node $u$ be active in the time
interval $(2,4.2)$. We go over the schedule of activations:
at $t=1$,  $v$ is not influenced by node $u$ as $u$ is inactive;  
at $t=3.2$, if $v$ is not active it will be activated by $u$; 
and nothing will happen at $t=5.8$. 
\end{example}
}{}We will use these semantics to prove monotonicity and submodularity
of the spread under the \model model in Section~\ref{sec:InflMax}.

\subsection{Advantages of \model over \dtime}
\label{sec:compare}
If we correctly map the rate parameters in \model model 
to the probabilities in \dtime model, the simulation results of two models will be 
similar. We note that the models are not equivalent, but have 
similar accuracy in terms of the expected spread, when the following mapping 
holds. 
In \model, for any edge $(u,v)$ where $u$ is active but $v$ is not, the probability that $u$ 
activates $v$ within the next time unit is equal to  the $\mathrm{CDF}(1, \gamma)$, 
where $\mathrm{CDF}$ is the cumulative distribution function of the exponential distribution, 
$\gamma$ is the rate parameter associated with $(u,v)$, and 1 is the time unit. 
The corresponding edge probabilities in the \dtime model would be  $\mathrm{CDF}(1, \gamma)$.
Similarly, we map the deactivation rates in \model to deactivation probabilities in \dtime.
Then, the resulting \dtime model will be a discrete-time approximation of the \model model.
Therefore, we expect the accuracy of \model and \dtime to be similar.  
\eat{\model is far more efficient in terms of computation time.
Since both the discrete-time and continuous-time models are 
designed to model  the same influence propagation process, 
the number of activations and deactivations in a given time horizon should be 
(almost) the same. }
We now compare the two models in terms of the computational cost 
incurred at each activation  and deactivation.
In the discrete time case, for each active node, we need to sample from a 
uniform distribution once at each timestamp to determine whether or not the node deactivates.
\techrep{For example, let the deactivation probability for the node be 0.001, then the expected 
number of samplings for a deactivation is  $1000$. }{}In the continuous-time setting, however, we first need to randomly 
choose the event that occurs with probability governed by Property 1, then we need to 
sample the exponential distribution to get the time at which it occurs, using Property 2. 
\techrep{Therefore, one activation or  deactivation happens for each pair of samplings. 
Revisiting  our example above, each deactivation requires sampling an 
exponential distribution twice, a 
significant improvement (500 times) over the expected number of samplings 
in the discrete-time case.}{}\eat{  
For each active node in the discrete-time setting, we need to sample from a 
uniform distribution once at each timestamp, whether or not the node deactivates
at that timestamp. 
In contrast, in the continuous-time setting, 
there is no need to visit an active node if a deactivation event does not
occur at that node. }Therefore, for nodes that do not deactivate in the time window, their 
cost of (attempted) deactivation is zero in the continuous-time setting, again, a significant 
saving from the discrete-time regime. 

\techrep{Since sampling is the main action that is repeatedly performed, we delve into further improving 
the efficiency of the sampling process even further in the next section. These methods are
similar to those described in~\cite{matias93}.
}{}

\eat{
The next question is that whether this situation is common or not.
From experiment, we know that both activation and deactivation rates are really
low,  which means when a Process starts, there
is small chance that an event will happen in the time window. I will use a
sample  simulation in Figure 2 to illustrate this. This is a simulation
of a spread from last.fm dataset in one year. As you can see, most nodes get
activated  at the end of the year. If one node get activated at 200th day and assume its
deactivation rate is 0.005, the pr
obability it will get deactivated in the time
window  is F(365-200,0.0005)=0.0792.
From the above example, we know that lots of active nodes do not deactivate in the time window.
\begin{figure}
\includegraphics[width=3in,height=2.5in]{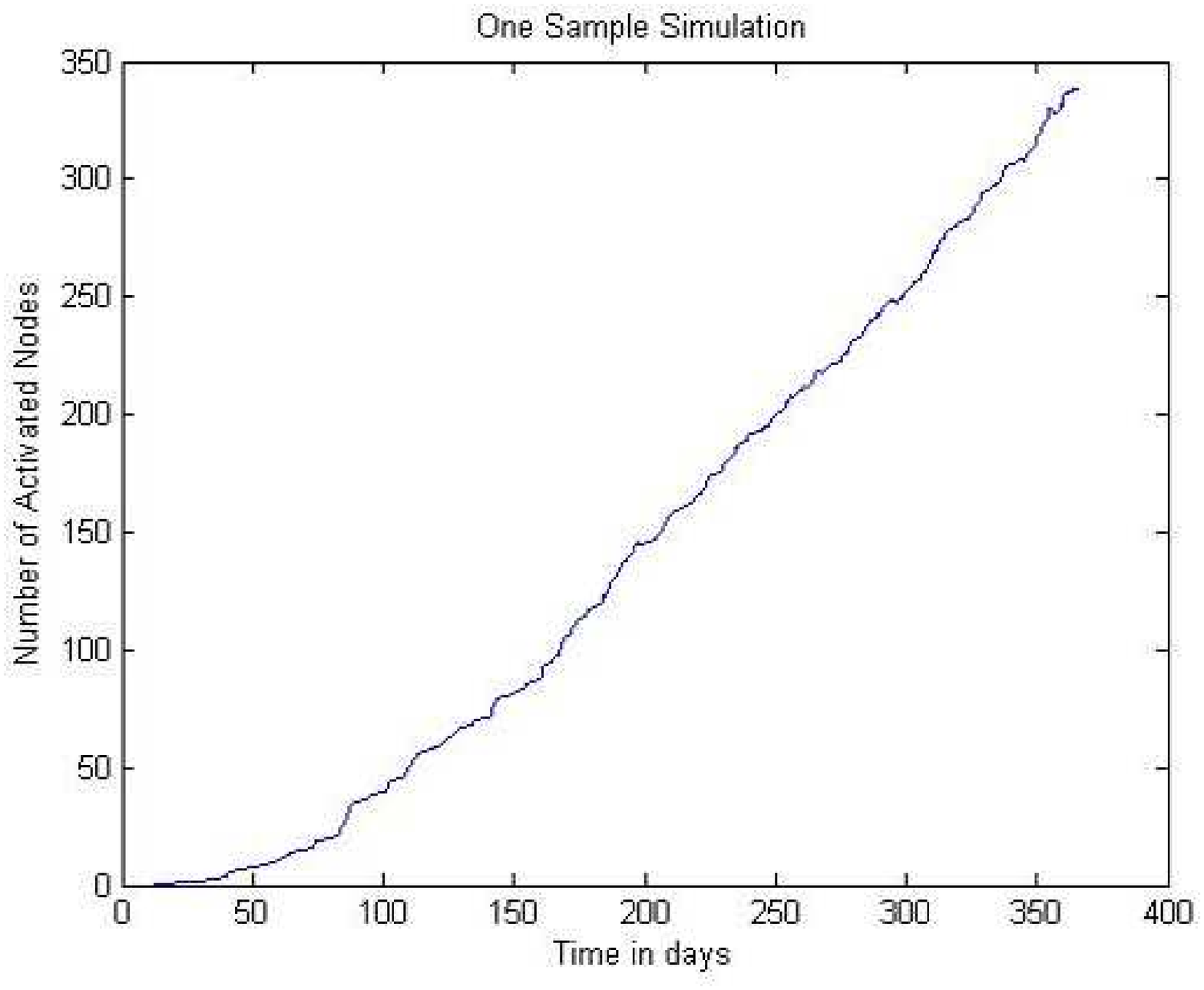}
\caption{An example for influence propagation process}
\end{figure}
}

\eat{
at start one starts with a set of seeds A. they are activated instantly by def. 
\forall node u that activated at time t, 
a random propagation time \tau from each of its out-arcs u-->v is sampled from the corresponding distribution. the semantics is that v will activate no later than t+\tau.  
a random deactivation time \delta is sampled from the corresponding distribution. the semantics is that u will deactivate at time t+\delta. 
an inactive node x that is reachable from one or more active nodes activates at a time equal to the shortest distance from those active nodes, where distance is defined as sum of sampled propagation times of the arcs forming the path. 
}

\makeatletter{}
\techrep{
\section{Fast Sampling Algorithm}
\label{sec:sampling}
The most basic action performed by our method is that of sampling a 
categorical distribution. The categorical distribution is one of all rate parameters
associated with the nodes and edges of the social network.
The sampled  value, determined by the rate parameter, decides the next event that will occur, 
which may be an activation or a deactivation. 
Furthermore, this sampling is repeatedly performed. 
Therefore, any improvements made to the speed of a 
single sampling action will greatly improve the efficiency of our method. 
Conceptually, a typical sampling action for picking the next event would work as follows.
A line segment of length equal to the rate parameter is drawn for each event. Then, 
all such segments are laid out in an arbitrary order. 
A point $x$ is sampled uniformly at random from this line, and the event
associated with that point is chosen. 
The process may be repeated for sampling many events. 
\begin{figure}[t!]
\centering
\includegraphics[width=\figwidth]{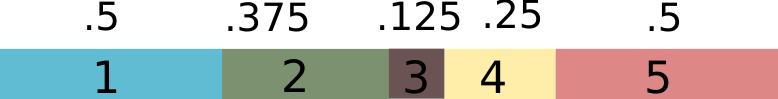}
\caption{Categorical distribution for Example~\ref{eg:dist}}
\label{fig:line}
\vspace{-6pt}
\end{figure}
\begin{figure}[t!]
\centering
\includegraphics[width=0.5\figwidth]{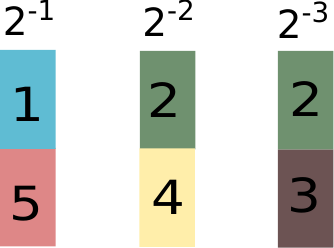}
\caption{Arrays of shards corresponding to Table~\ref{tab:ex}}
\label{fig:seg_line}
\vspace{-6pt}
\end{figure}
\begin{example}
Consider a categorical distribution with rate 
parameters $\gamma_1, \ldots, \gamma_5$ corresponding
to five different events 0.5, 0.375, 0.125, 0.25 and 0.5. The sum of the rate parameters 
$\sum_{k=1}^n\gamma_k=1.75$. Figure~\ref{fig:line} illustrates the line segments for the events. 
Say the point $x$ sampled on the line is 0.6, then, the event chosen will be the second 
(green in Figure~\ref{fig:line}) one corresponding to the rate parameter 0.375. 
\label{eg:dist}
\end{example}

A key challenge that we face in using this standard approach is that, 
in our setting the sampling is {\sl without replacement}.
That is, as events occur, the rate parameter corresponding to that event is removed, 
and cannot be sampled again. 
Therefore, an update equivalent to removing a line segment would leave a ``gap'' or ``hole''. 
A simple hack of filling the hole with the last end point would not work, as the line segments have 
varying lengths. It would require shifting all subsequent line segments to cover the gap, which 
is computationally expensive. This na\"ive  approach turns out to be extremely inefficient in our setting. 
To ensure an efficient implementation of our model, our goal is to be able to sample in \BigO1 time. 

Therefore, we need to design a new, efficient method for sampling a categorical distribution which 
undergoes frequent updates, owing to sampling without replacement. In our setting, these updates reflect the addition and 
removal of potential activation and deactivation events. 
We make two simple but important observations that motivate the design  of our sampling method.\\
\indent 1) If we divide each line segment into smaller ``shards'', the sampling process 
      would be statistically equivalent to that from the original categorical 
      distribution, as long as the sum of the lengths of these shards is equal to the original segment. \\
\indent 2) If the length of the line segment for each event is the same, we can simply replace an event 
  that we want to remove with the last one. 

\para{Data Structure}
We choose to represent the categorical distribution as a set of $m$ arrays. Each array corresponds
to a fixed shard length from $2^{-1}, \ldots, 2^{-m}$. For instance, say $m=3$, the 3 arrays
would correspond to shards of length $0.5,0.25,0.125$ respectively.  
Intuitively,  $m$ can be viewed as the precision of the rate parameter when using
the array representation. 
In particular, $m$ decides the granularity of shards, since the smallest shard 
that a line segment is divided into is of length $2^{-m}$. 
A line segment corresponding to a rate parameter may be divided into 
shards of different lengths, and hence may be present in multiple arrays. 
Furthermore, since the elements of each array are of equal length (length $2^{-i}$ for the $t^{th}$
array), the total length of the array can be easily determined by simply maintaining the 
count of elements in the array. 
Let $c_i$ denote the count or number of elements in array $i$. Then, the length of 
array $i$ is $2^{-i}\times c_i$.

\begin{example}
Figure~\ref{fig:seg_line} shows the array representation of the 
categorical distribution from Example~\ref{eg:dist}, 
for $m=3$ arrays (to ensure fine granularity we use $m=20$ in our implementation). 
Event 1 with $\gamma_1=0.5$ belongs completely to array corresponding to $2^{-1}$, 
while Event 2 with $\gamma_1=0.375$ is broken into shards of length $0.25,0.125$ and hence is 
present in arrays corresponding to  $2^{-2},2^{-3}$. 
Further, we can consider array 1 as covering the interval (0,1] as it contains 
two shards ($c_1=2$) of length  $2^{-1}$. Similarly, arrays 2 and 3 cover 
the intervals (1,1.5] and (1.5,1.75], respectively. 
Purely for clarity, we can think of sharding as mapping the $m$ arrays to a table of $m$ 
columns and $n$ rows as illustrated by Table~\ref{tab:ex} for the categorical 
distribution in Example~\ref{eg:dist}. The arrays are a sparse representation of the table. 
\end{example}

\begin{table}
\centering
\small
\begin{tabular}{|c|l|| c c c |}
\hline
Event Id & $\gamma$& \multicolumn{3}{|c|}{Columns} \\
& & $2^{-1}$ & $2^{-2}$ & $2^{-3}$  \\ \hline
1 & 0.5     & 1 & 0 & 0  \\
2 & 0.375 & 0 & 1 & 1 \\
3 & 0.125 & 0 & 0 & 1 \\
4 & 0.25   & 0 & 1 & 0  \\
5 &  0.5     & 1 & 0 & 0  \\ \hline
Total & $\sum_k\gamma_k=1.75$ &  $c_1$=2 & $c_2$=2  &$c_3$= 2  \\  \hline
\end{tabular}
\caption{Tabular representation of arrays of shards in Figure~\ref{fig:seg_line}}
\label{tab:ex}
\vspace{-10pt}
\end{table}

\para{Sampling}
As before the first step in sampling is to pick a number $x$ uniformly at random from the interval 
$(0,\sum_k\gamma_k]$. Notice that this interval is equal to $(0,\sum_i2^{-i}\times c_i]$ in the array representation. We have a two step process to determine which event the sampled point $x$ maps to.
First, find the $i^{th}$ array in which $x$ lies. Second, find the $j^{th}$ event in that array to be chosen 
as the sampled event. 

As described above, the length of the $i^{th}$ array is $2^{-i}\times c_i$. 
Assuming $c_0=0$, the $i^{th}$ array to which $x$ belongs is the one that satisfies: 
\begin{equation}
2^{-(i-1)}\times c_{i-1}< x \leq2^{-i}\times c_i
\label{eqn:col}
\end{equation}

Next, we find the index $j$ of the event chosen from array $i$ as:
\begin{equation}
j = \Bigg\lceil \frac{x - 2^{-(i-1)}\times c_{i-1} }{2^{-i}\times c_i-2^{-(i-1)}\times c_{i-1}}  \times c_i \Bigg\rceil
\label{eqn:row}
\end{equation}
\begin{example}
Say $x=1.3$, then it lies in the interval (1,1.5] covered by array 2, resulting in $i=2$. 
Then, $j=\big\lceil \frac{1.3 - 1 }{1.5-1} \times 2 \big\rceil = 2$. The sampled event is then 
the second event in the second array in Figure~\ref{fig:seg_line}, i.e., event 4.
\end{example}
\para{Updates}
Now, we show how our array representation helps prevent the ``holes'' created 
by updates with the na\"ive approach. 
First, it is easy to see that an update involving an addition of an event is trivial. 
Simply divide the line segment corresponding to the rate parameter of the event into 
shards, and append the shards to the end of the corresponding arrays. 
Now, if the update involves an event being removed because it has occurred, we 
need to find elements in all arrays that correspond to that event, and replace those
elements with the last element in the corresponding arrays. 
In order to efficiently find all the indices at which an event lies in different arrays, 
a mapping of positions of an event in different arrays can be maintained in memory. 
Finally, $c_i$ of each array $i$ that was updated is changed to reflect the new counts. 
Notice that  all shards in an array are of equal length, so shards can be replaced without resulting in gaps in the arrays. 
\begin{example}
Say the event 2 was sampled from the distribution by determining that $i=2$ and $j=1$. 
We keep track of each event's associated positions in all arrays, so we know that 
event 2 lies in arrays $2,3$, and corresponds to index 1 for both arrays, as 
shown in this example. We replace the first element of arrays 2 and 3 with events 4 and 3 respectively, 
and update $c_2=1, c_3=1$. 
\end{example}

\begin{algorithm}[t]
\caption{\algo}
\label{algo:sampling}
   \begin{algorithmic}[1]
	\STATE $x\leftarrow$ a number sampled uniformly at random from $(0,\sum_{k=1}^n\gamma_k]$
	\STATE $i\leftarrow$ the array containing $x$ using Equation (\ref{eqn:col})
	\STATE $j\leftarrow$ the index in array $i$ determined using Equation (\ref{eqn:row})
	\STATE $e\leftarrow$ the chosen event at $j^{th}$ index in $i^{th}$ array
	\STATE Update arrays by replacing occurrences of $e$ with last element in respective arrays; update 
	$c_i$s accordingly. 
	\STATE Update the array positions for the set of events associated with the moved elements
          \RETURN $e$
   \end{algorithmic}
\end{algorithm}

Algorithm~\ref{algo:sampling} summarizes the proposed sampling procedure, that 
includes the sampling and update steps.

\para{Running Time}
Sampling $x$, and computing $j$ using Equation (\ref{eqn:row}) takes constant time. 
We need \BigO{\log m} time to compute $i$. 
Updating the arrays takes \BigO{m} time. 
We store each event's associated positions in all arrays, so step 4 takes \BigO{1}.
The time to update events' associated positions each time an event 
is updated is \BigO{m}. 
Therefore, sampling one event using our \algo algorithm takes \BigO{m} time. 
Obviously, the number of sampling actions performed in a propagation depend on 
several factors, such as, the size of the graph, the time horizon, and the seed set size.
We discuss these factors in detail in our evaluation section.

{}

}

\makeatletter{}

\section{Influence Maximization} \label{sec:InflMax}
Next, we discuss influence maximization, i.e., the process of seed selection 
to maximize the spread of influence under the \model model. 
The influence maximization problem for non-progressive 
models is similar to that described in~\cite{kempe03}. However, since
nodes can deactivate, the {\em spread}, traditionally defined as the expected 
number of active nodes, changes with time. 
Thus, maximizing the expected number of active nodes at a 
given timestamp, or at the time horizon may not be ideal from the point of view 
of a company initiating a viral marketing campaign. We start by proposing 
an intuitive objective function for spread under a non-progressive model. 
Importantly, we show that our proposed spread function is monotone and submodular, 
hence the greedy approach yields a $(1-1/e)$-approximation to the optimal solution.

\techrep{\subsection{Objective Function}}{\para{Objective Function}}
\label{sec:spread}
In a non-progressive world, an intuitive objective from the point of view of 
a marketer is to maximize the ``active time'' of its customers in a given social network. 
That is, maximize the total amount of time that nodes in the network are active, in expectation. 
Given a seed set $A$, 
$$ spread_A=\sum_{v\in V} \tau_v $$
where $\tau_v$ is the sum of time intervals within T for which node $v$ is active. 
Then, the influence maximization problem~\cite{kempe03} is defined as: select a 
seed set of nodes $A\subseteq V$ to be activated such that the expected $spread_A$
is maximized over a chosen time horizon $T$, given 
the non-progressive influence propagation model. 

\techrep{\subsection{Monotonicity and Submodularity}}{\para{Monotonicity and Submodularity}}
As an important step towards solving the influence maximization problem, we
 show that the expected $spread$ is monotone and submodular. 
Then, we can use the state-of-the-art greedy algorithm, such as 
CELF \cite{Leskovec07} and CELF++  \cite{Celf}, to guarantee a 
$(1-1/e)$-approximation. 
It is easy to see that, 
$$\E[spread_A]=\sum _ {v \in V}\E[\tau_v] = \int_{t=0}^T \E[\sigma(A,t)]dt$$
where $\sigma(A,t)=|S|$, $S$ is the set of nodes activated from the seed set $A$ at 
timestamp $t$, and $\sigma(A,t)$ is the number such nodes or the cardinality of set $S$. 
Therefore, we can prove monotonicity and submodularity of the expected spread, 
by showing that these properties hold for $\E[\sigma(A,t)]$. 
For this, we follow the proof guidelines in~\cite{kempe03} to 
construct a deterministic possible world from the random process that we are modeling. 
Let $X$ be the set of all possible worlds, and given $x\in X$, let $pdf(x)$ denote the 
probability density function of $x$. Then, $$ \E[\sigma(A,t)]= \int_{x \in X} pdf(x)\times \sigma_x(A,t)  dx $$
Thus, we only need to prove that $\sigma_x(A,t)$ is monotone and 
submodular. 
Note, that we need to integrate over the possible worlds, 
as opposed to a summation performed in~\cite{kempe03},
because the number of deterministic possible worlds is uncountable in our setting. 

\begin{lemma}
Additivity of spreads: Given two sets of seed nodes $A,B$, timestamp $t$, and a possible world $x$, 
$$S_x(A \cup B,t) = S_x(A,t) \cup S_x(B,t)$$
where $S_x(A,t)$ denotes the set of nodes activated by seed set $A$ in possible 
world $x$ at timestamp $t$. 
\label{additivity}
\end{lemma}
\eat{
For clarity and flow as well as lack of space, the proof that indeed the additivity of spreads holds  
in our construction of all possible worlds corresponding to our \model model 
is omitted here and will be included in a technical report to soon appear in the arXiv. }
\begin{theorem}
Given lemma~\ref{additivity}, $\sigma_x(A,t)=|S_x(A,t)|$ is 
monotone and submodular. 
\end{theorem}
\techrep{\begin{proof}
Assuming the additivity of spreads holds for sets $A$ and $B$, 
then by definition, $S_x(A \cup B,t) = S_x(A,t) \cup S_x(B,t)$. 
On adding a set $B$ to an initial set $A$,  
$ |S_x(A \cup B,t)| = |S_x(A,t) \cup S_x(B,t)| \geq  |S_x(A,t)|$
Hence, $|S_x(A,t)|$, equivalently $\sigma_x(A,t)$ is monotone.

For all sets $A \subseteq B$ and all nodes $v$, we need to show that,
$$|S_x(A \cup \{v\},t)-S_x(A,t)| \geq |S_x(B \cup \{v\},t)-S_x(B,t)|$$
By the additivity of spreads, $S_x(A \cup \{v\},t) = S_x(A ,t) \cup  S_x(\{v\} ,t)$. Then,
\begin{align*}
S_x(A \cup \{v\},t) -S_x(A ,t) & = S_x(A ,t) \cup  S_x(\{v\} ,t) -S_x(A,t) \\
   & = S_x(\{v\} ,t) -  S_x(\{v\} ,t) \cap S_x(A,t) 
\end{align*}
Similarly, 
$S_x(B \cup \{v\},t)-S_x(B,t) =  S_x(\{v\} ,t) -  S_x(\{v\} ,t) \cap S_x(B,t)$.
Since $A\subseteq B$, $S_x(A ,t) \subseteq S_x(B ,t)$, applying an intersection 
with $S_x(\{v\} ,t)$ on both sides, then subtracting from $S_x(\{v\} ,t)$, we get
\begin{align*}
S_x(\{v\} ,t)-S_x(\{v\} ,t) \cap S_x(A,t) &\supseteq S_x(\{v\} ,t)-S_x(\{v\} ,t) \cap S_x(B,t)\\
                 S_x(A \cup \{v\},t)-S_x(A,t)  &\supseteq S_x(B \cup \{v\},t)-S_x(B,t) \\
                |S_x(A \cup \{v\},t)-S_x(A,t)| &\geq          |S_x(B \cup \{v\},t)-S_x(B,t)|
\end{align*}
Hence submodularity.
\end{proof}
}{The proofs are presented in our tech report~\cite{techreport}.}

\techrep{
\makeatletter{}
\begin{figure*}[th!]
\centering
\subfigure[Deactivation rate - Flixster]{
\includegraphics[width=\figfour]{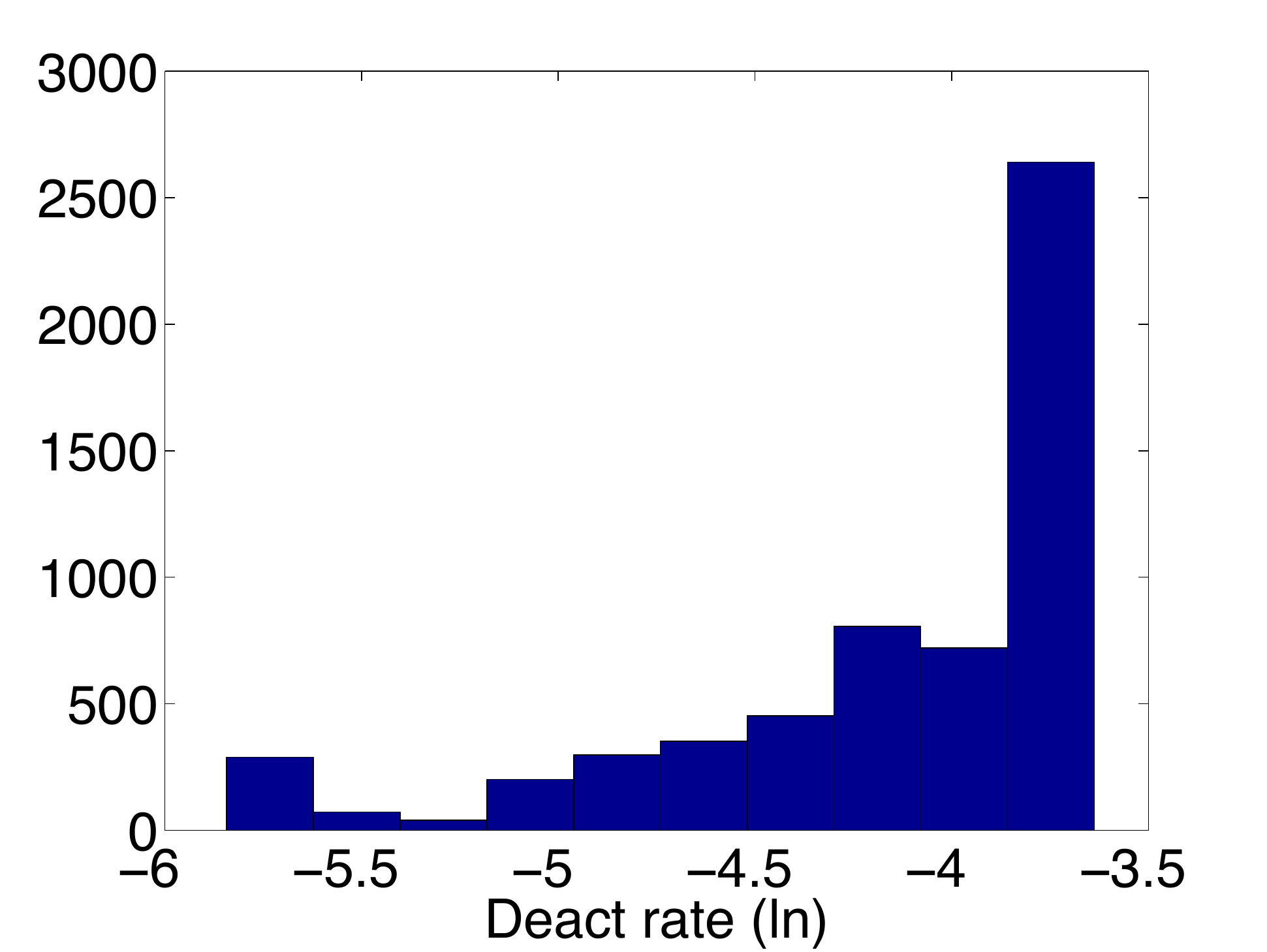}
\label{fig:fx_drate}
}\subfigure[Activation rate  -  Flixster]{
\includegraphics[width=\figfour]{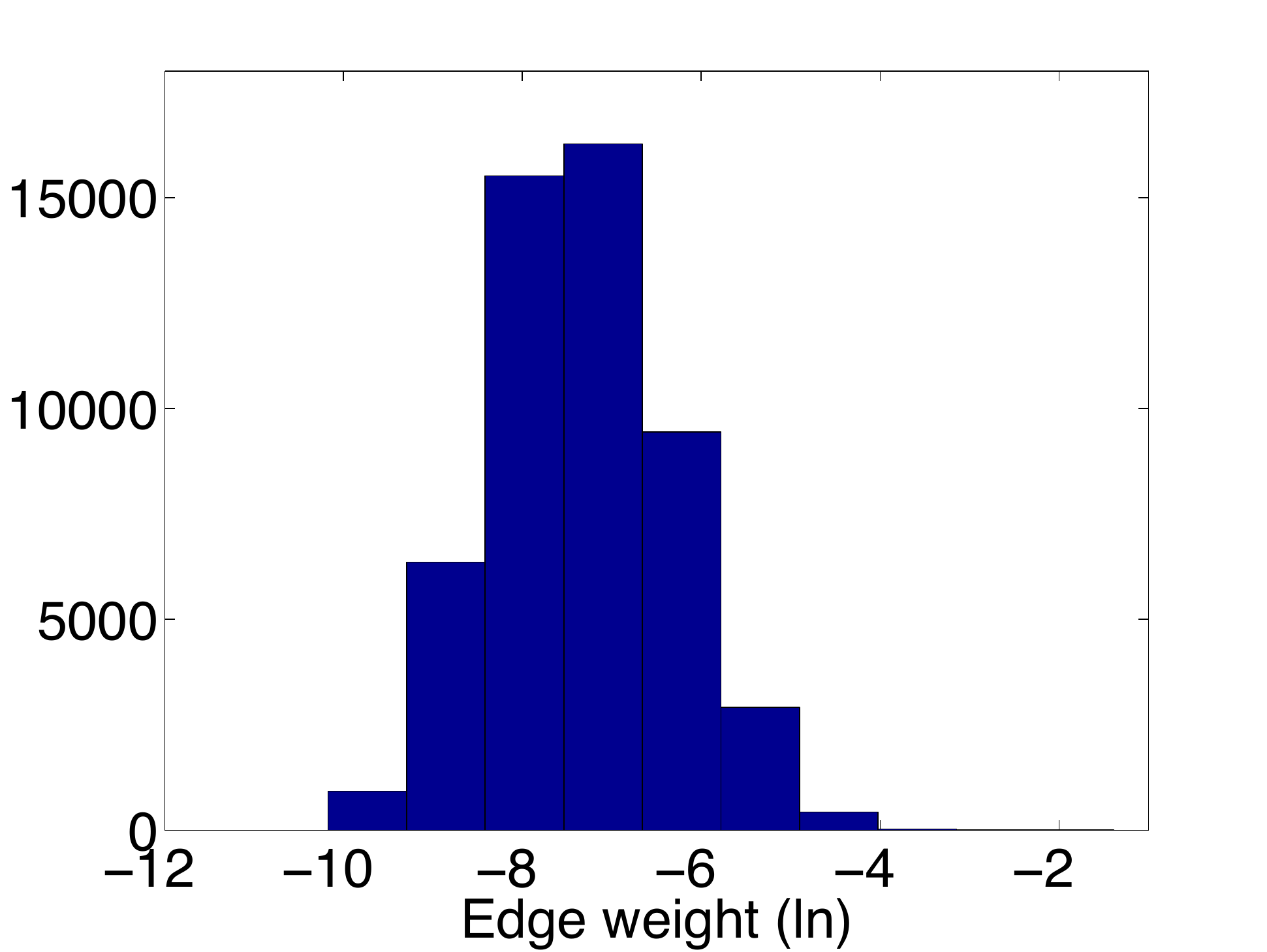}
\label{fig:fx_arate}
}\subfigure[Deactivation rate - Flickr]{
\includegraphics[width=\figfour]{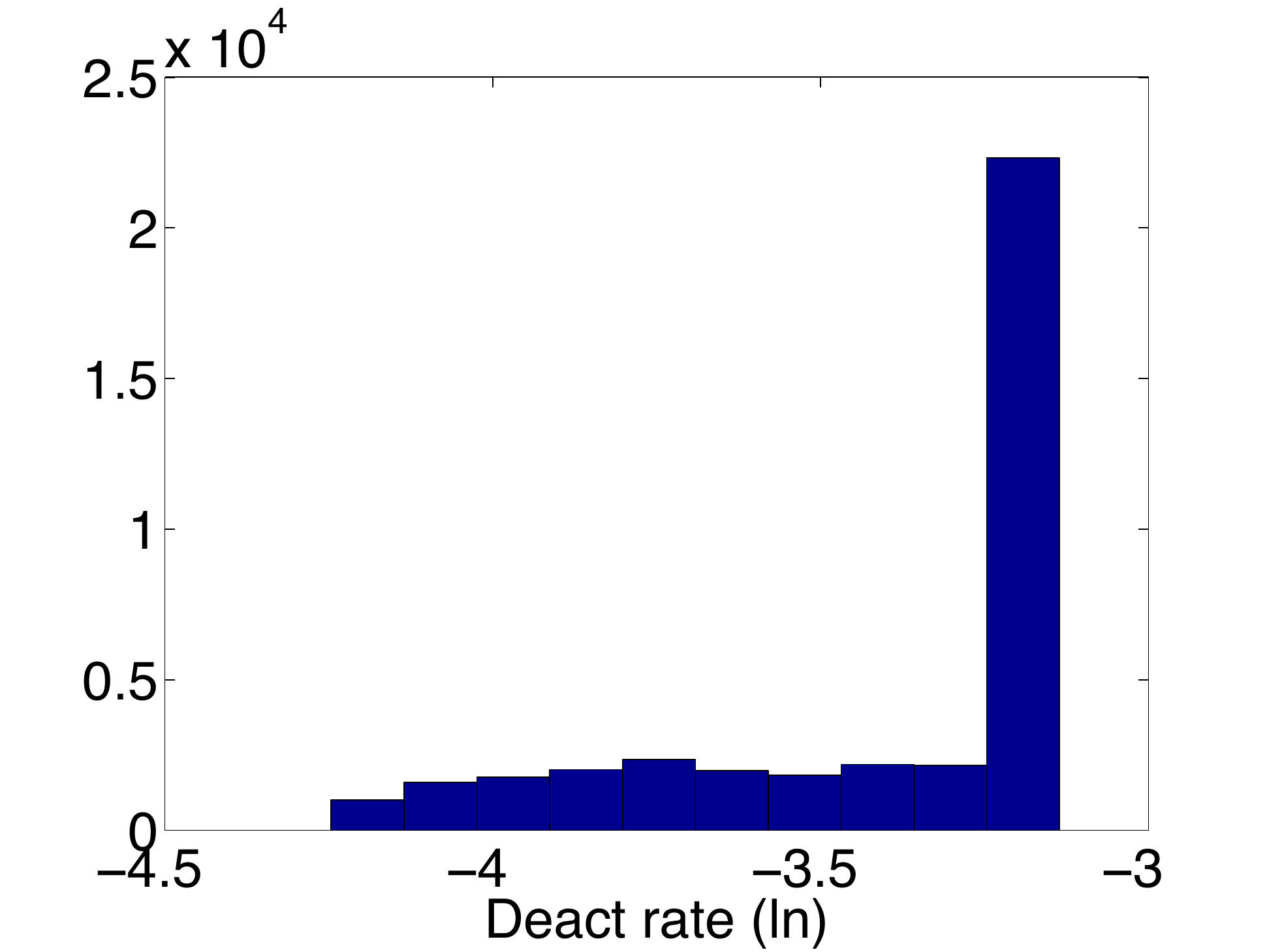}
\label{fig:fl_drate}
}\subfigure[Activation rate - Flickr]{
\includegraphics[width=\figfour]{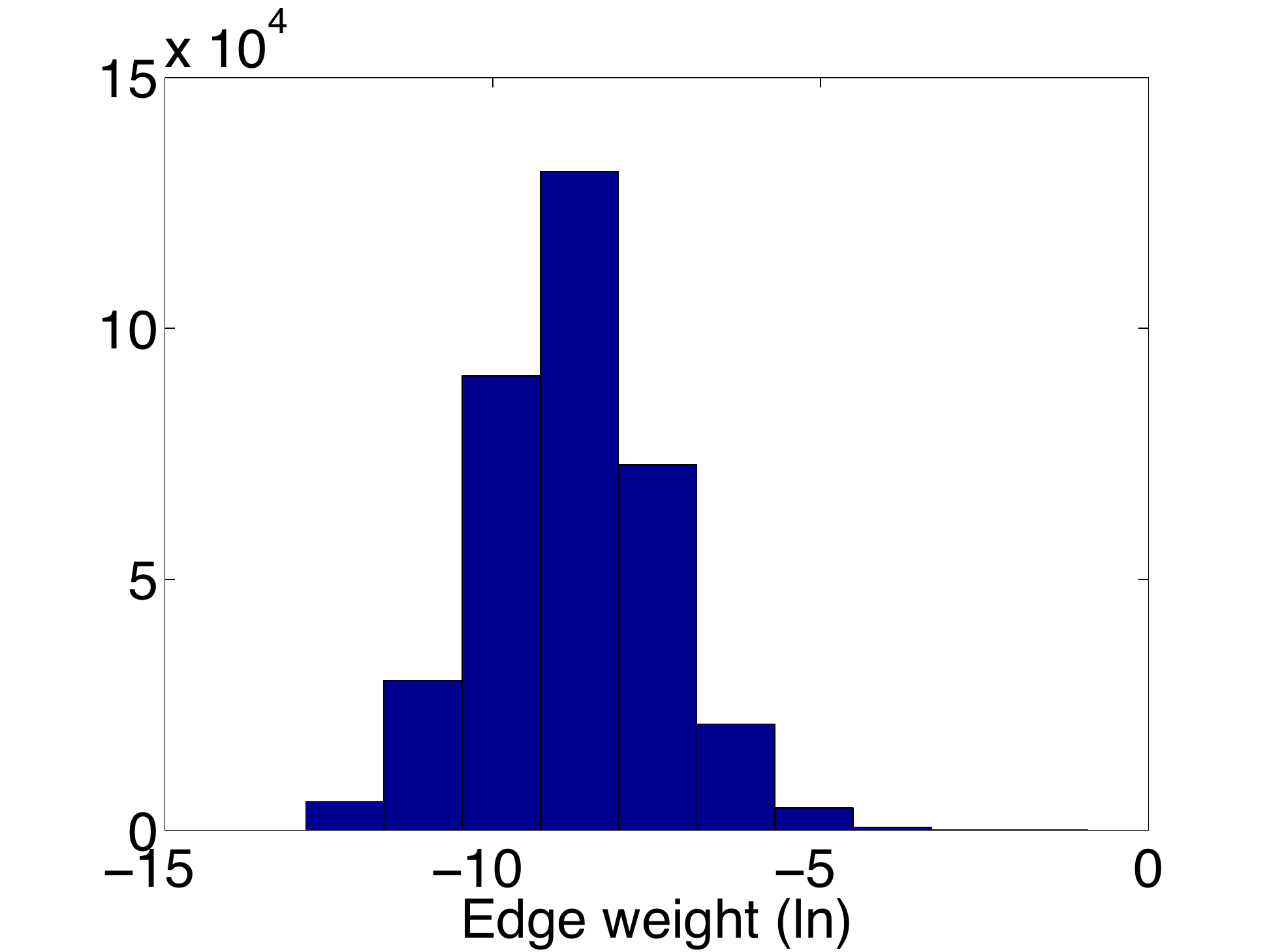}
\label{fig:fl_arate}
}\vspace{-6pt}
\caption{Model Parameters for Flixster and Flickr Datasets}
\label{fig:params}
\vspace{-10pt}
\end{figure*}

\section{Experimental Evaluation}
\label{sec:expn}

In this section we first describe how the edge weights of our model can be learned from data, 
and the details of our experimental setup. Next, we present the results of 
evaluating our model on real and synthetic datasets. We compare the 
accuracy and running time of: traditional \IC, state-of-the-art continuous time progressive model ConTinEst\cite{Du13}, 
\dtime and \model, for estimating the spread as defined in Section~\ref{sec:spread}.
\techrep{}{For implementing the sampling without replacement procedure for a 
categorical distribution, see methods described in~\cite{matias93}.}

\eat{
There are two major sets of experiment: accuracy experiments and computational performance experiments. In the accuracy experiments, we divide the datasets into training dataset and test dataset. We learn the edge weights and deactivation rates from the training dataset, and run simulation with traditional \IC model, \dtime and \model to estimate the global active time in the test dataset. We also obtain the ground truth (global active time) from the test dataset, and compare it with the estimations. In the computational performance experiments, we compare the running time for \dtime and \model with different graph sizes and different time horizon.
}

\subsection{Learning model parameters}
\label{sec:params}
We divide each dataset into training and test sets. We use the training set to learn the 
different model parameters, i.e., deactivation rates and edge weights. 
The first challenge we face is identifying deactivations in common datasets. 
If we had access to logs associated with each activations and deactivations, 
for instance, timestamps for service subscriptions and unsubscribe actions, this would be trivial. 
However, such service subscription datasets are not publicly available. 

\subsubsection{Activation and Deactivation}
Given a training set in the form of an action log with \textless user,action,timestamp\textgreater~tuples, 
we would like to find the timestamps for activations and deactivations. 
 We use the following proxy for defining events: 
 1) when a user performs an action we call it an activation event and mark the user active
 2) at each activation we start a timer, the event of the timer running out before another activation occurs, 
 is called a deactivation, and the user is said to be deactivated.  
 The timer is set for a length equal to the {\em deactivation time window}. 
 Similar definitions of active users are used by social networking services as a metric of the popularity 
 of these sites. For instance, Facebook maintains statistics of its daily (and monthly) active users,
 where a day (resp., month) is the deactivation time window. 

\para{Deactivation time window} 
To find the value of the deactivation time window from the action log, 
we first get all the time gaps between two consecutive 
actions of each user in the training dataset. We ignore time gaps of less than one day to 
avoid the bias of a user performing multiple actions in a single session.
We set the deactivation time window as the expected value (mean) of time gaps for all 
experiments. In addition, to be rigorous with our analysis, we also evaluate our model over
different deactivation time windows in the next section.

\para{Rate parameters}
Using the above definition of activation and deactivation events, we now define the 
{\em total active time} for a node $u$ as the sum of time intervals for which node $u$ is active in 
our training set. Then, the deactivation rate parameter $\gamma_{-,u}$ for node $u$ is
\begin{displaymath}deactivation\;rate(u)=\frac{number \; of \; deactivations (u)}{total\; active \;time \; (u)}\end{displaymath}

\noindent
Finally, we define the rate parameters corresponding to edges which are responsible for activations. 
At each activation, we find the set of active neighbors of that node at that timestamp. 
Say, the node has $k$ active neighbors when it activates. 
We assign a contribution score of $1/k$ to each of the $k$ edges from an active neighbor. Then, 
the activation rate parameter $\gamma_{+,u,v}$ for edge $(u,v)$ is
\begin{displaymath}activation\;rate(u,v)=\frac{\sum contribution \;score(u,v)}{total\; active \;time \; (u)}\end{displaymath}

We use these methods to learn the parameters of the \model model. As described in Section~\ref{sec:compare}, we can cast the rate parameters of \model to probabilities for the \dtime model,
 with the $\mathrm{CDF}$ of the exponential distribution.

\spara{\IC edge weights}
The existing methods for \IC models  are not scalable. Our approach is similar to the PCB method in~\cite{Goyal10}.
This is essentially a counting argument and a recent paper~\cite{lin13} uses PCB to learn the edge weights for the IC model and report good accuracy w.r.t to the true edge weights, even better than the likelihood method of Saito et al.~\cite{saito}.
$$Probability(u,v)=\frac{\sum contribution \;score(u,v)}{number\;  of \;activations \; (u)}$$

\subsubsection{Global Influence}
In real social networks, users may get influenced by external sources, such as the popular trends, 
news etc. We model such influences by inserting a ``global node'' and that has a directed edge
to all users. This global node is always active, and influences all nodes equally. 
Technically, the influence from this global node is used to explain activation of a 
node $u$ that has no other active neighbors. Thus, we can compute its total contribution score 
over all its outgoing edges, and define the global influence value as:
$$\frac{total\;  contribution\; score}{time \;horizon \times number \;of \;nodes \;in \;the\; graph}$$

\eat{
When we learn the model, we find that in many cases, there is no active neighbors when a node activates. Thus, we cannot assign the contribution score to any nodes, which doesn't sound right. In real world influence propagation, a person can be influenced from friends in the social network or from sources outside of the social network (for example, mass media). We need to model that influence to provide an accurate estimation of the global active time. We created an super global node, which is always active and connects to all nodes in the graph. Since we don't know anything outside of the social network, we assume the influence from the global node to any node in the graph is same.
The global node is just another node is the graph, we can easily get its total contribution score from the previous section.
For the influence from global node to any node y is:
}

\subsection{Datasets}
We evaluate our model on synthetically generated data and 
two real datasets: Flixster and Flickr, for which we have a social network, 
and an action log which contains the timestamps of users' actions. 

\para{Synthetic}
We generate a random 500 node unweighted forest fire network. We artificially generate cascades and deactivation events. For generating cascades, we pick a node at random and randomly generate a timestamp at which this node performs an action and hence becomes active. To avoid the dependence of cascade generation on any underlying diffusion model, we randomly choose a neighbor and sample a random activation time for it. We repeat this process recursively to generate the cascades. For the deactivation events too, we randomly select a node and make it inactive at a random timestamp. 

\para{Flixster} 
Flixster is a movie rating service, where users form an explicit social network. Our 
dataset~\cite{Flixster} has 1 million nodes (users) and 26.7 million edges (social connections). 
An activation event is the act of {\em rating a movie from a specific genre} (horror in our 
experiments). 
We divide the action log into one year training and one year test set. Using the expectation
of intervals between consecutive actions as the deactivation time window, we find it to be 38 days
for our dataset. The distribution of rate parameters are shown in 
Figures~\ref{fig:fx_drate} and~\ref{fig:fx_arate}, as
computed using the approach described above, and the global influence 
value for this dataset is $2.095\times10^{-5}$.

\para{Flickr} Our second dataset comes from the photo sharing service of Flickr. 
The dataset~\cite{Flickr} has 2.3 million nodes and 33.1 million edges. 
An action corresponds to using the ``favorite photo'' feature, i.e., marking a photo as favorite. 
The associated action log is over 138 days, of which we use the first 70 days as 
our training set for learning parameters, and 
the next 68 days as our test set. The deactivation time window is 23 days and the global influence 
value for our dataset is $1.415\times10^{-4}$. Figures~\ref{fig:fl_drate} and~\ref{fig:fl_arate} show the distribution of rate parameters for Flickr.

\begin{figure}[t]
\centering
\subfigure[Error in Spread estimation]{
\includegraphics[width=\figfour]{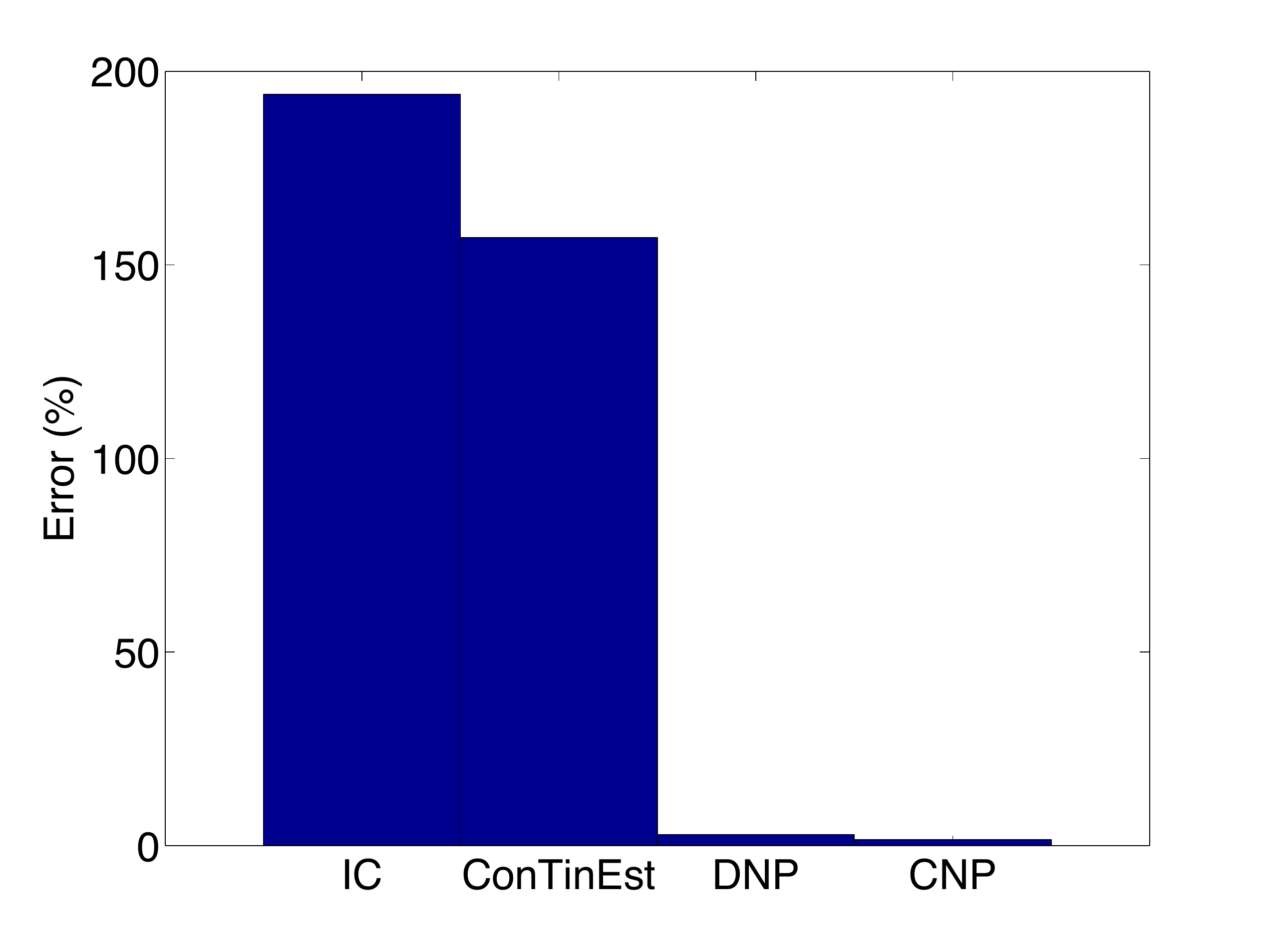}
\label{fig:err_tab}
}\subfigure[Running time]{
\includegraphics[width=\figfour]{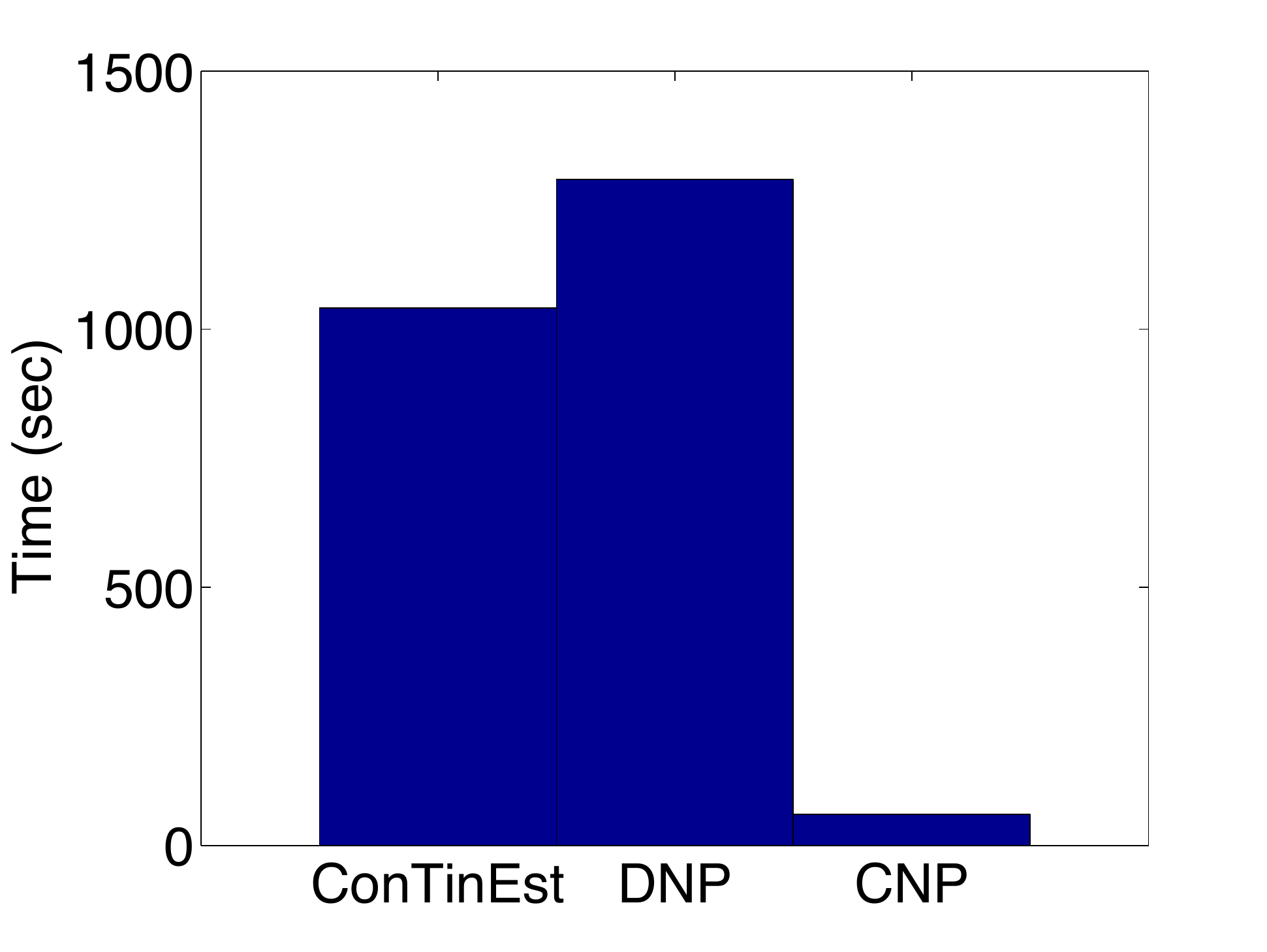}
\label{fig:time_tab}
}
\caption{Spread estimation error and time comparison for progressive (IC, ConTinEst) and non-progressive models (\dtime, \model) for Flixster dataset}
\label{fig:models}
\end{figure}

\begin{table}[t]
\centering
\small
\setlength\tabcolsep{3pt}
  \begin{tabular}{|c|c|c|c|c|c| }
    \hline
    Dataset & Ground & \model & \dtime & ConTinEst & IC \\
                    & Truth & & &\cite{Du13} &  \\ \hline
    Flixster & 964013 & 949750 & 991141 &2477678 & 2833860\\ 
	&& (1.5\%) & (2.8\%)& (157\%)  & (194\%) \\
	\hline
    Flickr & 2435663 & 2432053 & 2409695 & 4423372 & 4922860 \\ 
	&& (0.148\%) & (1.07\%)& (81\%) & (102\%) \\ 
	\hline
  \end{tabular}
  \caption{Spread estimated by progressive (IC, ConTinEst) and non-progressive models (\dtime, \model) and error percentage w.r.t. ground truth}
  \label{tab:acc}
 \vspace{-10pt}
\end{table}

\subsection{Comparison Across Models}
We start with presenting an overview of our results comparing different models: progressive vs. non-progressive models, and discrete vs. continuous time models, across two axes: accuracy of spread estimation and running time. Figure~\ref{fig:models} illustrates this comparison for Flixster dataset. See detailed numbers for both datasets in Table~\ref{tab:acc}. We make the following key observations, and substantiate these with details  through the remainder of this section.
\begin{itemize}
\item Progressive models, here classical \IC and state-of-the-art \nips, overestimate the spread and result in an error of 80-194\% compared with the ground truth. 
\item Non-progressive models, \dtime and \model, are highly accurate in estimating the spread with very small errors of 0.1-3\%. Notice that \dtime is the non-progressive counterpart of the progressive \IC model and 
improves the accuracy or error in estimating spread by a 100\%. \item Continuous models \nips and \model are slightly better at estimating accuracy than the discrete models IC and \dtime. 
\item Our model \model is not just more accurate but also an order of magnitude faster than the state-of-the-art continuous time model \nips (Figure~\ref{fig:time_tab}). These results are for running 100 Monte-Carlo simulations.
\end{itemize}

\begin{figure}[t]
\centering
\subfigure[Error in Spread estimation for synthetic data]{
\includegraphics[width=\figfour]{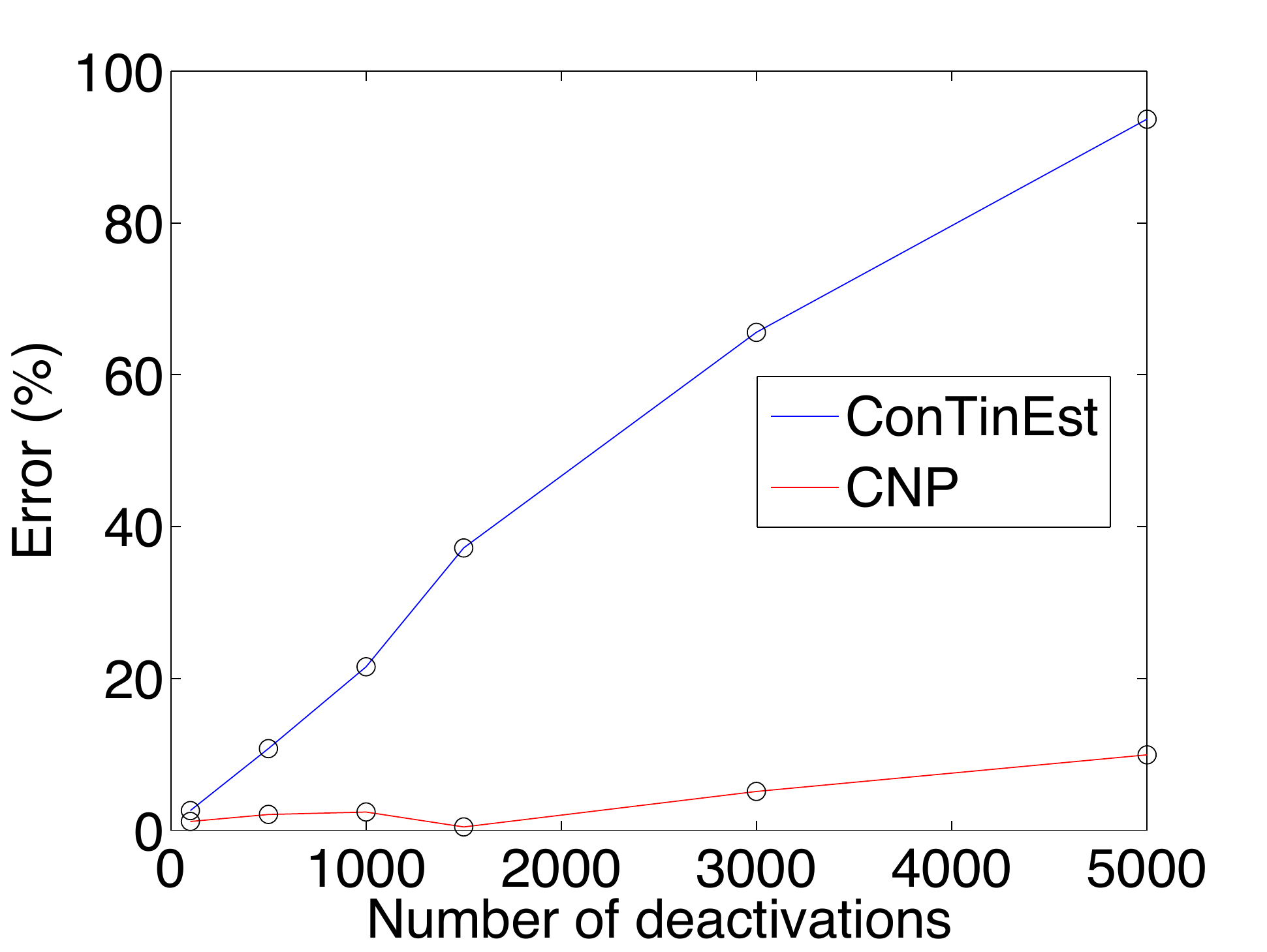}
\label{fig:syn_acc}
}\subfigure[Running time of ConTinEst vs. \model on real data]{
\includegraphics[width=\figfour]{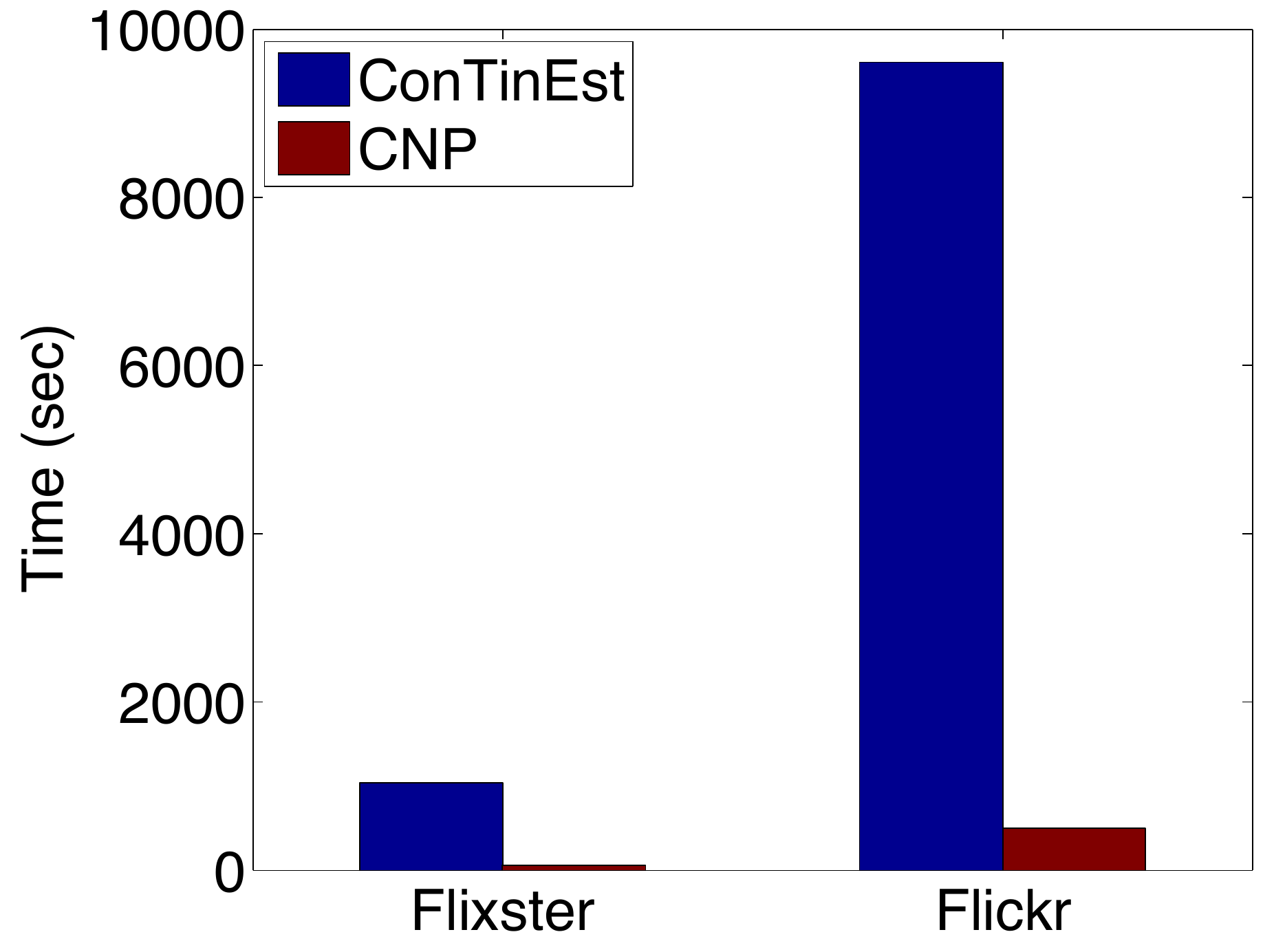} \label{fig:nips_speed}
}\vspace{-3pt}
\caption{Accuracy comparison for progressive (IC, ConTinEst) and non-progressive models (\dtime, \model)}
\label{fig:acc}
\vspace{-6pt}
\end{figure}

\para{Evaluating Accuracy}
We evaluate the accuracy of the estimation of spread of our model by simulating the propagation, 
starting at the state of the network at the last timestamp in our training set, and evaluating 
against the ground truth of spread achieved in the test set. 
In other words, the nodes active at the end of the training set are treated as the seed set, and 
the propagation is run for the time horizon equal to the length of the test set. 
The {\em ground truth} of spread is computed as the total active time of all nodes for the test set. 
Table~\ref{tab:acc} shows the spread as estimated by our model compared with 
the ground truth. For our model, error in spread estimation is just 1.5\% and 0.1\% over the Flixster 
and Flickr datasets resp. 
The difference between \IC and non-progressive models is two orders of magnitude. 
This validates that deactivation occurs in real datasets, and that modeling 
deactivation properly is critical for a reasonable estimation of influence spread 
in non-progressive settings.

Next, we perform an experiment on synthetic data to show the impact of 
number of deactivations on the spread estimates by a progressive model (\nips~\cite{Du13}) and our
\model model. Figure~\ref{fig:syn_acc} shows this the error (\%) in estimating the spread. 
As the deactivations increase, the gap in the accuracy of the two methods increase, with \model 
performing over 83\% better than the competitor, establishing that in the 
presence of deactivations, non-progressive phenomena are modeled accurately by \model.

\para{Evaluating Computational Cost}
We compare the running time of ConTinEst with \model for our two real datasets Flickr and Flixster for running
100 Monte-Carlo simulations. As seen in Figure~\ref{fig:nips_speed}, 
our model is an order of magnitude faster than its progressive competitor.

\begin{figure*}[t]
\centering
\subfigure[Varying $\#$ deactivations Synthetic data]{
\includegraphics[width=\figfour]{figs/synthetic_accuracy}
\label{fig:syn_acc}
}\subfigure[Varying deactivation rate Flixster]{
\includegraphics[width=\figfour]{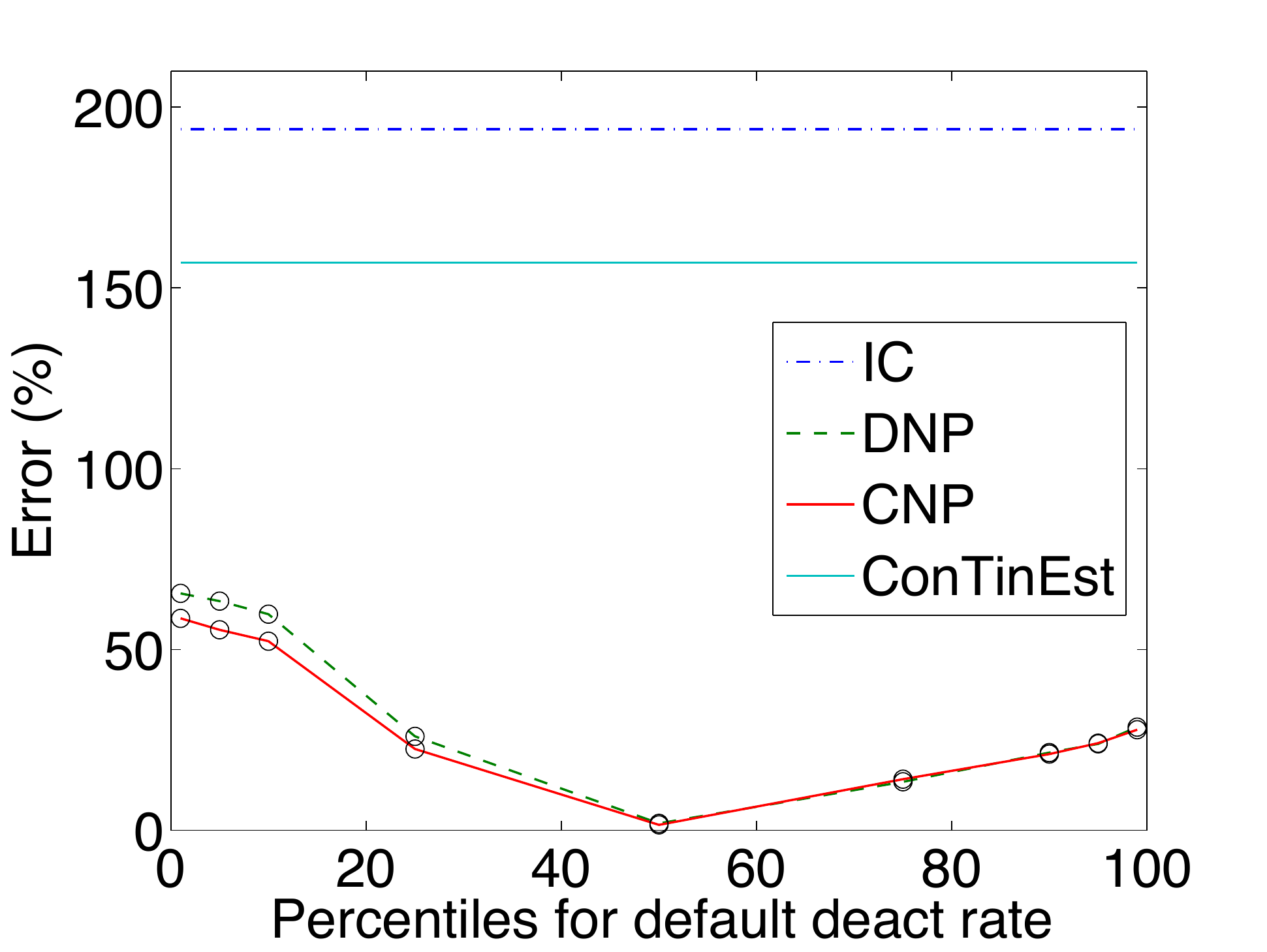}
\label{fig:fx_accuracy}
}\subfigure[Varying deactivation rate Flickr]{
\includegraphics[width=\figfour]{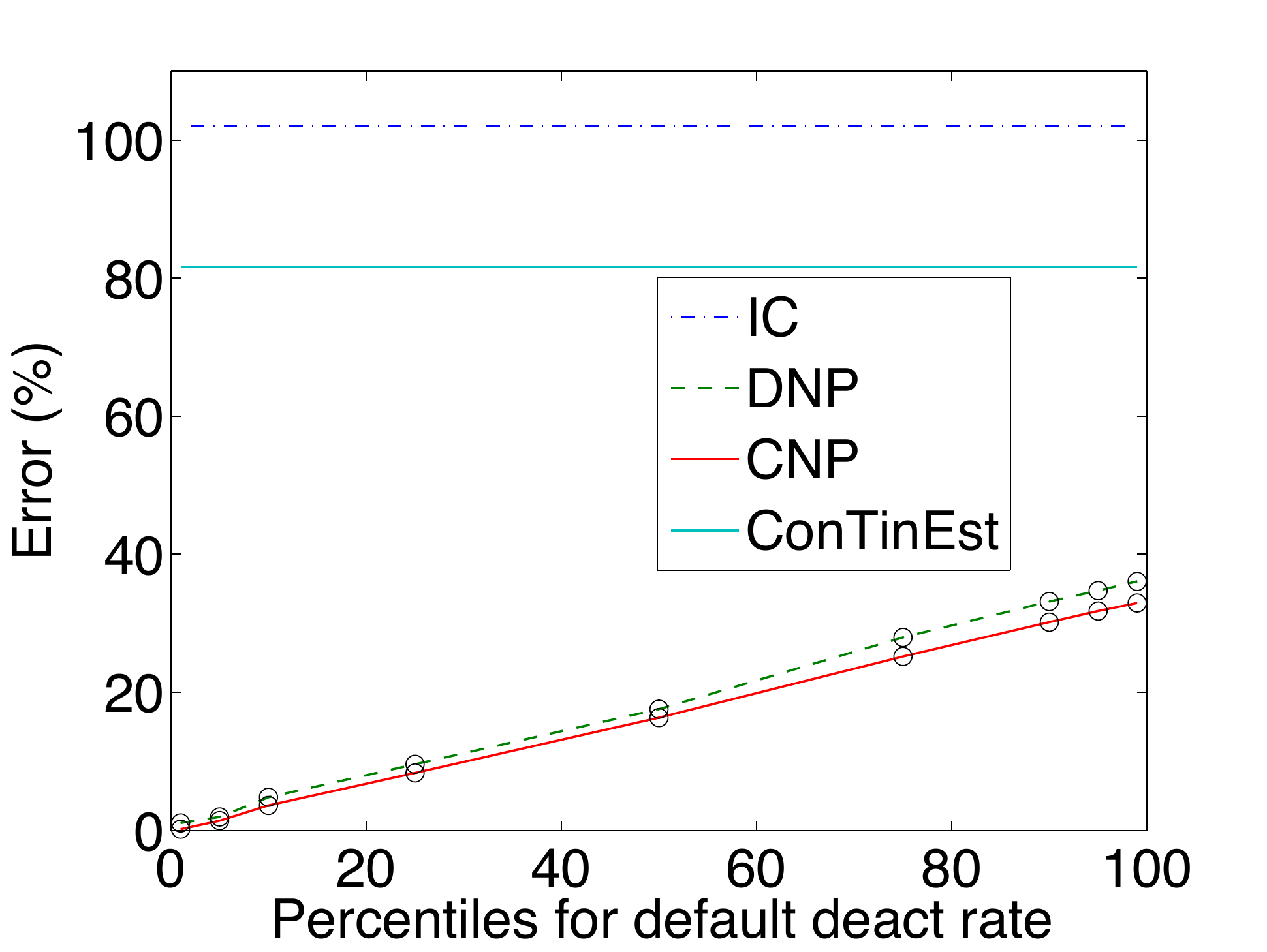}
\label{fig:fl_accuracy}
}\subfigure[Varying deactivation time window Flixster]{
\includegraphics[width=\figfour]{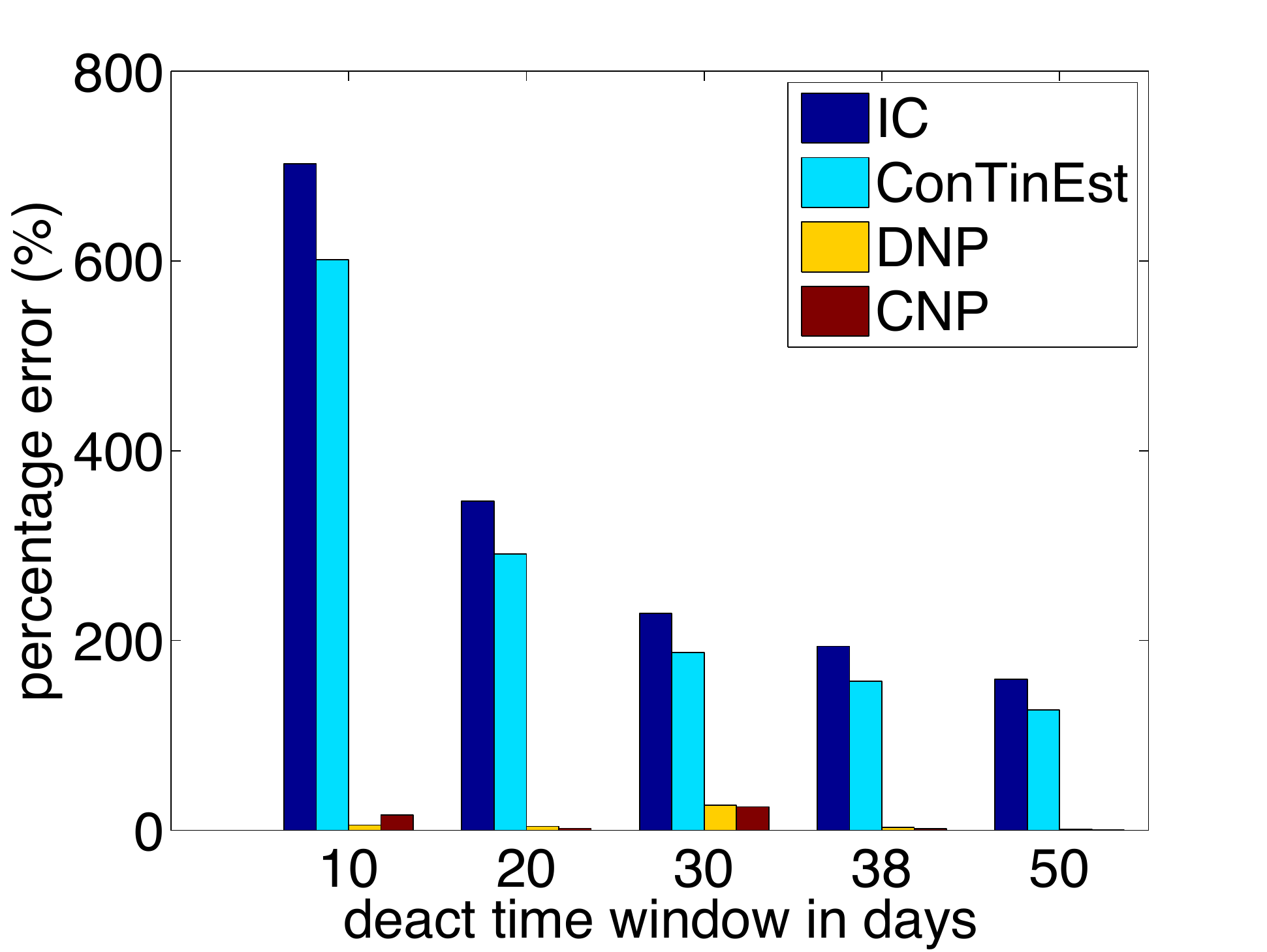}
\label{fig:fx_deact}
}\vspace{-3pt}
\caption{Accuracy comparison for progressive (IC, ConTinEst) and non-progressive models (\dtime, \model) 
with varying deactivation parameters}
\label{fig:acc}
\vspace{-6pt}
\end{figure*}

\subsection{Varying Parameter Values}
\para{Effect of deactivation parameters on accuracy}
For Flickr, we observe that 98\% of the nodes perform no action in our training set, and hence 
get a zero deactivation rate. 
This is an artifact of the short timespan of the training data. 
Filtering those nodes would result in a disconnected graph.  To overcome this shortcoming of the 
data sample and to avoid overfitting, we  assign a {\em default deactivation rate} to all such nodes. 
We use the set of non-zero deactivation rates (as learned from the data) as a guideline, 
and test different percentiles of this set as the default deactivation rate. Instead of fixing the value, 
we evaluate its impact on the accuracy by varying it, as shown in 
Figure~\ref{fig:fx_accuracy} for Flixster dataset.
As seen in the figure, the default deactivation rate does impact the accuracy slightly, still the 
estimates by \model are orders of magnitude more accurate than \IC and ConTinEst. 
Also notice that the estimated  spread for \dtime and \model is very similar validating  
our argument in Section~\ref{sec:compare} that \dtime is an approximation of \model.
The plots for Flickr are skipped for brevity, but the methodology adopted and results were similar. For the remainder of the experiments we set the default deactivation rate for Flixster and Flickr to the best obtained, i.e., $50^{th}$ and 1 percentile resp. of the unique non-zero deactivation rates learned. 

Next, we show using Figure~\ref{fig:fx_deact} that changing the deactivation 
time window does not significantly impact the accuracy of \model model. Although the progressive models are 
unaffected by the deactivation window, the ground truth computed is different across windows, 
and this change is reflected in the error percentage.

\eat{ Lots of the nodes never deactivates in the training datasets. According to section~\ref{sec:params}, the deactivation rates for these nodes are 0, which is obviously an overfit. We cannot learn the deactivation rates directly, so we need to set an default deactivation rate for those nodes.  We know other nodes' deactivation rates, which can provide us with a reasonable range for default deactivation rate. We try different percentiles of the set of unique non-zero deactivation rates as the default deactivation rates, and calculate the percentage error compared to the ground truth. As you can see, default deactivation rate does influence the accuracy, but the difference between \IC and non-progressive models is far more significant. The figure shows deactivation happens in real dataset, and modeling deactivation properly is critical for a reasonable estimation of the real world situation. Moreover, the estimated global active time for \dtime and \model is very similar, which confirms our argument in section 4.3 that \dtime is an approximation of \model.
}

\para{Effect of varying parameters on running time}
\begin{figure*}[t]
\centering
\subfigure[Running time of ConTinEst vs. \model]{
\includegraphics[width=\figfour]{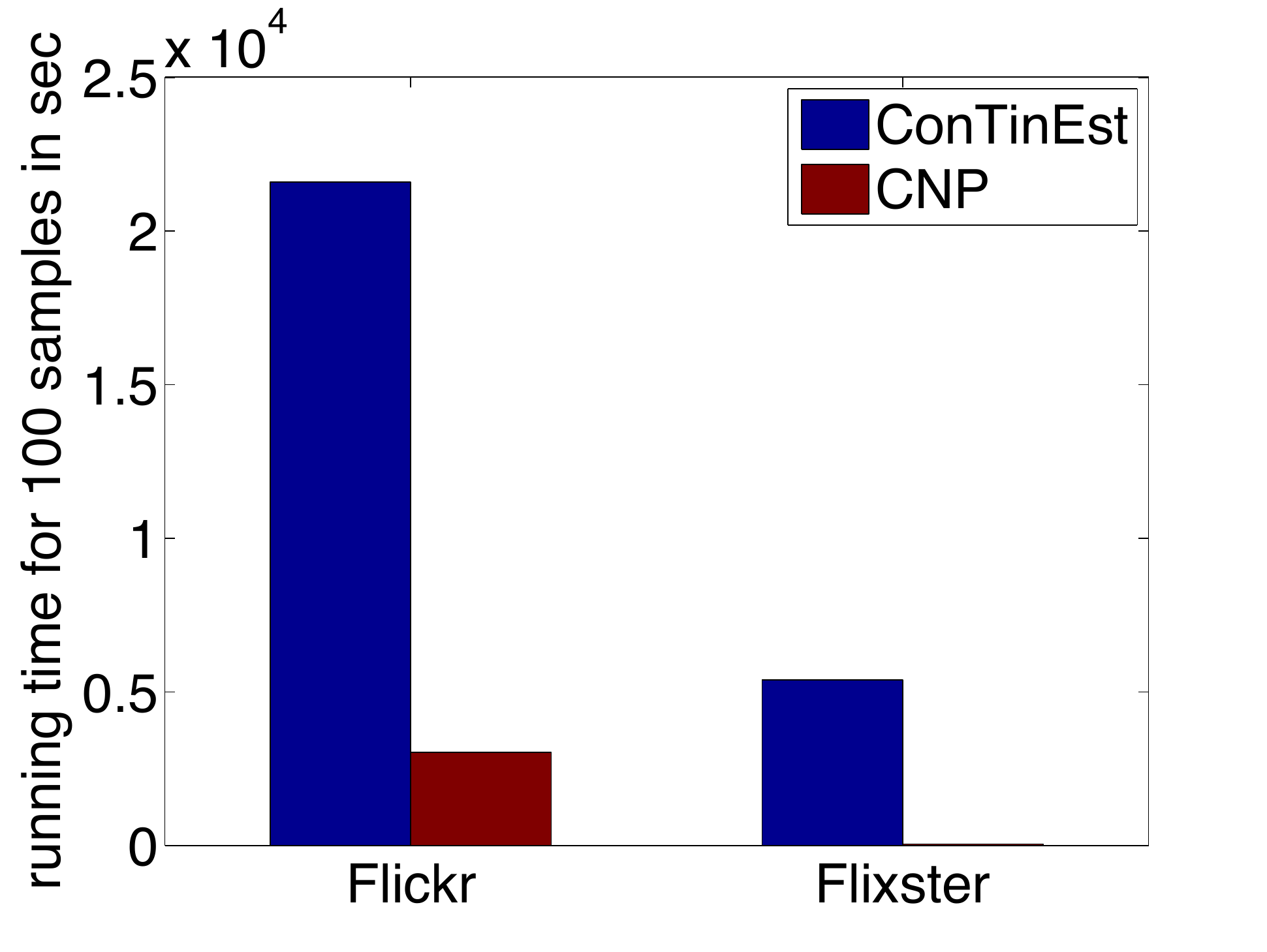}
\label{fig:nips_speed}
}\subfigure[Increasing time horizon Flickr]{
\includegraphics[width=\figfour]{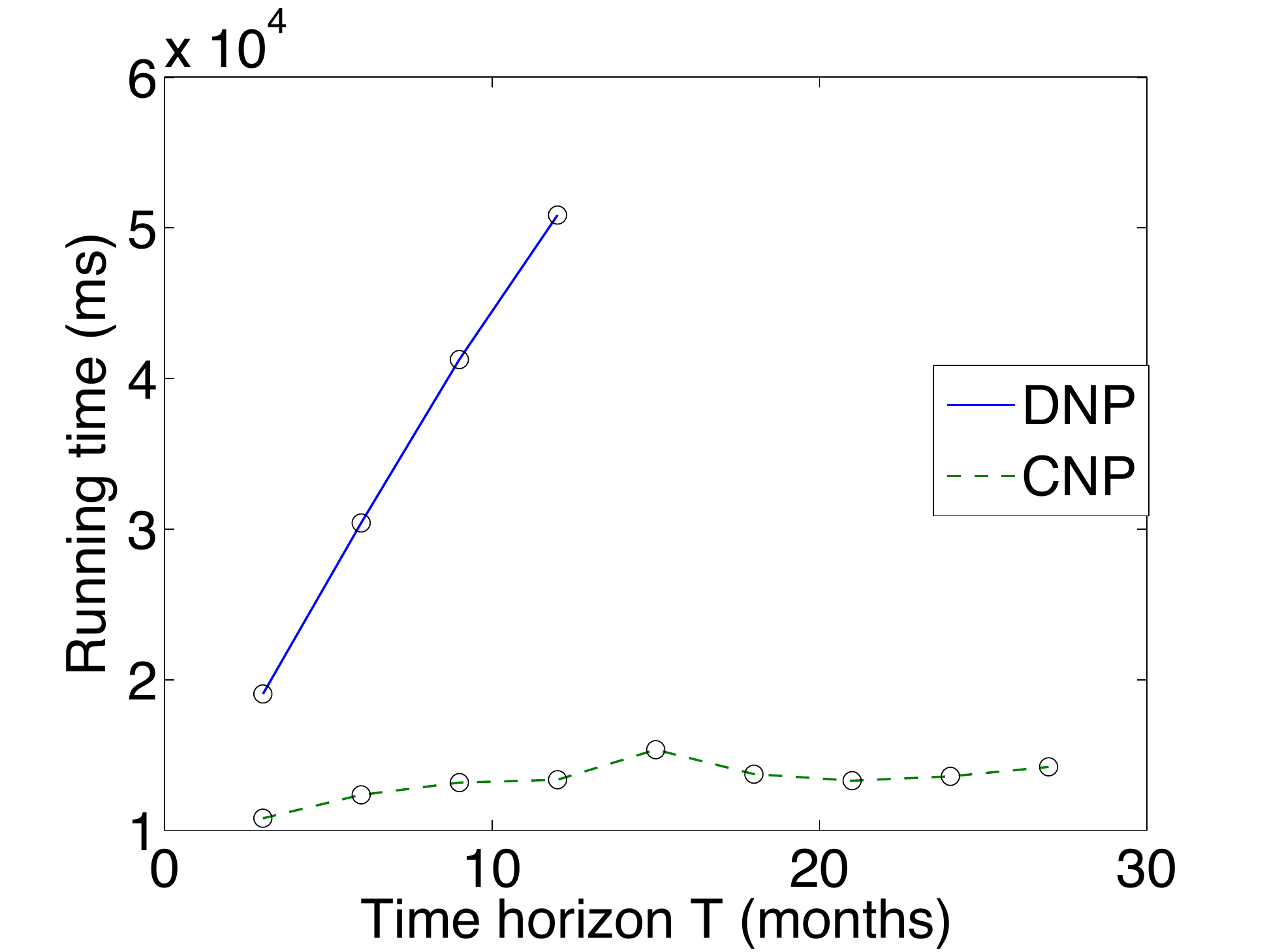}
\label{fig:fl_speed}
}\subfigure[Increasing time horizon Flixster]{
\includegraphics[width=\figfour]{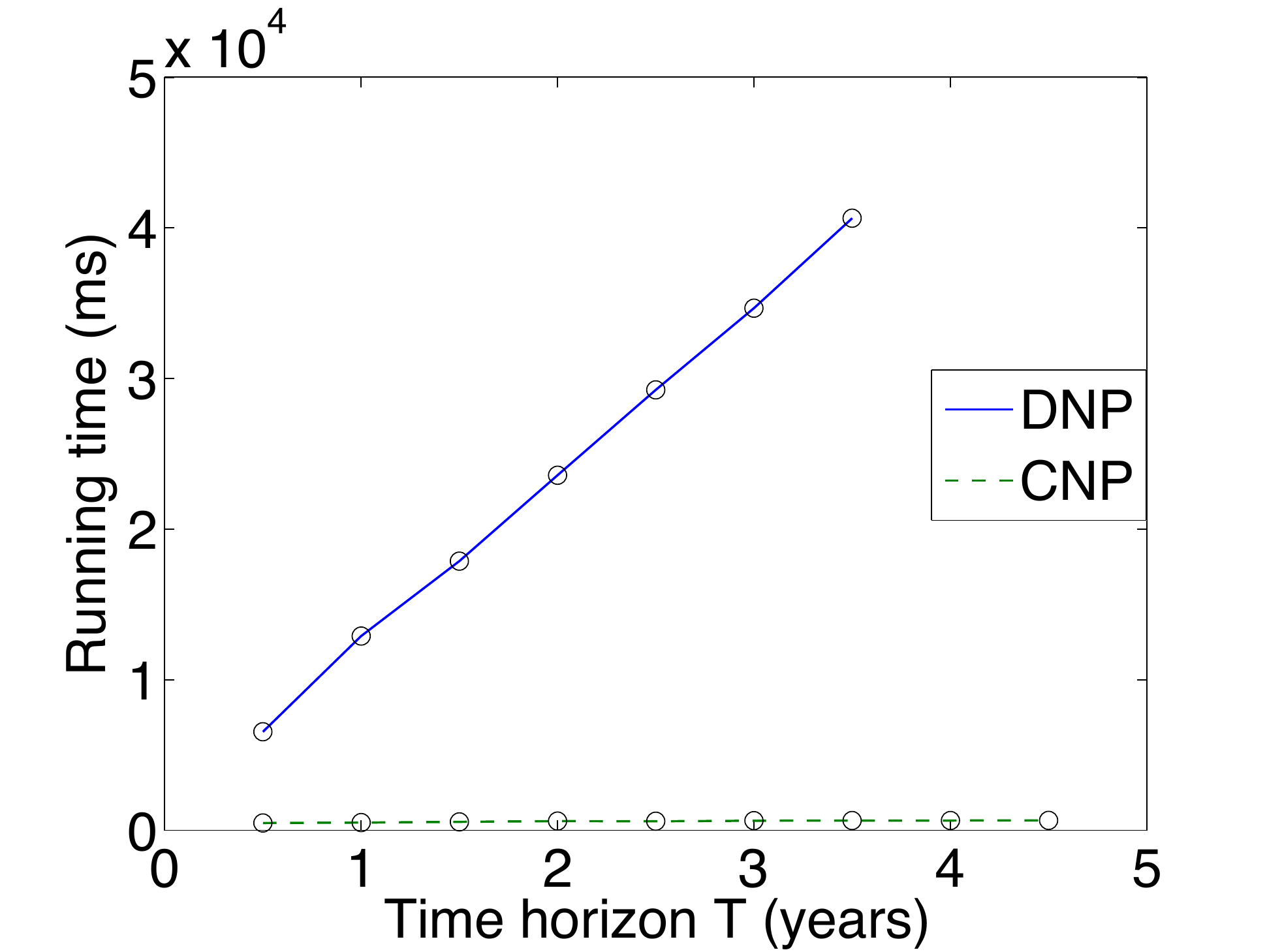}
\label{fig:fx_speed}
}\subfigure[Increasing graph size Flixster]{
\includegraphics[width=\figfour]{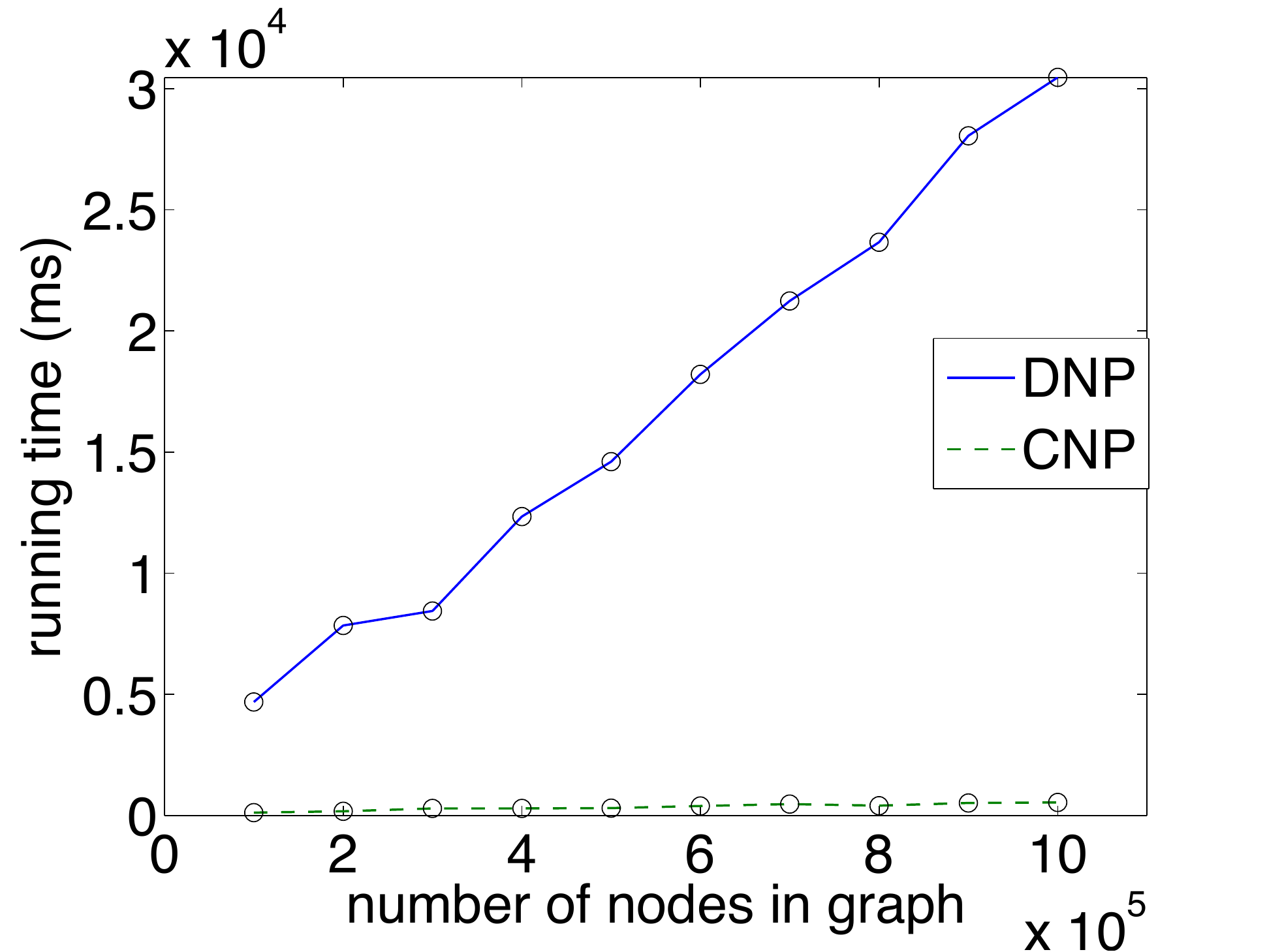}
\label{fig:fx_graphsize}
}
\vspace{-6pt}
\caption{Running time comparisons for non-progressive models: \dtime and \model}
\label{fig:time}
\vspace{-10pt}
\end{figure*}
First, we compare the running time of ConTinEst with \model, as seen in Figure~\ref{fig:nips_speed}, 
our model is an order of magnitude faster than its progressive competitor. 
Next, comparing the non-progressive models. 
The two factors that affect the computational cost of simulating the non-progressive 
models are: time horizon and graph size. 
We compare the computational cost for \model and \dtime for increasing time 
horizon on the full graphs of the two datasets. 
As seen in Figure~\ref{fig:fx_speed}, the running time for 
\dtime increases linearly over increasing time horizon, while that for \model changes only slightly. 
For instance, for Flickr, the running time for \model is 73\% less than \dtime at 27 month time horizon. 
Finally, we set time horizon as 2 years for Flixster and show the running time 
with increasing graph size in Figure~\ref{fig:fx_graphsize}. Again, \model scales up very well. 
The results for Flickr were similar and skipped for brevity.

\para{Progressive Setting}
We ask the question, ``What if the world is progressive, i.e., there are no deactivations, how well would \model perform?''
To this end we perform an experiment of setting the deactivation time window to the end of the time horizon, essentially saying no nodes deactivate. We then compare this model we call CP for the continuous progressive version of our proposed model against \nips and \model. 
We observe that the running time on Flixster for CP, \model and \nips are 60, 64 and 1041s resp. 
The running time results for Flickr for the CP, \model and \nips were 498, 606 and 9605s resp. 
This illustrates that 
despite the data being progressive in nature, our model is 17-20 times faster than the state-of-the-art progressive continuous model.

       \eat{
The computational cost of the simulation models increases with larger graph size and longer time horizon. 
We first use the full graph for both Flixster and Flickr, and show the running time with increasing time horizon. The running time for DNP increases linearly with the time horizon, while the increasing rate of CNP is much slower. For instance, when time horizon is 3 months, the running time for CNP is 43\% smaller than the running time for DNP; when the time horizon is 27 months, the running time for CNP is 73\% smaller. The reason is that at timestamp 0, CNP needs to build the data structure to sample from categorical distribution. In other words, each node has an activation or deactivation process at the beginning; we need to insert all these millions of processes into the data structure at timestamp 0. Once the data structure is built, sampling from categorical distribution and updating the data structure are efficient.
}

\eat{
\begin{figure}
\centering
\includegraphics[width=\figwidth]{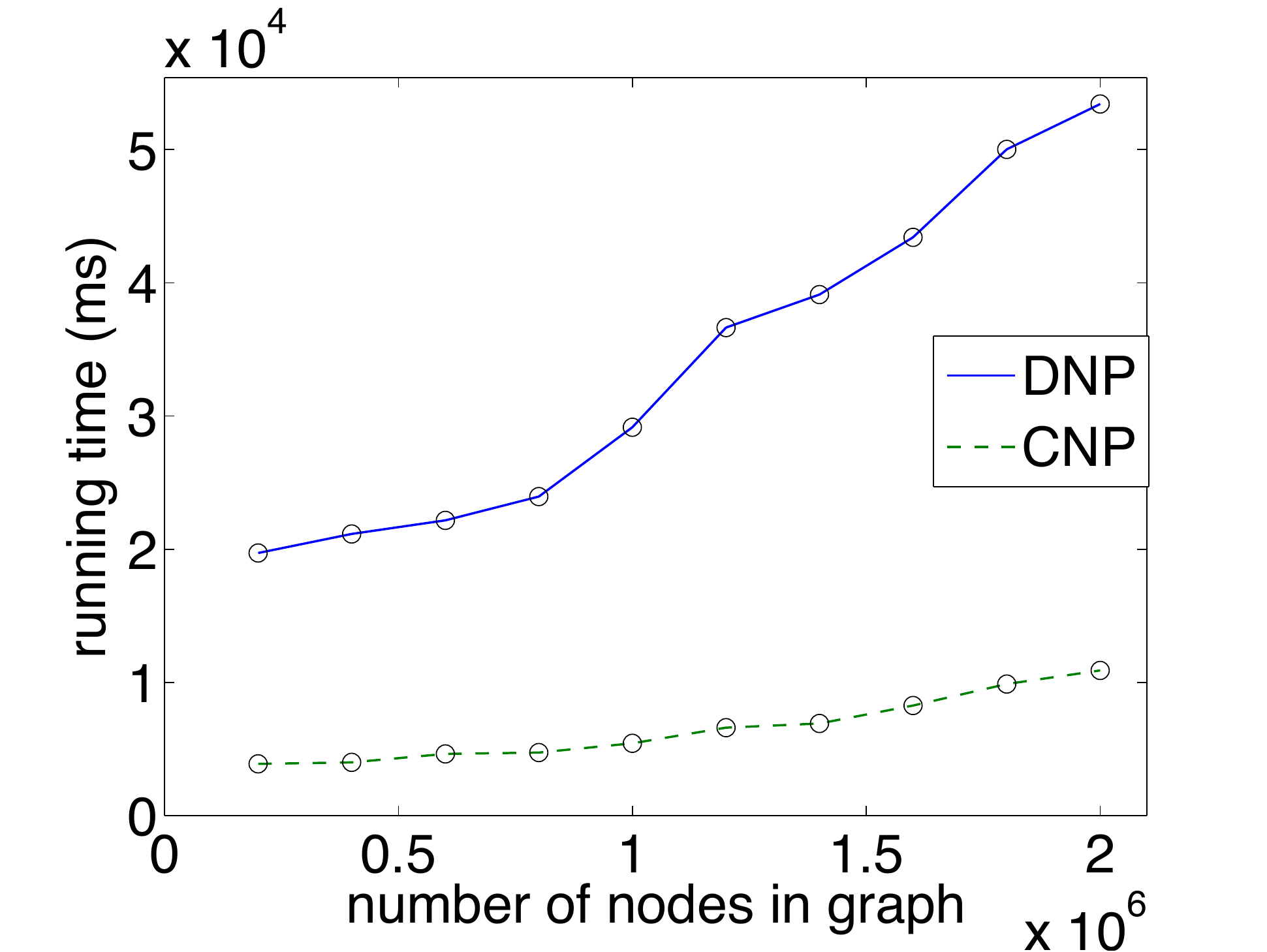}
\caption{Running time on Flickr dataset}
\label{fig:line}
\end{figure}
\begin{figure}
\centering
\includegraphics[width=\figwidth]{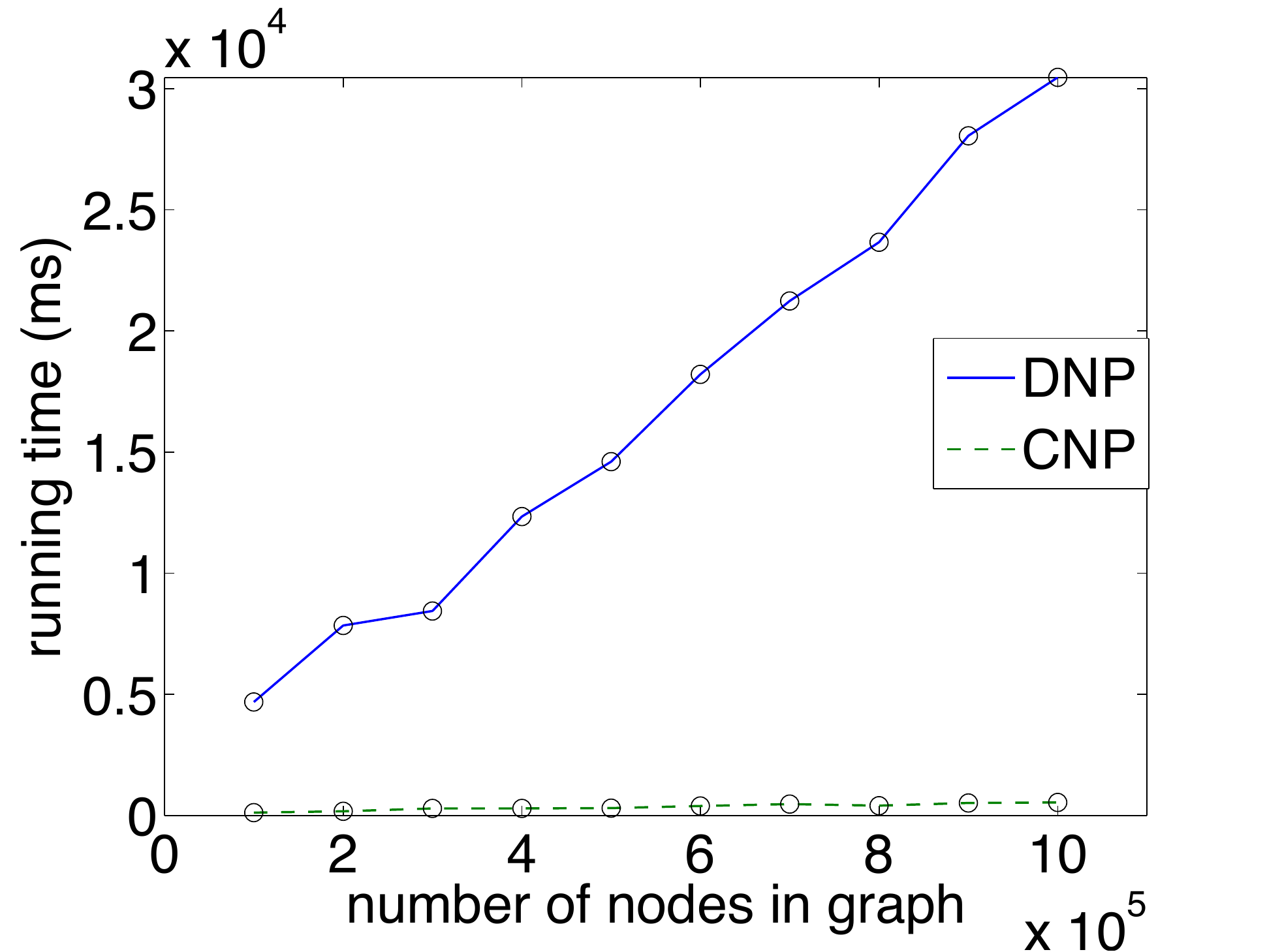}
\caption{Running time on Flixster dataset}
\label{fig:line}
\end{figure}

We also plot the running with different deact time windows to evaluate the influence of deact time window.

\begin{figure}
\centering
\includegraphics[width=\figwidth]{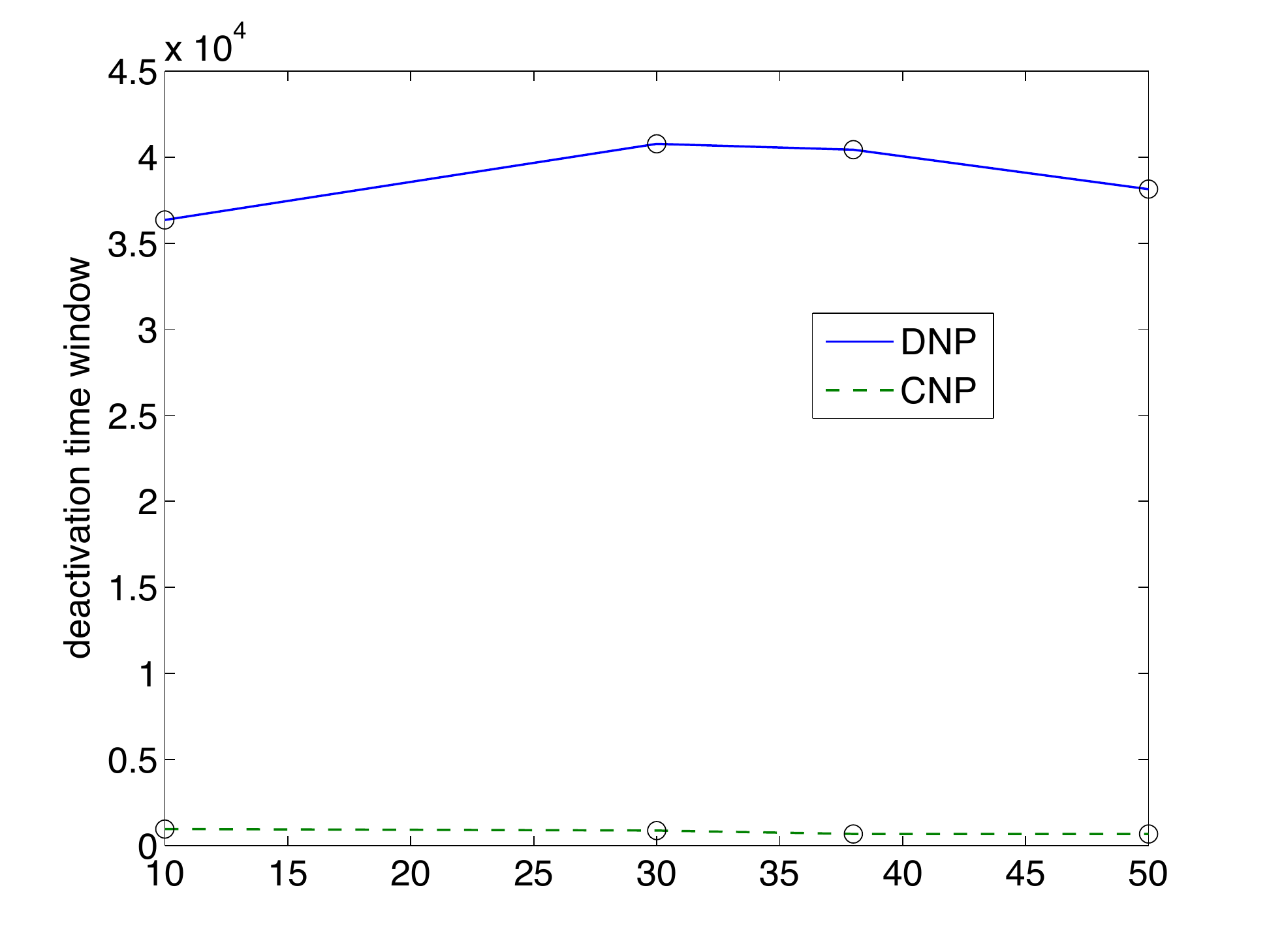}
\caption{Running time on Flixster dataset}
\label{fig:line}
\end{figure}
}
  
}{
\makeatletter{}

\section{Experimental Evaluation}
\label{sec:expn}

\techrep{
\begin{figure*}[th!]
\centering
\subfigure[Deactivation rate - Flixster]{
\includegraphics[width=\figfour]{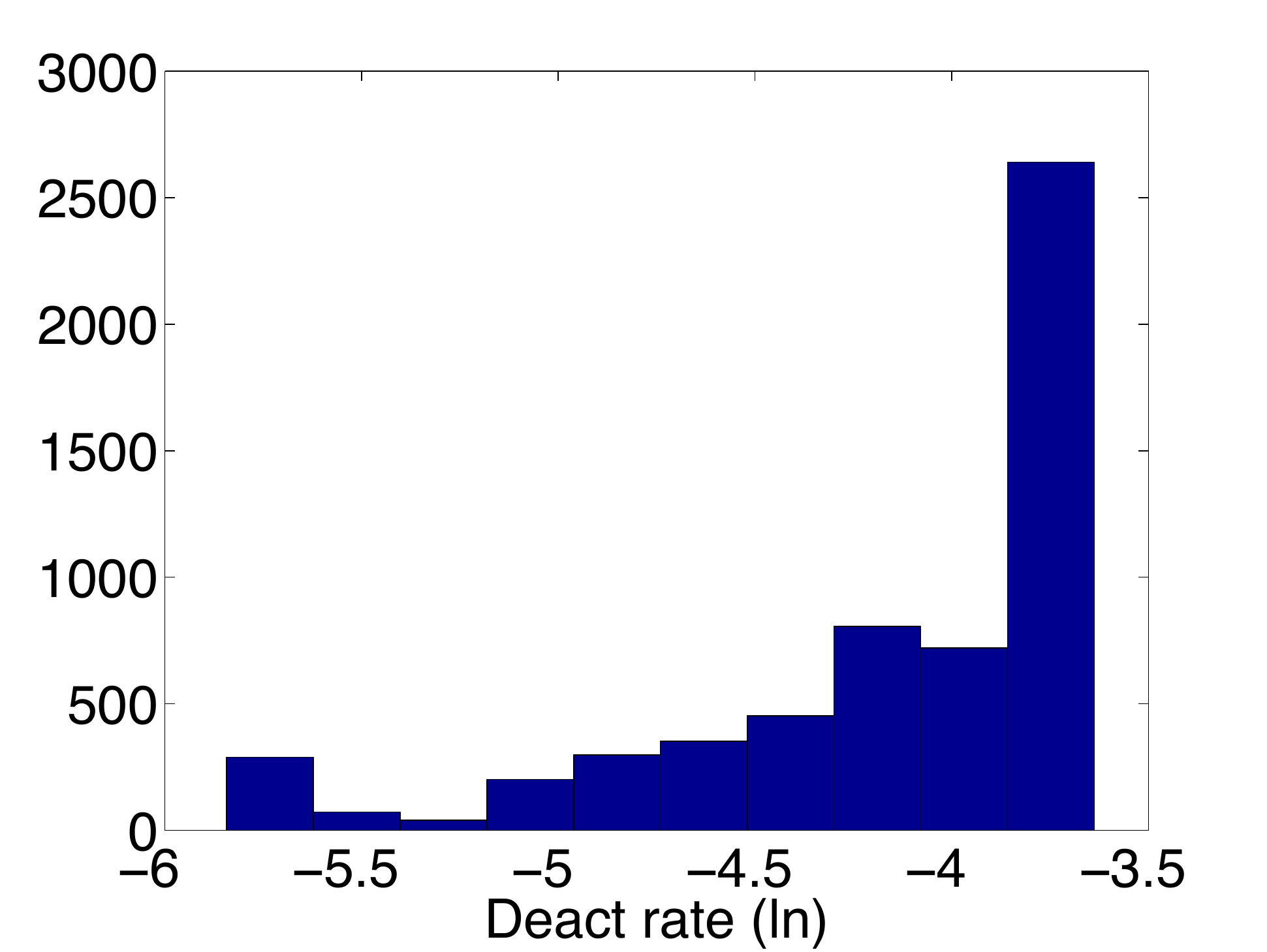}
\label{fig:fx_drate}
}\subfigure[Activation rate  -  Flixster]{
\includegraphics[width=\figfour]{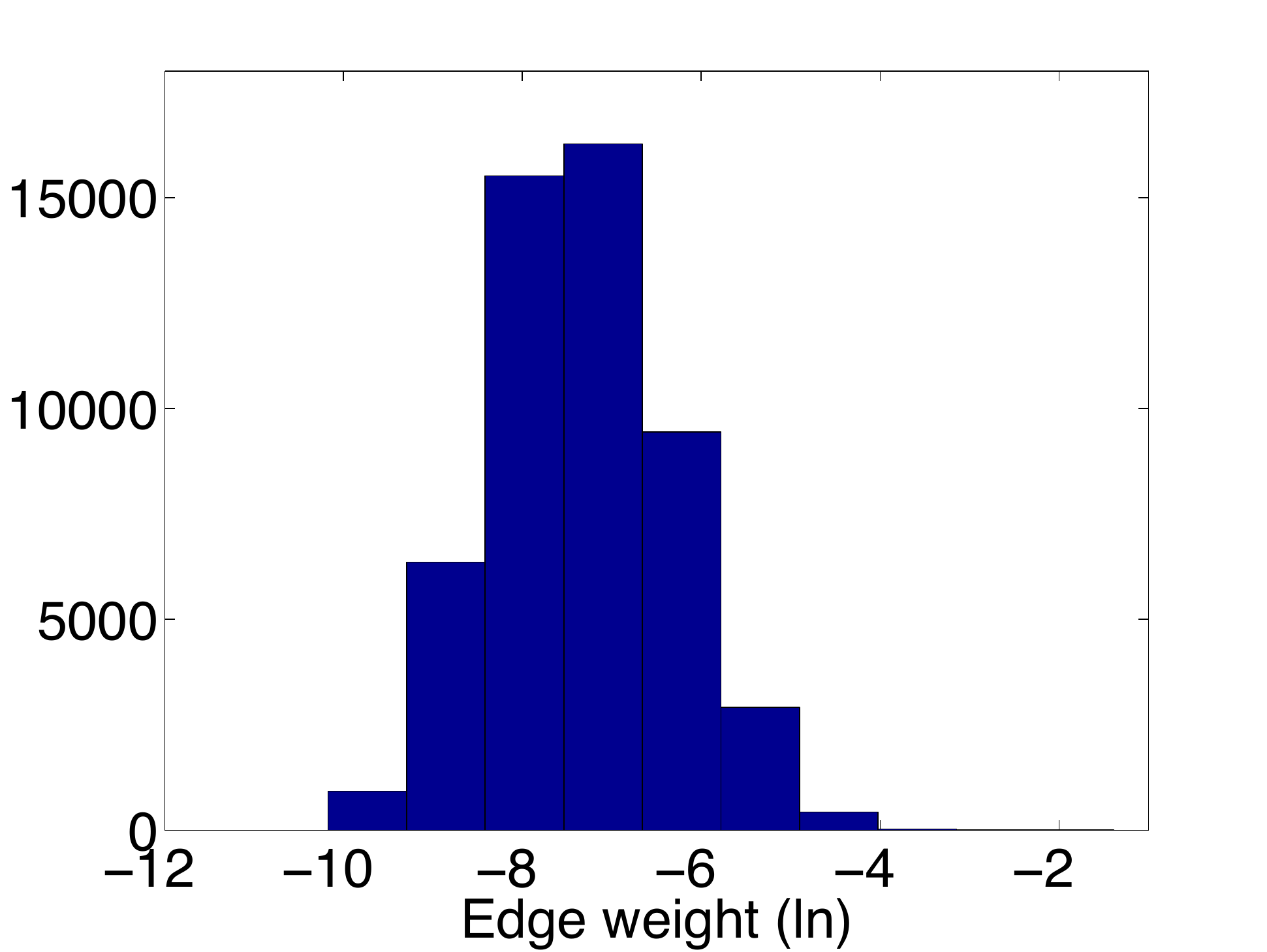}
\label{fig:fx_arate}
}\subfigure[Deactivation rate - Flickr]{
\includegraphics[width=\figfour]{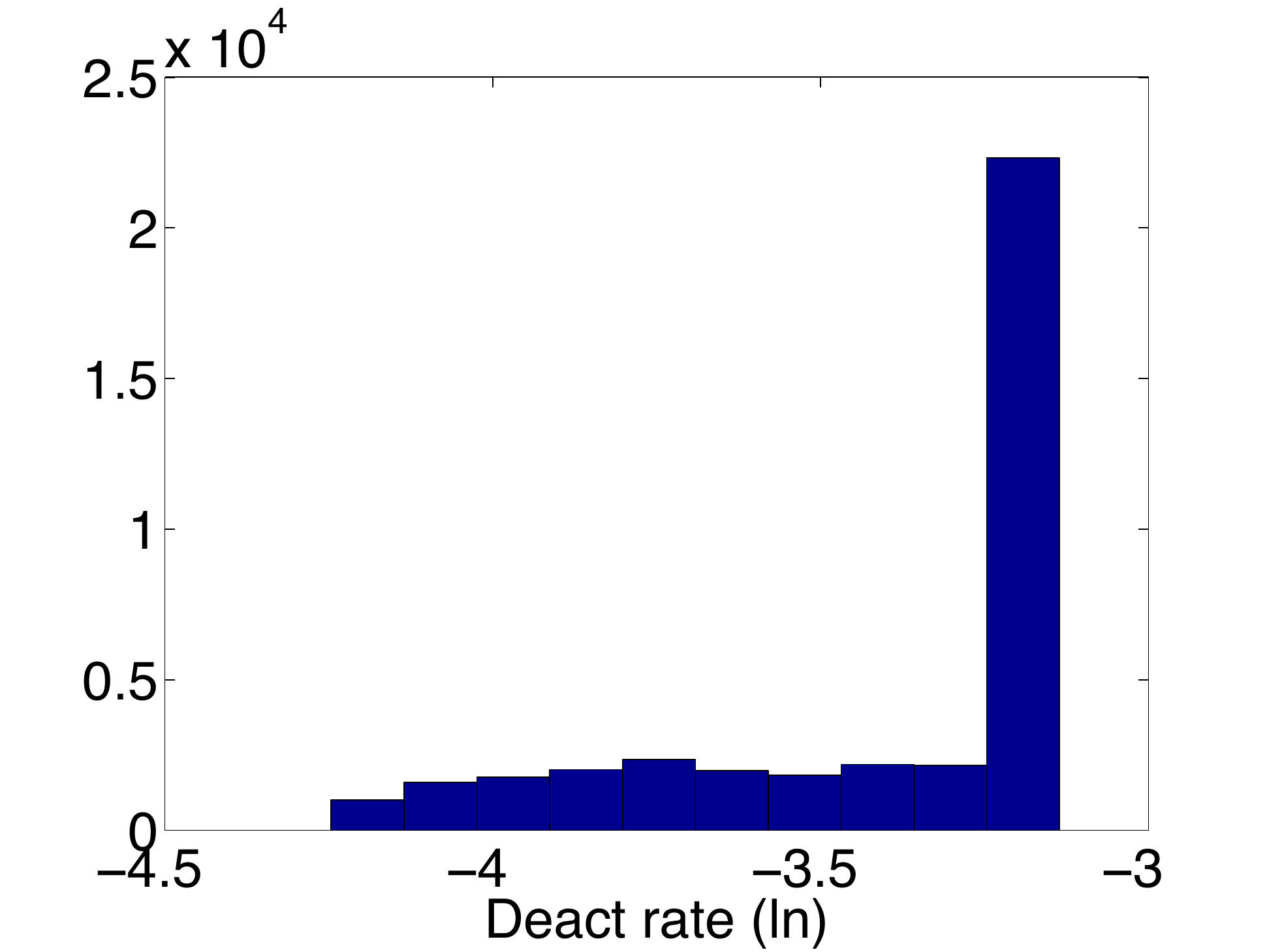}
\label{fig:fl_drate}
}\subfigure[Activation rate - Flickr]{
\includegraphics[width=\figfour]{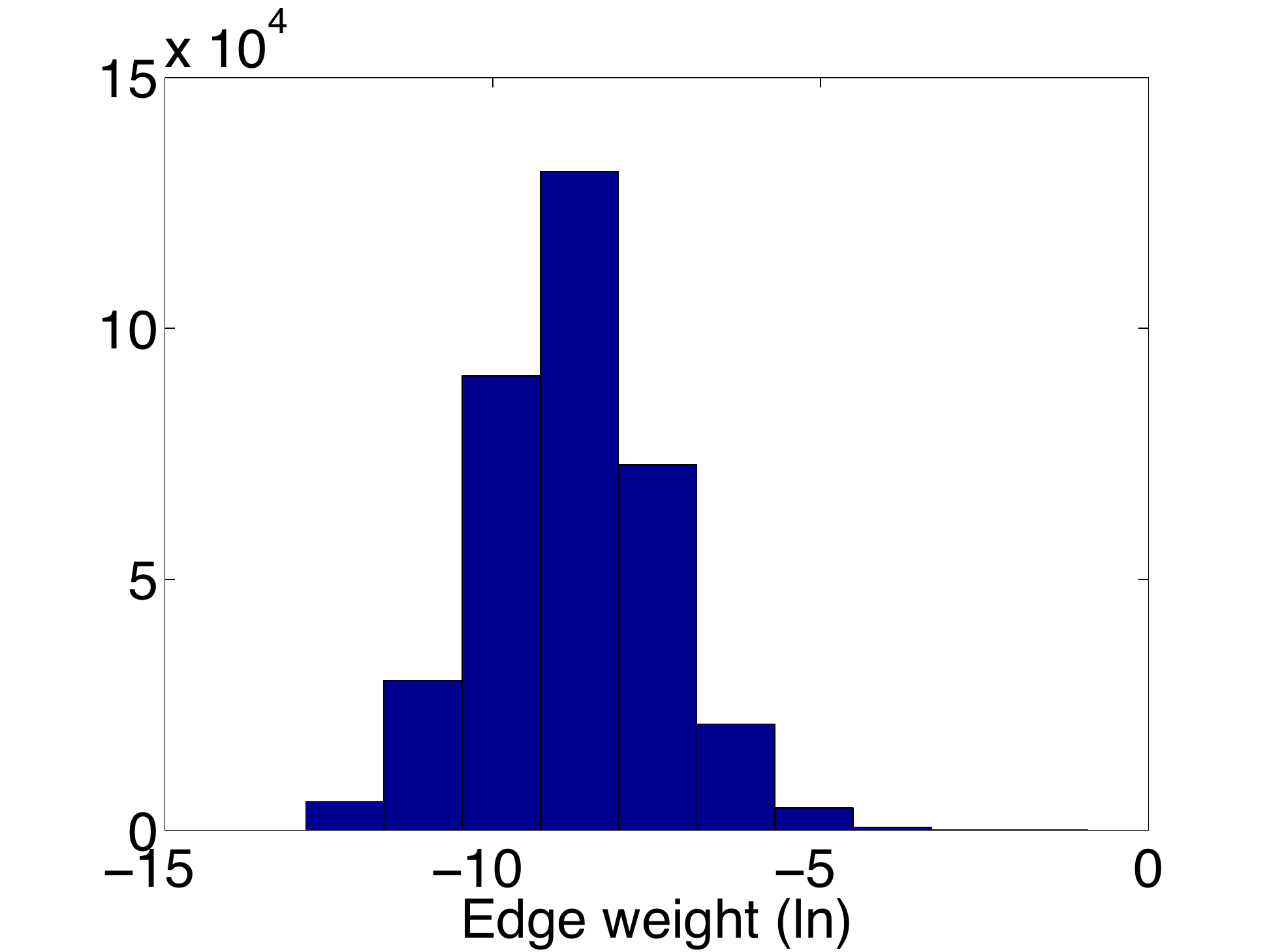}
\label{fig:fl_arate}
}\vspace{-6pt}
\caption{Model Parameters for Flixster and Flickr Datasets}
\label{fig:params}
\vspace{-10pt}
\end{figure*}

In this section we first describe how the edge weights of our model can be learned from data, 
and the details of our experimental setup. Next, we present the results of 
evaluating our model on real and synthetic datasets. We compare the 
accuracy and running time of: traditional \IC, state-of-the-art continuous time progressive model ConTinEst\cite{Du13}, 
\dtime and \model, for estimating the spread as defined in Section~\ref{sec:spread}.

\subsection{Learning model parameters}
\label{sec:params}
We divide each dataset into training and test sets. We use the training set to learn the 
different model parameters, i.e., deactivation rates and edge weights. 
The first challenge we face is identifying deactivations in common datasets. 
If we had access to logs associated with each activations and deactivations, 
for instance, timestamps for service subscriptions and unsubscribe actions, this would be trivial. 
However, such service subscription datasets are not publicly available. 

\subsubsection{Activation and Deactivation}
Given a training set in the form of an action log with \textless user,action,timestamp\textgreater~tuples, 
we would like to find the timestamps for activations and deactivations. 
 We use the following proxy for defining events: 
 1) when a user performs an action we call it an activation event and mark the user active
 2) at each activation we start a timer, the event of the timer running out before another activation occurs, 
 is called a deactivation, and the user is said to be deactivated.  
 The timer is set for a length equal to the {\em deactivation time window}. 
 Similar definitions of active users are used by social networking services as a metric of the popularity 
 of these sites. For instance, Facebook maintains statistics of its daily (and monthly) active users,
 where a day (resp., month) is the deactivation time window. 

\para{Deactivation time window} 
To find the value of the deactivation time window from the action log, 
we first get all the time gaps between two consecutive 
actions of each user in the training dataset. We ignore time gaps of less than one day to 
avoid the bias of a user performing multiple actions in a single session.
We set the deactivation time window as the expected value (mean) of time gaps for all 
experiments. In addition, to be rigorous with our analysis, we also evaluate our model over
different deactivation time windows in the next section.

\para{Rate parameters}
Using the above definition of activation and deactivation events, we now define the 
{\em total active time} for a node $u$ as the sum of time intervals for which node $u$ is active in 
our training set. Then, the deactivation rate parameter $\gamma_{-,u}$ for node $u$ is
\begin{displaymath}deactivation\;rate(u)=\frac{number \; of \; deactivations (u)}{total\; active \;time \; (u)}\end{displaymath}

\noindent
Finally, we define the rate parameters corresponding to edges which are responsible for activations. 
At each activation, we find the set of active neighbors of that node at that timestamp. 
Say, the node has $k$ active neighbors when it activates. 
We assign a contribution score of $1/k$ to each of the $k$ edges from an active neighbor. Then, 
the activation rate parameter $\gamma_{+,u,v}$ for edge $(u,v)$ is
\begin{displaymath}activation\;rate(u,v)=\frac{\sum contribution \;score(u,v)}{total\; active \;time \; (u)}\end{displaymath}

We use these methods to learn the parameters of the \model model. As described in Section~\ref{sec:compare}, we can cast the rate parameters of \model to probabilities for the \dtime model,
 with the $\mathrm{CDF}$ of the exponential distribution.

\spara{\IC edge weights}
The existing methods for \IC models  are not scalable. Our approach is similar to the PCB method in~\cite{Goyal10}.
This is essentially a counting argument and a recent paper~\cite{lin13} uses PCB to learn the edge weights for the IC model and report good accuracy w.r.t to the true edge weights, even better than the likelihood method of Saito et al.~\cite{saito}.
$$Probability(u,v)=\frac{\sum contribution \;score(u,v)}{number\;  of \;activations \; (u)}$$

\subsubsection{Global Influence}
In real social networks, users may get influenced by external sources, such as the popular trends, 
news etc. We model such influences by inserting a ``global node'' and that has a directed edge
to all users. This global node is always active, and influences all nodes equally. 
Technically, the influence from this global node is used to explain activation of a 
node $u$ that has no other active neighbors. Thus, we can compute its total contribution score 
over all its outgoing edges, and define the global influence value as:
$$\frac{total\;  contribution\; score}{time \;horizon \times number \;of \;nodes \;in \;the\; graph}$$

\subsection{Datasets}
We evaluate our model on synthetically generated data and 
two real datasets: Flixster and Flickr, for which we have a social network, 
and an action log which contains the timestamps of users' actions. 

\para{Synthetic}
We generate a random 500 node unweighted forest fire network. We artificially generate cascades and deactivation events. For generating cascades, we pick a node at random and randomly generate a timestamp at which this node performs an action and hence becomes active. To avoid the dependence of cascade generation on any underlying diffusion model, we randomly choose a neighbor and sample a random activation time for it. We repeat this process recursively to generate the cascades. For the deactivation events too, we randomly select a node and make it inactive at a random timestamp. 

\para{Flixster} 
Flixster is a movie rating service, where users form an explicit social network. Our 
dataset~\cite{Flixster} has 1 million nodes (users) and 26.7 million edges (social connections). 
An activation event is the act of {\em rating a movie from a specific genre} (horror in our 
experiments). 
We divide the action log into one year training and one year test set. Using the expectation
of intervals between consecutive actions as the deactivation time window, we find it to be 38 days
for our dataset. The distribution of rate parameters are shown in 
Figures~\ref{fig:fx_drate} and~\ref{fig:fx_arate}, as
computed using the approach described above, and the global influence 
value for this dataset is $2.095\times10^{-5}$.

\para{Flickr} Our second dataset comes from the photo sharing service of Flickr. 
The dataset~\cite{Flickr} has 2.3 million nodes and 33.1 million edges. 
An action corresponds to using the ``favorite photo'' feature, i.e., marking a photo as favorite. 
The associated action log is over 138 days, of which we use the first 70 days as 
our training set for learning parameters, and 
the next 68 days as our test set. The deactivation time window is 23 days and the global influence 
value for our dataset is $1.415\times10^{-4}$. Figures~\ref{fig:fl_drate} and~\ref{fig:fl_arate} show the distribution of rate parameters for Flickr.
}{In this section we compare the 
accuracy and running time of: traditional \IC, state-of-the-art continuous time 
progressive model ConTinEst\cite{Du13}, 
\dtime and \model, for estimating the spread as defined in Section~\ref{sec:spread}.
We evaluate our model on synthetically generated data and 
two real datasets: Flixster and Flickr, for which we have a social network, 
and an action log which contains the timestamps of users' actions. 
The synthetically generated dataset consists of a 500 node graph, and randomly 
generated cascades and deactivation events. The Flixster 
dataset~\cite{Flixster} has 1 million nodes (users) and 26.7 million edges (social connections). 
An activation event is the act of {\em rating a movie from a specific genre}. 
We divide the action log into one year training and one year test set. 
Finally, the Flickr dataset~\cite{Flickr} has 2.3 million nodes and 33.1 million edges. 
An action corresponds to using the ``favorite photo'' feature. 
The associated action log is over 138 days, of which we use the first 70 days as 
our training set for learning parameters, and the next 68 days as our test set. 
Given a training set in the form of an action log with \textless user,action,timestamp\textgreater~tuples, 
we define the activation and deactivation events as: 
 1) when a user performs an action we call it an activation event and mark the user active
 2) at each activation we start a timer, the event of the timer running out before another activation occurs, 
 is called a deactivation, and the user is said to be deactivated.  
The timer is set for a length equal to the {\em deactivation time window}. 
We provide a detailed description of how our model parameters can be learned from data, 
and our experimental setup in our tech report~\cite{techreport}. 
For implementing the sampling without replacement procedure for a 
categorical distribution, see methods described in~\cite{matias93,techreport}.
}

\subsection{Comparison Across Models}
We start with presenting an overview of our results comparing different models: progressive vs. non-progressive models, and discrete vs. continuous time models, across two axes: accuracy of spread estimation and running time. Figure~\ref{fig:models} illustrates this comparison for Flixster dataset. See detailed numbers for both datasets in Table~\ref{tab:acc}. We make the following key observations, and substantiate these with details  through the remainder of this section.
\begin{itemize}
\item Progressive models, here \IC and state-of-the-art \nips, overestimate the spread and result in an error of 80-194\% compared with the ground truth. 
\item Non-progressive models, \dtime and \model, are highly accurate in estimating the spread with very small errors of 0.1-3\%. Notice that \dtime is the non-progressive counterpart of the progressive \IC model and 
improves the accuracy of estimating spread by a 100\%. \item Continuous models \nips and \model are slightly better at estimating accuracy than the discrete models IC and \dtime. 
\item Our model \model is not just more accurate but also an order of magnitude faster than the state-of-the-art continuous time model \nips (Figure~\ref{fig:time_tab}). These results are for running 100 Monte-Carlo simulations.
\end{itemize}

\begin{figure}[t]
\subfigure[Error in Spread estimation]{
\includegraphics[width=\figfour]{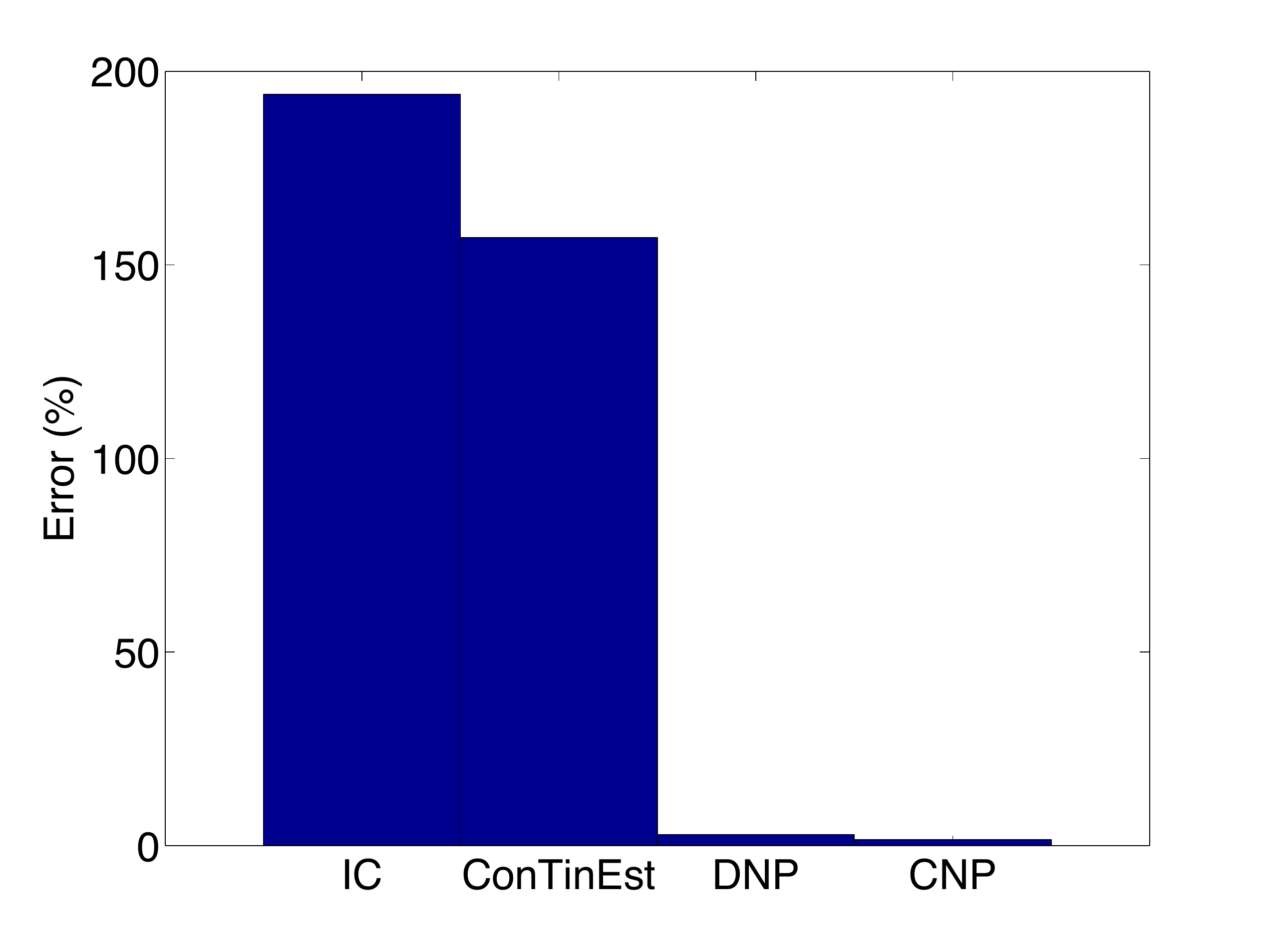}
\label{fig:err_tab}
}\subfigure[Running time]{
\includegraphics[width=\figfour]{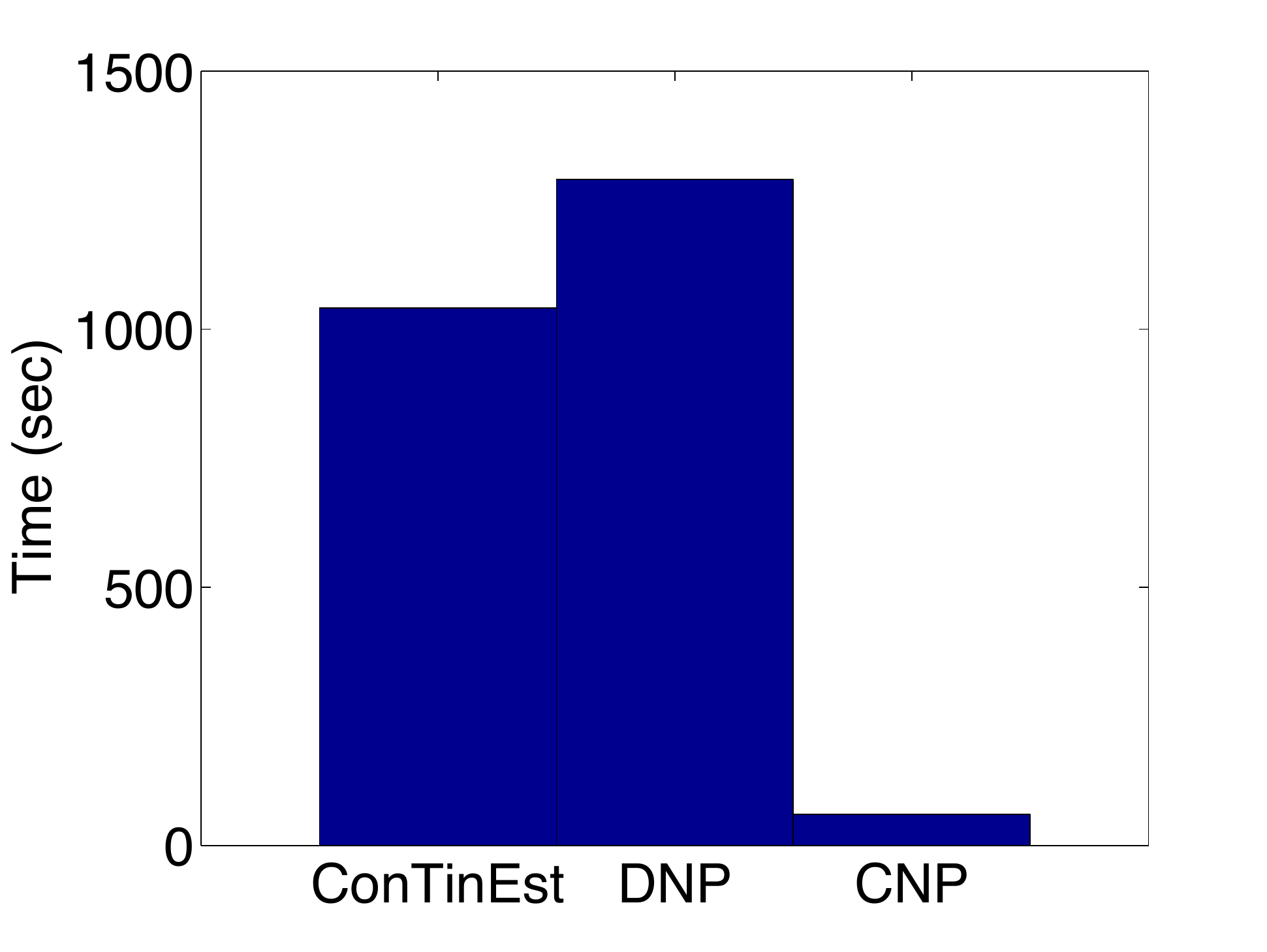}
\label{fig:time_tab}
}
\caption{Spread estimation error and time comparison for progressive (IC, ConTinEst) and non-progressive models (\dtime, \model) for Flixster dataset}
\label{fig:models}
\end{figure}

\begin{table}[t]
\centering
\small
\setlength\tabcolsep{3pt}
  \begin{tabular}{|c|c|c|c|c|c| }
    \hline
    Dataset & Ground & \model & \dtime & ConTinEst & IC \\
                    & Truth & & &\cite{Du13} &  \\ \hline
    Flixster & 964013 & 949750 & 991141 &2477678 & 2833860\\ 
	&& (1.5\%) & (2.8\%)& (157\%)  & (194\%) \\
	\hline
    Flickr & 2435663 & 2432053 & 2409695 & 4423372 & 4922860 \\ 
	&& (0.148\%) & (1.07\%)& (81\%) & (102\%) \\ 
	\hline
  \end{tabular}
  \caption{Spread estimated by progressive (IC, ConTinEst) and non-progressive models (\dtime, \model) and error percentage w.r.t. ground truth}
  \label{tab:acc}
 \vspace{-10pt}
\end{table}
\begin{figure}[t]
\subfigure[Error in Spread estimation for synthetic data]{
\includegraphics[width=\figfour]{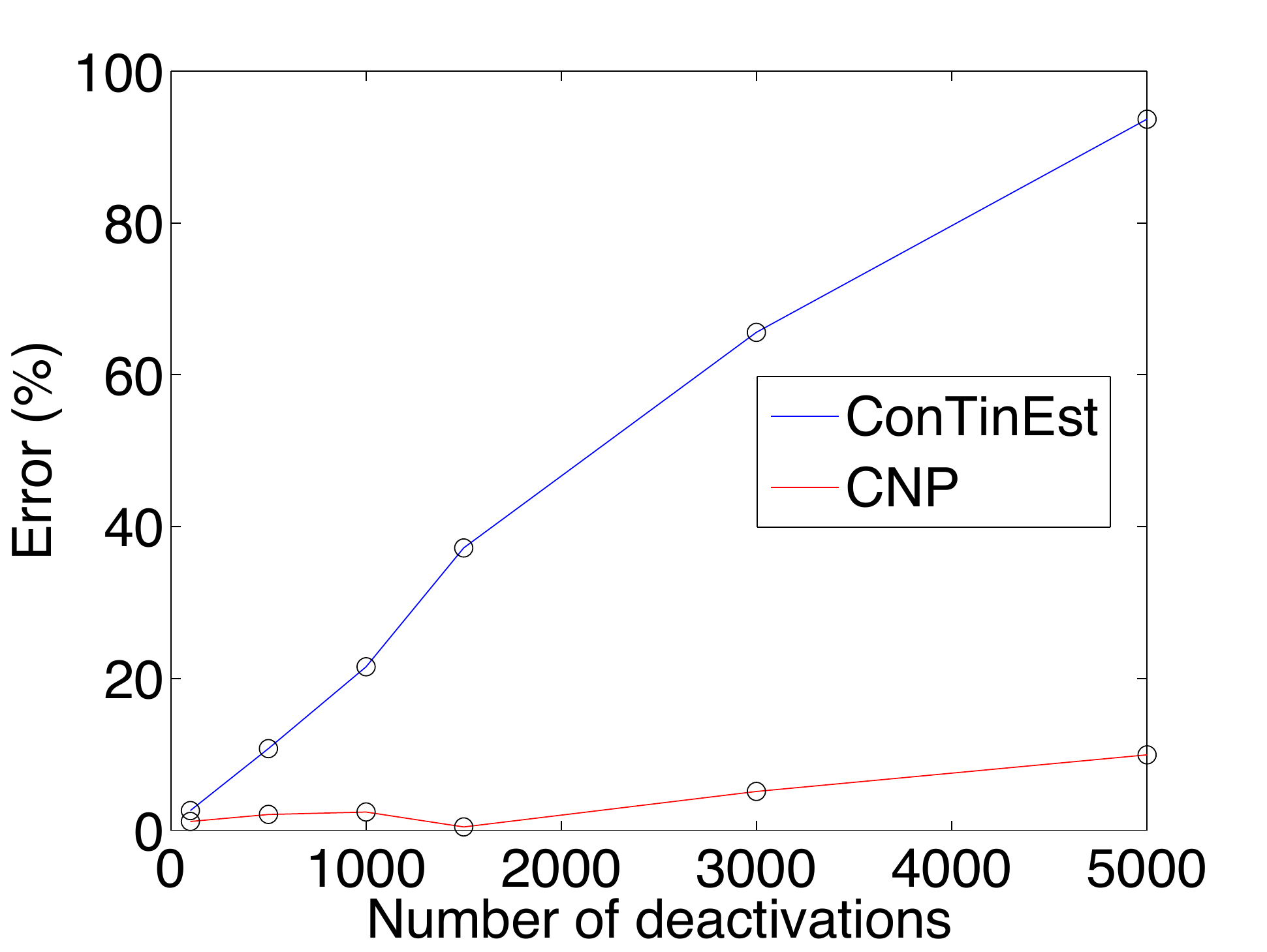}
\label{fig:syn_acc}
}\subfigure[Running time of ConTinEst vs. \model on real data]{
\includegraphics[width=\figfour]{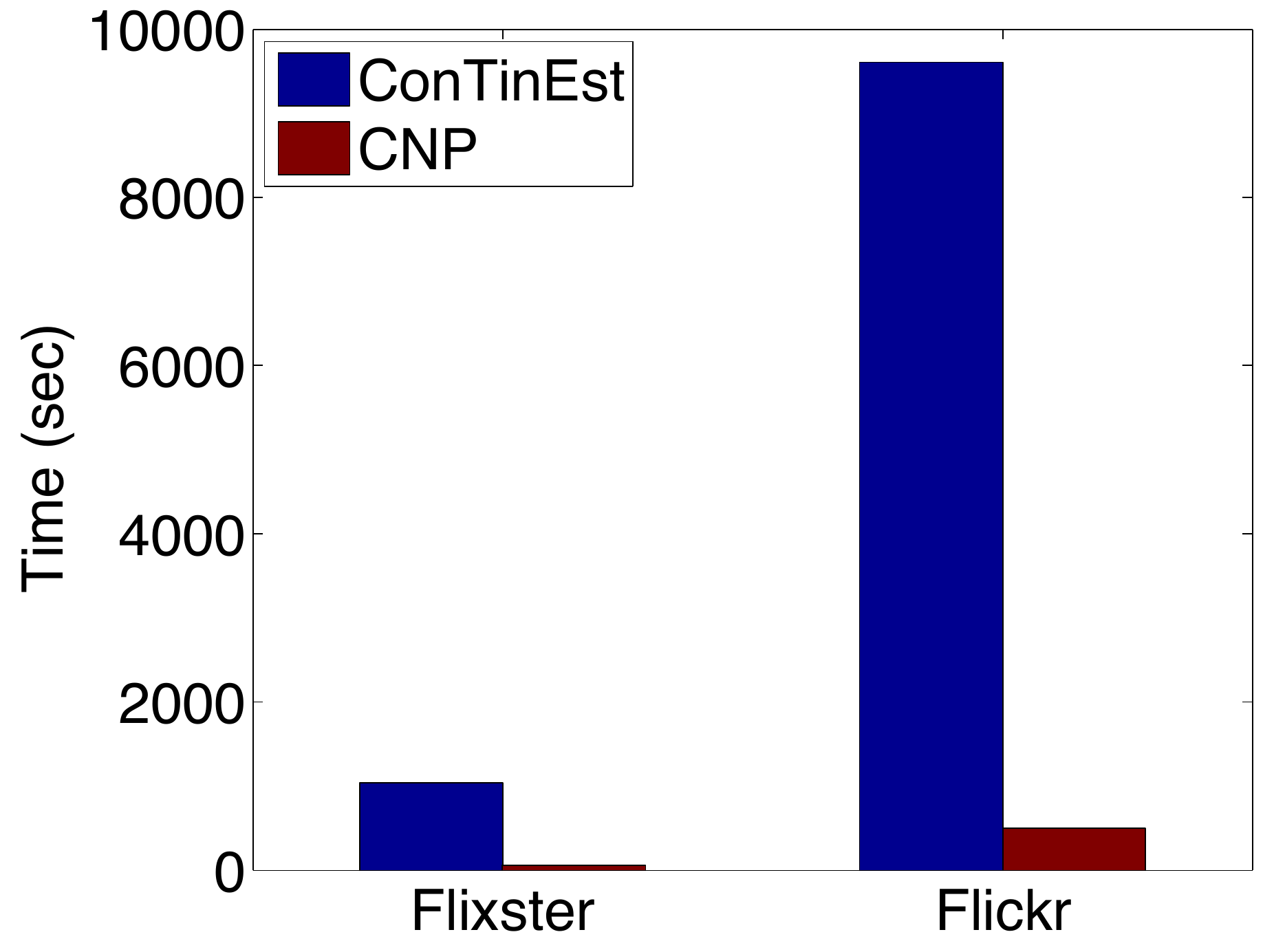} \label{fig:nips_speed}
}\vspace{-3pt}
\caption{Accuracy comparison for progressive (IC, ConTinEst) and non-progressive models (\dtime, \model)}
\label{fig:acc}
\vspace{-6pt}
\end{figure}

\para{Evaluating Accuracy}
We evaluate the accuracy of the estimation of spread of our model by simulating the propagation, 
starting at the state of the network at the last timestamp in our training set, and evaluating 
against the ground truth of spread achieved in the test set. 
In other words, the nodes active at the end of the training set are treated as the seed set, and 
the propagation is run for the time horizon equal to the length of the test set. 
The {\em ground truth} of spread is computed as the total active time of all nodes for the test set. 
Table~\ref{tab:acc} shows the spread as estimated by our model compared with 
the ground truth. For our model, error in spread estimation is just 1.5\% and 0.1\% over the Flixster 
and Flickr datasets resp. 
The difference between \IC and non-progressive models is two orders of magnitude. 
This validates that deactivation occurs in real datasets, and that modeling 
deactivation properly is critical for a reasonable estimation of influence spread 
in non-progressive settings.

Next, we perform an experiment on synthetic data to show the impact of 
number of deactivations on the spread estimates by a progressive model (\nips~\cite{Du13}) and our
\model model. Figure~\ref{fig:syn_acc} shows this the error (\%) in estimating the spread. 
As the deactivations increase, the gap in the accuracy of the two methods increase, with \model 
performing over 83\% better than the competitor, establishing that in the 
presence of deactivations, non-progressive phenomena are modeled accurately by \model.

\para{Evaluating Computational Cost}
We compare the running time of ConTinEst with \model for our two real datasets Flickr and Flixster for running
100 Monte-Carlo simulations. As seen in Figure~\ref{fig:nips_speed}, 
our model is an order of magnitude faster than its progressive competitor.

\begin{figure}[t]
\subfigure[Varying deactivation rate]{
\includegraphics[width=\figfour]{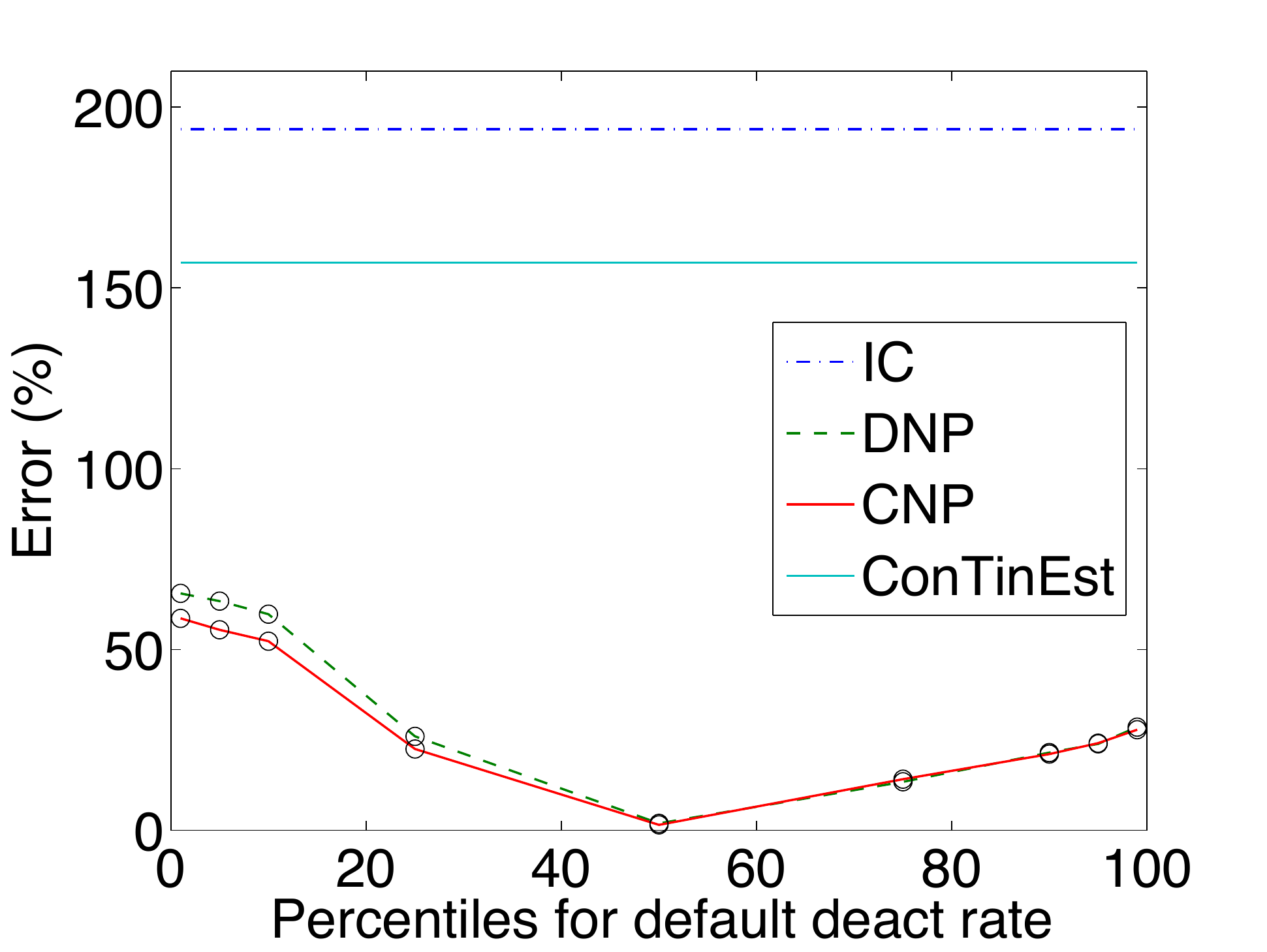}
\label{fig:fx_accuracy}
}\subfigure[Varying deactivation time window for Flixster]{
\includegraphics[width=\figfour]{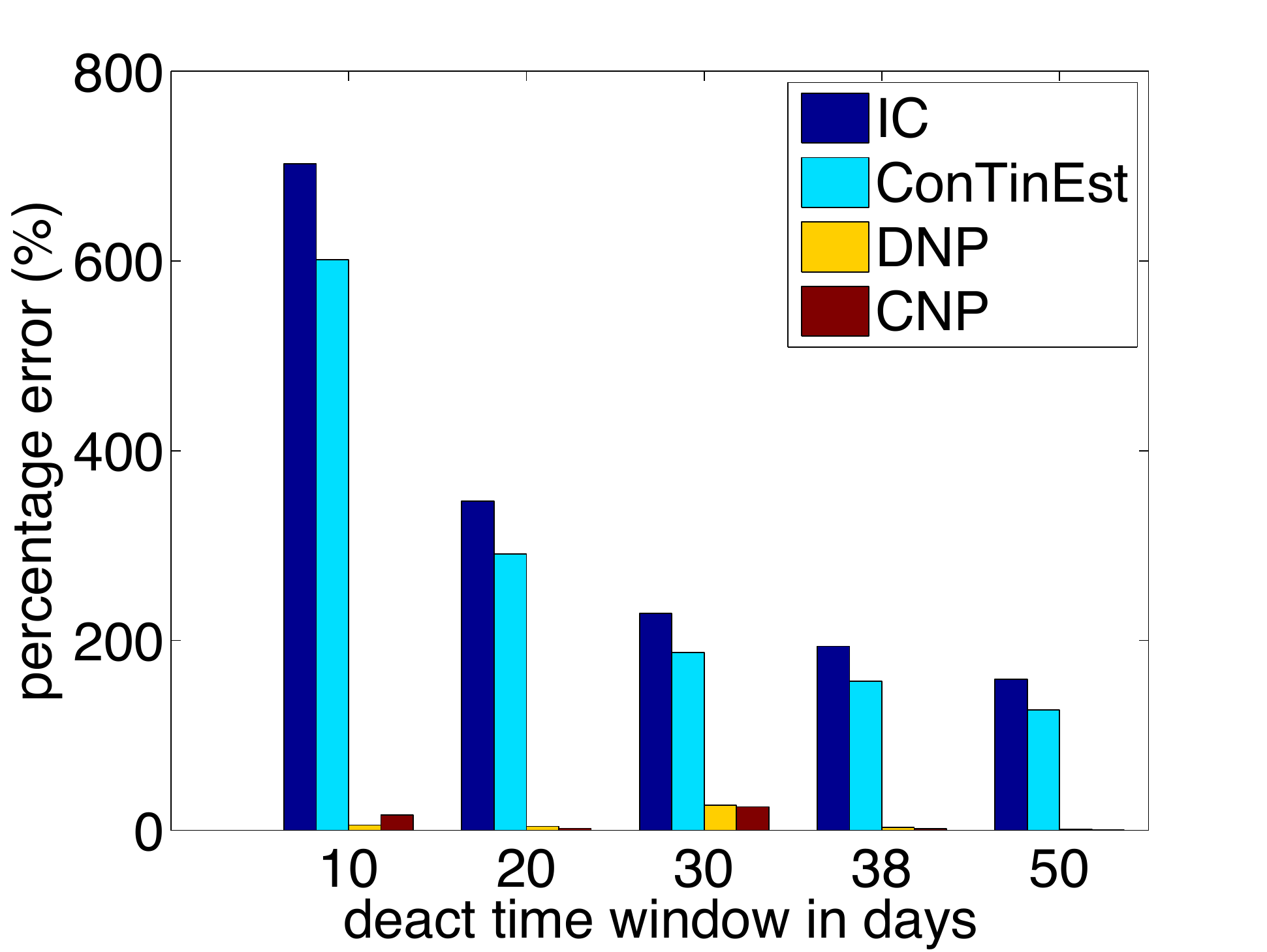}
\label{fig:fx_deact}
}\vspace{-3pt}
\caption{Accuracy comparison for progressive (IC, ConTinEst) and non-progressive models (\dtime, \model) 
with varying deactivation parameters}
\label{fig:acc}
\vspace{-6pt}
\end{figure}

\subsection{Varying Parameter Values}
\para{Effect of deactivation parameters on accuracy}
For Flickr, we observe that 98\% of the nodes perform no action in our training set, and hence 
get a zero deactivation rate. 
This is an artifact of the short timespan of the training data. 
Filtering those nodes would result in a disconnected graph.  To overcome this shortcoming of the 
data sample and to avoid overfitting, we  assign a {\em default deactivation rate} to all such nodes. 
We use the set of non-zero deactivation rates (as learned from the data) as a guideline, 
and test different percentiles of this set as the default deactivation rate. Instead of fixing the value, 
we evaluate its impact on the accuracy by varying it, as shown in 
Figure~\ref{fig:fx_accuracy} for Flixster dataset.
As seen in the figure, the default deactivation rate does impact the accuracy slightly, still the 
estimates by \model are orders of magnitude more accurate than \IC and ConTinEst. 
Also notice that the estimated  spread for \dtime and \model is very similar validating  
our argument in Section~\ref{sec:compare} that \dtime is an approximation of \model.
The plots for Flickr are skipped for brevity, but the methodology adopted and results were similar. For the remainder of the experiments we set the default deactivation rate for Flixster and Flickr to the best obtained, i.e., $50^{th}$ and 1 percentile resp. of the unique non-zero deactivation rates learned. 

Next, we show using Figure~\ref{fig:fx_deact} that changing the deactivation 
time window does not significantly impact the accuracy of \model model. Although the progressive models are 
unaffected by the deactivation window, the ground truth computed is different across windows, 
and this change is reflected in the error percentage.

\eat{ Lots of the nodes never deactivates in the training datasets. According to section~\ref{sec:params}, the deactivation rates for these nodes are 0, which is obviously an overfit. We cannot learn the deactivation rates directly, so we need to set an default deactivation rate for those nodes.  We know other nodes' deactivation rates, which can provide us with a reasonable range for default deactivation rate. We try different percentiles of the set of unique non-zero deactivation rates as the default deactivation rates, and calculate the percentage error compared to the ground truth. As you can see, default deactivation rate does influence the accuracy, but the difference between \IC and non-progressive models is far more significant. The figure shows deactivation happens in real dataset, and modeling deactivation properly is critical for a reasonable estimation of the real world situation. Moreover, the estimated global active time for \dtime and \model is very similar, which confirms our argument in section 4.3 that \dtime is an approximation of \model.
}

\para{Effect of varying parameters on running time}
\begin{figure}[t]
\subfigure[Increasing time horizon]{
\includegraphics[width=\figfour]{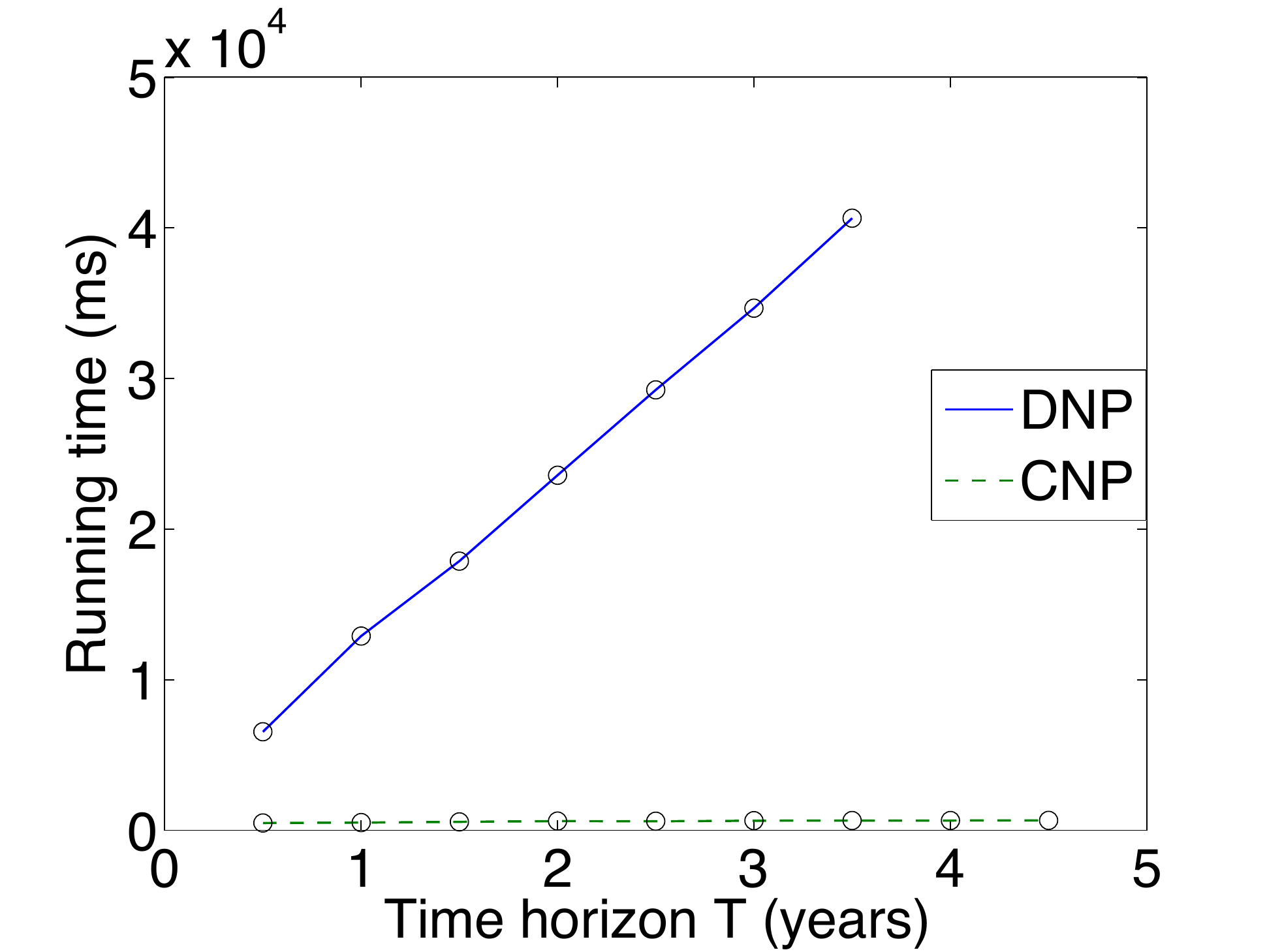}
\label{fig:fx_speed}
}\subfigure[Increasing graph size]{
\includegraphics[width=\figfour]{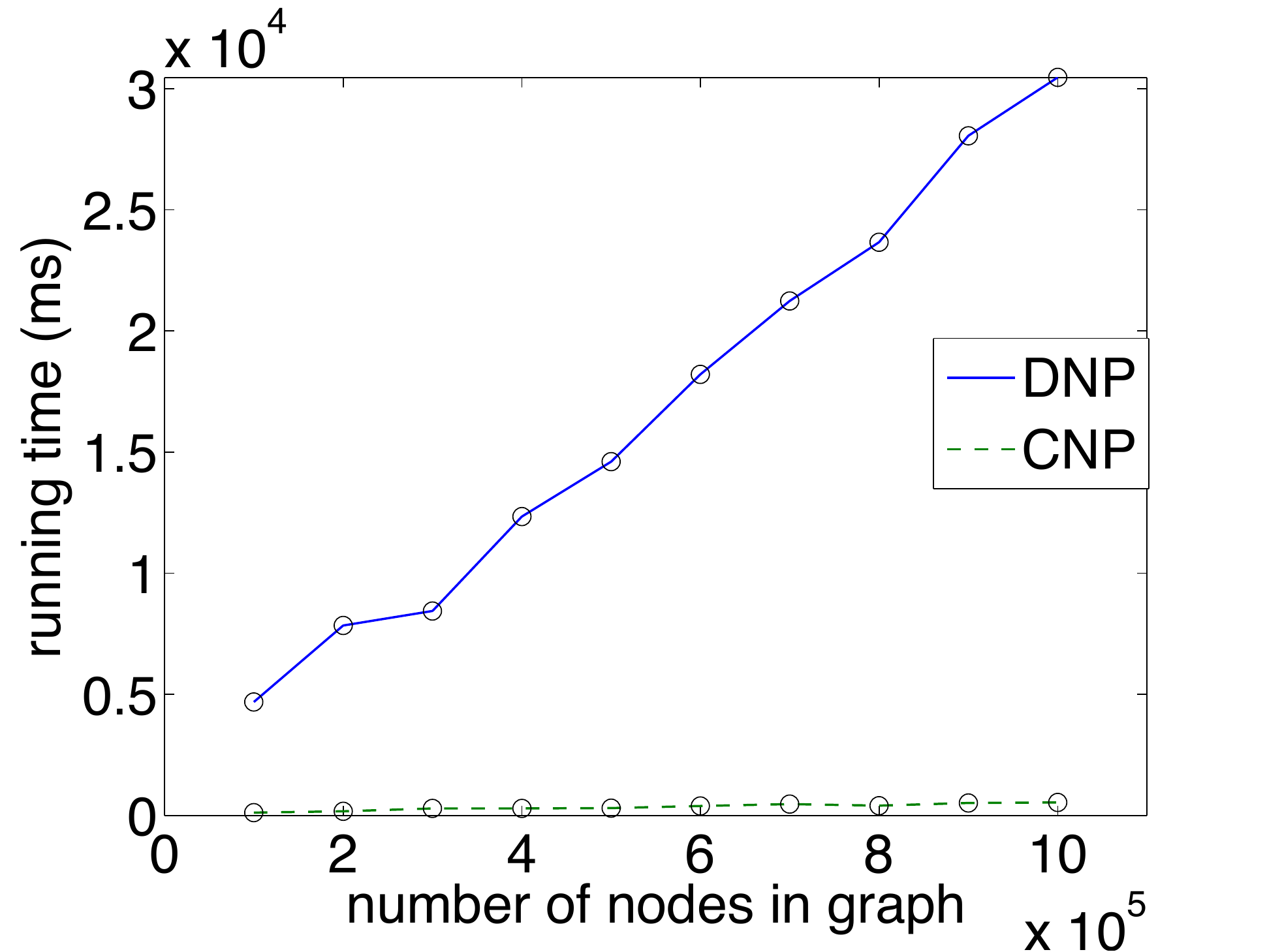}
\label{fig:fx_graphsize}
}
\vspace{-6pt}
\caption{Running time comparisons on Flixster dataset for non-progressive models: \dtime and \model}
\label{fig:time}
\vspace{-10pt}
\end{figure}
The two factors that affect the computational cost of simulating the non-progressive 
models are: time horizon and graph size. 
We compare the computational cost for \model and \dtime for increasing time 
horizon on the full graphs of the two datasets. 
As seen in Figure~\ref{fig:fx_speed}, the running time for 
\dtime increases linearly over increasing time horizon, while that for \model changes only slightly. 
For instance, for Flickr, the running time for \model is 73\% less than \dtime at 27 month time horizon. 
Finally, we set time horizon as 2 years for Flixster and show the running time 
with increasing graph size in Figure~\ref{fig:fx_graphsize}. Again, \model scales up very well. 
The results for Flickr were similar and skipped for brevity.

\para{Progressive Setting}
We ask the question, ``What if the world is progressive, i.e., there are no deactivations, how well would \model perform?''
To this end we perform an experiment of setting the deactivation time window to the end of the time horizon, essentially saying no nodes deactivate. We then compare this model we call CP for the continuous progressive version of our proposed model against \nips and \model. 
We observe that the running time on Flixster for CP, \model and \nips are 60, 64 and 1041s resp. 
The running time results for Flickr for the CP, \model and \nips were 498, 606 and 9605s resp. 
This illustrates that 
despite the data being progressive in nature, our model is 17-20 times faster than the state-of-the-art progressive continuous model.

       \eat{
The computational cost of the simulation models increases with larger graph size and longer time horizon. 
We first use the full graph for both Flixster and Flickr, and show the running time with increasing time horizon. The running time for DNP increases linearly with the time horizon, while the increasing rate of CNP is much slower. For instance, when time horizon is 3 months, the running time for CNP is 43\% smaller than the running time for DNP; when the time horizon is 27 months, the running time for CNP is 73\% smaller. The reason is that at timestamp 0, CNP needs to build the data structure to sample from categorical distribution. In other words, each node has an activation or deactivation process at the beginning; we need to insert all these millions of processes into the data structure at timestamp 0. Once the data structure is built, sampling from categorical distribution and updating the data structure are efficient.
}

\eat{
\begin{figure}
\centering
\includegraphics[width=\figwidth]{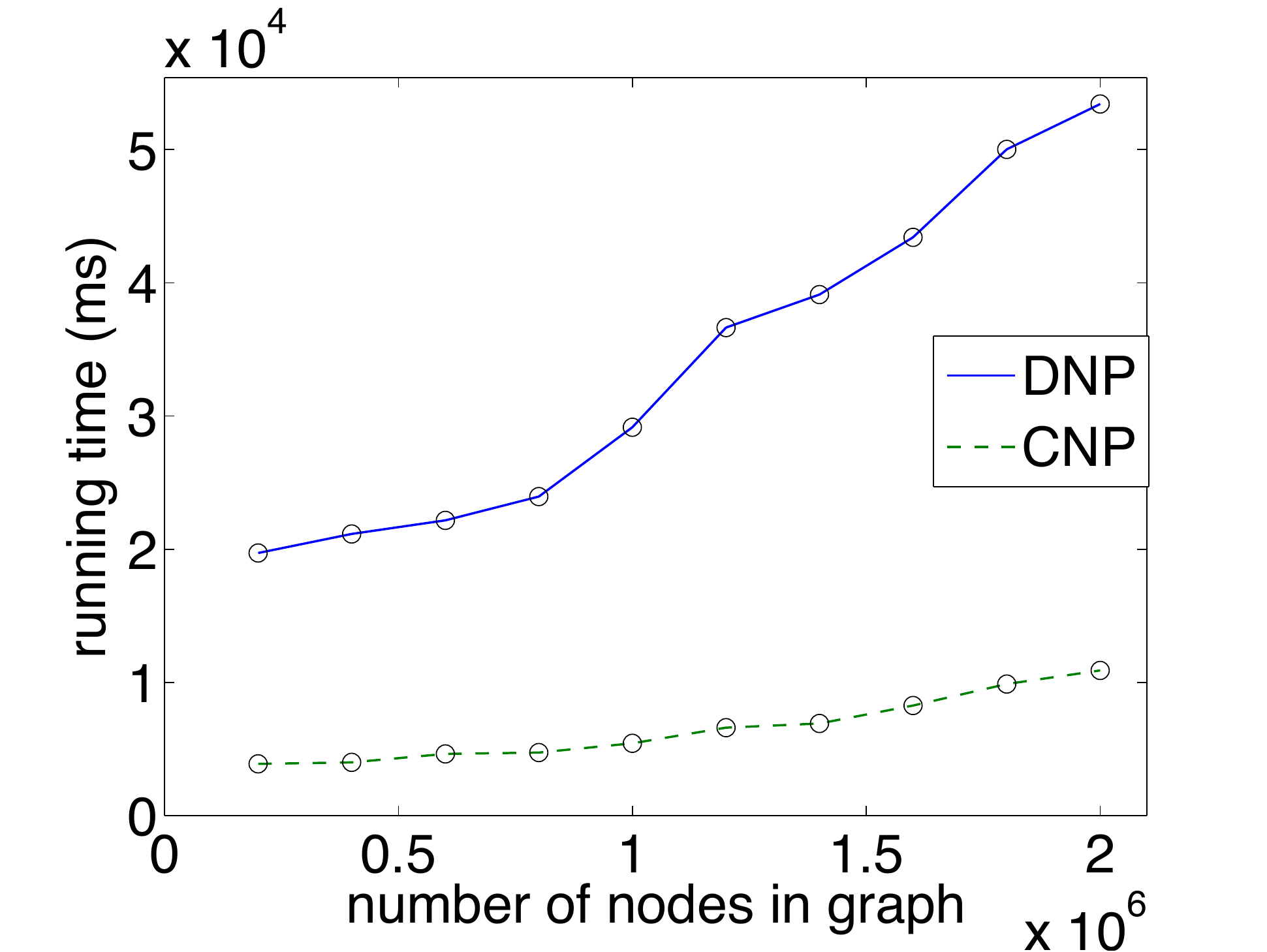}
\caption{Running time on Flickr dataset}
\label{fig:line}
\end{figure}
\begin{figure}
\centering
\includegraphics[width=\figwidth]{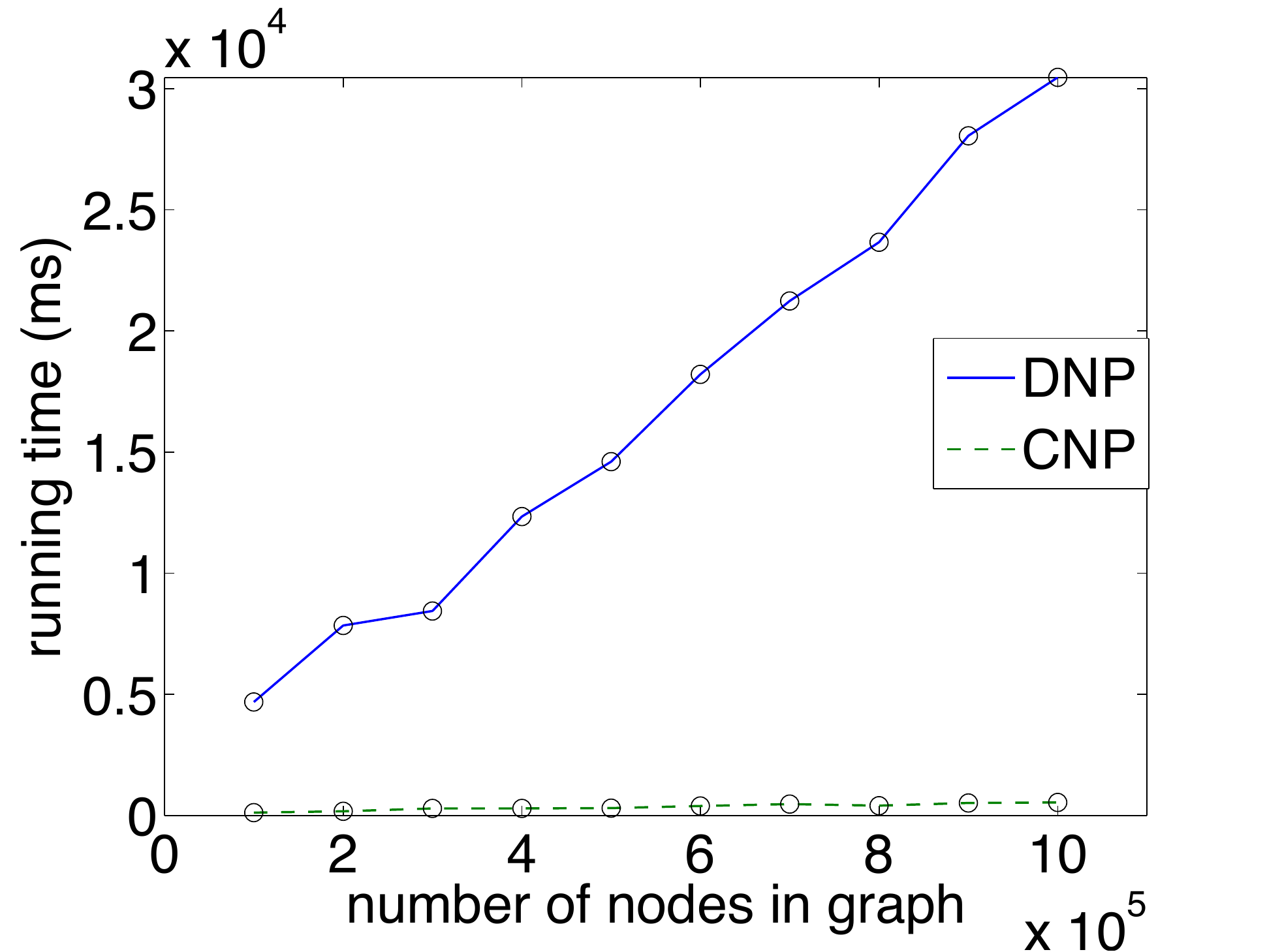}
\caption{Running time on Flixster dataset}
\label{fig:line}
\end{figure}

We also plot the running with different deact time windows to evaluate the influence of deact time window.

\begin{figure}
\centering
\includegraphics[width=\figwidth]{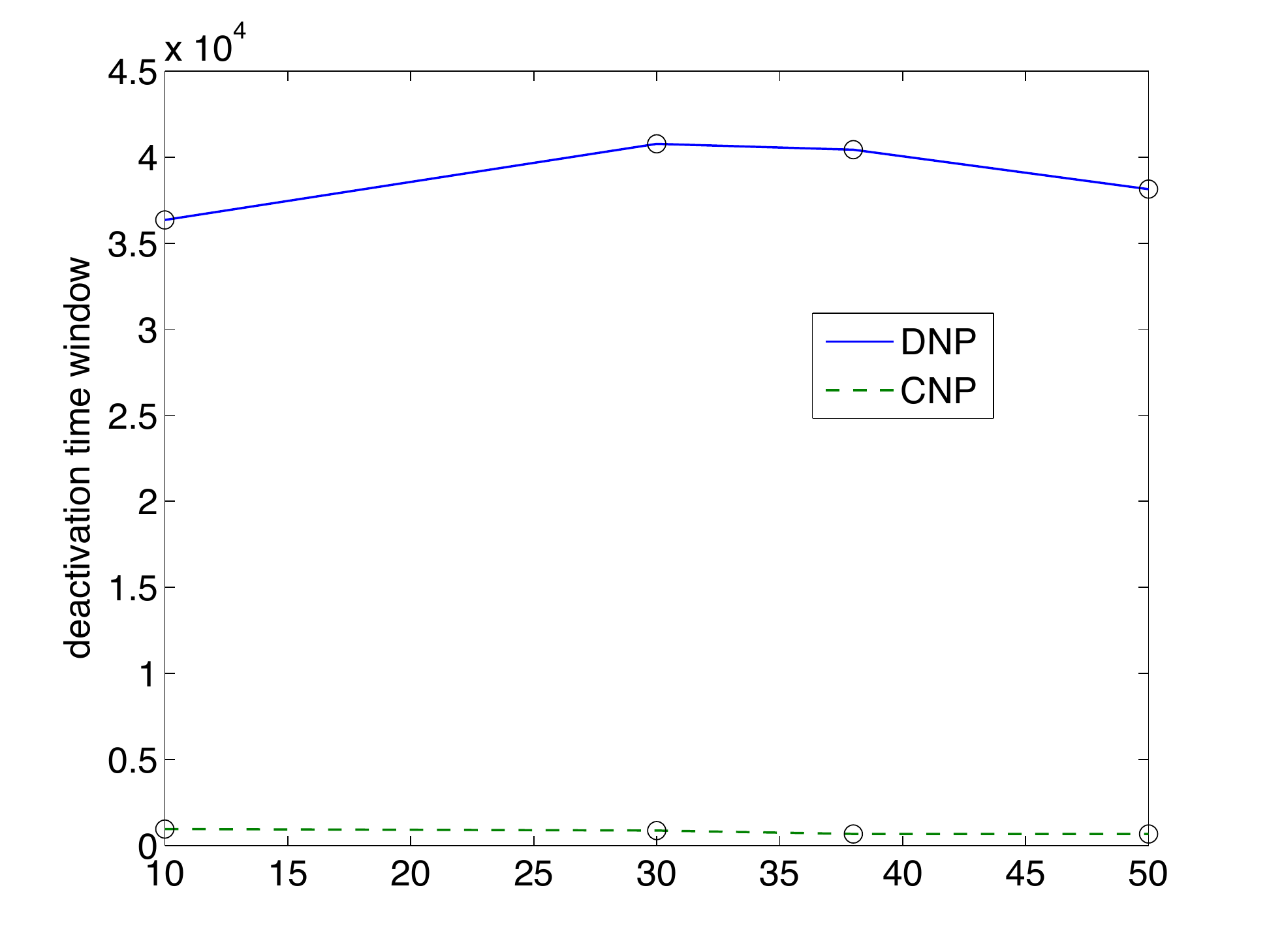}
\caption{Running time on Flixster dataset}
\label{fig:line}
\end{figure}
}
 
}

\makeatletter{}\section{Conclusions}
\label{sec:concl}
There are applications where the propagation phenomena are more accurately captured 
using non-progressive models. In their seminal paper, Kempe et al. \cite{kempe03} proposed a 
non-progressive LT model and showed that over any finite time horizon of interest, its 
behavior can be effectively simulated by a progressive model with the given 
social graph replicated at every timestamp in the horizon. 
Inspired by this, we proposed a non-progressive model and showed that its behavior over a 
time horizon can be simulated without any need for graph 
replication. The resulting discrete time non-progressive model is still not scalable owing to the 
prohibitive number of samplings necessary in order to monitor the state of nodes at 
every time. We proposed an alternative continuous time non-progressive model and showed that 
it permits a highly efficient implementation. 
\techrep{We developed an efficient sampling strategy to further improve the efficiency of our continuous 
time model.}{}In place of expected number of active nodes, for our continuous time model, we 
motivated the expected total amount of time the nodes in the network are active, as the right 
notion of spread, which a seed selection algorithm should optimize. We showed that this 
objective function is monotone and submodular in the set of seed nodes. 
By extensive experiments on two data sets, we show that our model significantly outperforms the state of the art progressive model ConTinEst \cite{Du13} both 
on accuracy of spread estimating and on running time. 
It would be interesting to study non-progressive continuous time models in the competitive setting, 
where competitors may be adversarial.

\techrep{
\makeatletter{}
\appendix
{\bf Proof of Lemma~\ref{additivity}.}
We now show that in the construction of
possible worlds, lemma~\ref{additivity} holds for any two spreads. 
Given a possible world $x$, a seed set $A$, the influence propagation is 
deterministic. That is, each activation and deactivation event has a timestamp 
associated with it, and the propagation can be viewed as a sequence of 
activations and deactivations. In this deterministic setting, we can use 
mathematical induction to prove this lemma.

First, we consider the base case at time zero,  $S_x(A ,0) = A$.
Then, $S_x(A \cup B ,0) = A \cup B = S_x(A,0) \cup S_x(B,0)$. Hence, 
the property holds for this base case. 

Next, we find the next smallest  number in all the schedules. Let 
$t$ be the timestamp when the next event happens. Say, 
$S_x(A \cup B ,t- \delta)  = S_x(A,t-\delta) \cup  S_x(B,t-\delta)$ holds,
where $\delta$ is a very small number. Then, at timestamp $t$, an event  happens. 
The event can either be an activation or a deactivation. 
We consider each of the 10 possible scenarios, the first 7 
correspond to activations and the next 3 to deactivations. 
For any edge $v \rightarrow u$ , in the possible world $x$, we
sampled an activation at timestamp $t$, then,

\spara{Case 1} when $u$ is in both $S_x(A,t-\delta)$ and  
$S_x(B,t-\delta)$, there is no activation in the possible world x.

\spara{Case 2} when $v$ is in both $S_x(A,t-\delta)$ and $S_x(B,t-\delta)$, 
i.e., $v$ is in $S_x(A\cup B,t-\delta)$, but  
$u$ is in neither of them, then after timestamp $t$, $u$ is in $S_x(A,t), S_x(B,t)$
and $S_x(A\cup B,t)$. Thus,
$ S_x(A \cup B ,t) = S_x(A,t) \cup S_x(B,t) $ after the event.

\spara{Case 3}  $v$ is in both $S_x(A,t-\delta)$ and $S_x(B,t-\delta)$, 
however, $u$ is in only one of them, without loss of generality, say $u$ is in 
$S_x(A,t-\delta)$.
Since $S_x(A \cup B ,t- \delta)  = S_x(A,t-\delta) \cup S_x(B,t-\delta) $, 
$v$ and $u$ are in $S_x(A \cup B ,t- \delta)$.
After the event,  $u$ is in  $S_x(A,t), S_x(B,t)$
and $S_x(A\cup B,t)$, therfore, 
$S_x(A \cup B ,t) = S_x(A,t) \cup S_x(B,t)$. 
\spara{Case 4}  when $v$ is in one of $S_x(A,t-\delta)$ and  
$S_x(B,t-\delta)$, without loss of generality, say $v$ is
in $S_x(A,t-\delta)$ ,  however, $u$ is in neither of them. 
Now, $v$ is in $S_x(A \cup B ,t- \delta)$ and after the event, $u$ is in $S_x(A,t)$
and $S_x(A \cup B ,t)$. Then after the event, $S_x(A \cup B ,t) = S_x(A,t) \cup S_x(B,t) $.

\spara{Case 5} when both $v$ and $u$ are in $S_x(A,t-\delta)$ but not in 
$S_x(B,t-\delta)$. 
Then, both $v$ and $u$ are also in $S_x(A\cup B,t-\delta)$.
The event does not result in any changes.

\spara{Case 6} when $v$ is in $S_x(A,t-\delta)$ but not in  $S_x(B,t-\delta)$, 
and $u$ is in $S_x(B,t-\delta)$ but not in $S_x(A,t-\delta)$, i.e., $u$ is in $S_x(A\cup B,t-\delta)$.
After the event, $u$ is in $S_x(A,t)$,  $S_x(B,t)$ and  $S_x(A \cup B ,t) $, thus, 
$S_x(A \cup B ,t) = S_x(A,t) \cup S_x(B,t) $. 
\spara{Case 7} when $v$ is in neither of $S_x(A,t-\delta)$ and $S_x(B,t-\delta)$. 
No change happens after the event, and $S_x(A \cup B ,t) = S_x(A,t) \cup S_x(B,t) $ holds.

\spara{Case 8}  when $v$ is in both  $S_x(A,t-\delta)$ and $S_x(B,t-\delta)$.
After the event, $v$ is not in  $S_x(A,t)$, $S_x(B,t)$ and $S_x(B\cup A,t)$,
thus, $S_x(A \cup B ,t) = S_x(A,t) \cup S_x(B,t) $. 
\spara{Case 9} when $v$ is in one of $S_x(A,t-\delta)$ or $S_x(B,t-\delta)$.
After the event, $v$ is not in  $S_x(A,t)$, $S_x(B,t)$ and $S_x(B\cup A,t)$, 
hence, $S_x(A \cup B ,t) = S_x(A,t) \cup S_x(B,t). $ 
\spara{Case 10}  when $v$ is neither in  $S_x(A,t-\delta)$ nor $S_x(B,t-\delta)$.
No change happens after the event.

In each of the above cases, $S_x(A \cup B ,t) = S_x(A,t) \cup S_x(B,t) $  after the event.
By mathematical induction, $S_x(A \cup B ,t) =  S_x(A,t) \cup S_x(B,t) $.
Therefore, the objective function in our model is monotone and submodular. 
\qed

}{}

\bibliographystyle{abbrv}
\bibliography{sigproc}

\end{document}